\documentclass[12pt,tightenlines,eqsecnum,floats,showpacs,nofootinbib,amsmath,amssymb,aps,prd]{revtex4}
\usepackage{graphicx,verbatim}
\usepackage{amsmath}
\usepackage{amsfonts}
\usepackage{amssymb}
\usepackage[colorinlistoftodos]{todonotes}
\usepackage{epstopdf}
\def\Lie{\mathcal{L}}
\def\scri{\mathcal{I}}
\def\scrip{\scri^{+}}
\def\scrim{\scri^{-}}
\def\scrimr{\scri^{\,-}_{\rm Rel}}
\def\scriml{\scri^{\,-}_{\rm Loc}}
\def\scripl{\scri^{\,+}_{\rm Loc}}
\def\scrimz{\scri_{o}^{\,-}}
\def\scripz{\scri_{o}^{\, +}}
\def\omegaz{\mathring\omega}
\def\izl{i^{o}_{\rm Loc}}

\def\qo{\mathring{q}}
\def\qot{\mathring{\t{q}}}
\def\vo{\mathring{v}}
\def\lo{\mathring{\ell}}
\def\no{\mathring{n}}

\def\Dot{\mathring{\tilde{D}}}

\def\CLi{\mathcal{C}_{\rm isol}^{\Lambda}}
\def\Mr{M_{\rm Rel}}
\def\Ml{M_{\rm Loc}}
\def\Mo{M_{o}}
\def\Gcov{\Gamma_{\rm Cov}}

\def\V{V}
\def\U{U}
\def\rmd{\mathrm{d}}
\def\dso{\rmd \mathring{s}_{2}}

\def\scriab{{\mathcal{I}}_{\rm Abs}}
\def\scriabt{\tilde{\mathcal{I}}_{\rm Abs}}

\def\S{\mathcal{S}}
\def\T{\mathcal{T}}
\def\Im{{\rm Im}}
\def\Re{{\rm Re}}
\def\IH{\Delta}
\def\l{\ell}

\newcommand{\pb}[1]{\hbox{\lower0.5ex\hbox{${}_{\leftarrow}$}}\kern-1.9ex{#1}}

\def\R{\mathcal{R}}
\def\Diff{\rm Diff}
\def\G{\mathfrak{G}}
\def\LG{\mathfrak{g}}
\def\LS{\mathfrak{s}}
\def\Ver{\mathfrak{V}}
\def\LVer{\mathcal{V}}
\def\Lor{\mathfrak{L}}
\def\LLor{\mathcal{L}}
\def\B{\mathfrak{B}} 

\def\t{\tilde}
\def\h{\hat}
\def\b{\bar}
\def\ub{\underbar}
\def\ul{\underline}
\def\={\hat{=}}
\def\f{\frac}

\def\be{\begin{equation}}
\def\ee{\end{equation}}
\def\ba{\begin{eqnarray}}
\def\ea{\end{eqnarray}}

\def\SO(3){\rm SO(3)}
\def\so(3){\rm so(3)}
\def\SO(4){\rm SO(4)}
\def\so(4){\rm so(4)}
\def\SO(1,4){\rm SO(1,4)}
\def\so(1,4){\rm so(1,4)}
\def\SU(2){\rm SU(2)}

\begin{document}

\title{Asymptotics with a positive cosmological constant:\\ IV.  The `no-incoming radiation' condition} 
\author{Abhay Ashtekar}
\email{ashtekar.gravity@gmail.com} 
\author{Sina Bahrami}
\email{sina.bahrami.igc@gmail.com} \affiliation{Institute for Gravitation and the
Cosmos \& Physics Department, Penn State, University Park, PA 16802,
U.S.A.}

\begin{abstract}

Consider  compact objects --such as neutron star or black hole binaries--  in \emph{full, non-linear} general relativity.  In the case with zero cosmological constant $\Lambda$, the gravitational radiation emitted by such systems is described by the well established, 50+ year old framework due to Bondi, Sachs, Penrose and others.  However,  so far we do not have a satisfactory extension of this framework to include a \emph{positive} cosmological constant --or, more generally, the dark energy responsible for the accelerated expansion of the universe. In particular, we do not yet have an adequate gauge invariant characterization of gravitational waves in this context. As the next step in extending the Bondi et al framework to the $\Lambda >0$ case, in this paper we address the following questions:  How do we impose the `no incoming radiation' condition for such isolated systems in a gauge invariant manner? What is the relevant past boundary where these conditions should be imposed, i.e., what is the \emph{physically relevant} analog of  past null infinity  $\scrimz$ used in the $\Lambda=0$ case? What is the symmetry group at this boundary? How is it related to the Bondi-Metzner-Sachs  (BMS) group? What are the associated conserved charges? What happens in the $\Lambda \to 0$ limit? Do we systematically recover the Bondi-Sachs-Penrose structure at $\scrimz$ of the $\Lambda=0$ theory, or do some differences persist even in the limit?   We will find that while there are many close similarities, there are also some subtle but important differences from the asymptotically flat case.  Interestingly,  to analyze these issues one has to combine conceptual structures and mathematical techniques introduced by Bondi et al with those associated with \emph{quasi-local horizons}. 
The framework introduced in this paper will serve as the point of departure in the construction of the analog(s) of future null infinity, $\scripz$ where the radiation emitted by isolated systems can be analyzed systematically.

\end{abstract}

\maketitle

\section{Introduction}
\label{s1}

This is a continuation of a series of papers aimed at constructing the theory of gravitational radiation  emitted by isolated systems in full, non-linear general relativity with a positive cosmological constant $\Lambda$. The first paper in the series \cite{abk1} pointed out that there are unforeseen --and rather deep--  conceptual obstructions that prevent a direct generalization of the well developed $\Lambda=0$ theory due to Bondi \cite{bondi}, Sachs \cite{sachs}, Penrose \cite{rp1} and others. From a physical perspective these difficulties can be traced back to the fact that, if $\Lambda>0$, space-time curvature does not decay no matter how far one recedes from sources,  and its presence in the asymptotic region makes it difficult to extract gravitational waves in a gauge invariant manner.  From a geometrical perspective, in the $\Lambda=0$ case $\scri^{\pm}_{o}$ are null, and using their null normals one can extract radiation fields unambiguously.  By contrast, in the $\Lambda >0$ case, $\scri^{\pm}$ is space-like and, in absence of preferred null directions,  the notion of the radiation field becomes ambiguous \cite{rp1,kp,rp2}. Although we have formulated the discussion in terms of a positive $\Lambda$, the conceptual and technical issues that are relevant to this series of paper also arise if the observed accelerated expansion of the universe is because of another form of dark energy, so long as that the accelerated expansion continues indefinitely.

The subsequent two papers \cite{abk2,abk3} showed that these obstructions can be overcome for linearized gravitational waves on a de Sitter background, although subtleties still persist. For example, because all Killing fields in the de Sitter space-time are space-like near its boundaries $\scri^{\pm}$, the conserved de Sitter `energy' carried away by gravitational --or even electromagnetic waves-- across $\scri^{\pm}$ can be arbitrarily negative. Can time dependent isolated systems then emit large amounts of negative energy (thereby increasing their own energy by large amounts)? A natural setup to analyze such issues is provided by a time changing mass quadrupole, studied by Einstein over a century ago, using the first post-Minkowski, post-Newtonian approximation \cite{ae}.  In presence of a positive $\Lambda$, one can analyze the same problem using the first post-deSitter, post-Newtonian approximation. However, one immediately faces a number of non-trivial conceptual issues and technical difficulties in extending Einstein's  quadrupole formula \cite{abk3}. Fortunately, by now these issues have been resolved. One finds that a time changing quadrupole moment can only create gravitational waves with positive energy! Thus,  although a neighborhood of $\scrip$ does admit solutions to linearized Einstein's equations with negative energy, those waves cannot be produced by \emph{physical sources} \cite{abklett,abk3}. Thus, at least at the linearized level a careful analysis enables us to extend the $\Lambda=0$ theory to allow a positive $\Lambda$ and the extension leads to physically desirable results, just as one would hope.

Are there any observable consequences of this weak field analysis? Einstein's quadrupole formula does receive corrections that depend on $\Lambda$.  As one would expect, they go as powers of $T_{\rm dyn}/T_{\rm H}$, where $T_{\rm dyn}$ is the dynamical time scale associated with compact binaries and $T_{\rm H} = \sqrt{3/\Lambda}$ is the Hubble time scale of the background de Sitter space-time. $T_{\rm dyn}$ associated with compact binaries of interest to the current gravitational wave observatories is at most a few minutes.  The value of the Hubble parameter in our universe changes with time and the current value of $T_{\rm H}^{\,0}$ is huge.  For a rough estimate of the size of corrections, one could choose as our background the de Sitter space-time whose $T_{\rm H}$ equals $T_{\rm H}^{\,0}$. \hskip-0.25cm
\footnote{However, from the linearized analysis it is not clear whether this strategy is justified; the value of $H$ at the time of emission may be more appropriate \cite{bbthesis}. Then the corrections would be more significant, especially for the super-massive black holes created early in the history of the universe.} \hskip-0.1cm
Then the corrections  to Einstein's quadrupole formula are completely negligible for the LIGO-Virgo detectors. However,  this is now a \emph{conclusion of a systematic analysis} rather than assumption. Furthermore, the modifications are conceptually important as they bring out features of  general relativistic gravity that had remained unnoticed in the asymptotically flat case. (For a summary, see \cite{abklett}.)  In this sense, the overall situation is not dissimilar to what Einstein encountered with his quadrupole formula. At the time, his result was only of conceptual importance because it brought out a deep underlying contrast between general relativity and Newtonian gravity, although the result had no practical importance at all because of the then technological limitations.\vskip0.2cm

In this paper we will begin the analysis of gravitational waves emitted by isolated systems in \emph{full, nonlinear} general relativity with $\Lambda >0$,  using the experience and intuition gained from the weak field analysis. Specifically, we will introduce the analog of the past boundary $\scrimz$ of asymptotically flat space-times, now tailored to the study of isolated system such as oscillating stars or compact binaries that constitute interesting sources of gravitational radiation.  

The central issue we resolve is the following. For these isolated systems, one is interested in gravitational waves produced by sources themselves, \emph{not the ones that are incident from past infinity}. In the $\Lambda=0$, asymptotically flat case, the required `no incoming radiation' condition can be imposed in a gauge invariant fashion simply by requiring the vanishing of the Bondi news tensor $N_{ab}$ at $\scrimz$ \cite{bondi,sachs,rp1,aa-yau}. However, in the $\Lambda>0$ case, we do not yet have an unambiguous analog of $N_{ab}$. 
Therefore, one has to find other geometric structures that capture the `no incoming radiation condition' in a gauge invariant manner.  In the mathematical literature, there are powerful results on nonlinear stability of de Sitter space \cite{hf}. Can we not use them to introduce the notions needed to impose this condition? Unfortunately we cannot, at least not directly. Indeed, even in the asymptotically flat case with $\Lambda=0$, the mathematically powerful results on non-linear stability of Minkowski   space-time \cite{dcsk,pced,lb} do not by themselves provide us with  criteria  to characterize gravitational radiation, or to calculate energy-momentum carried by gravitational waves; these came from the independent and older Bondi-Sachs-Penrose framework. The non-linear stability results  do provide us confidence that the boundary conditions are satisfied by a large class solutions to Einstein's equations. However, there are important limitations even in this respect. First, in  both $\Lambda>0$ and $\Lambda=0$ cases, the primary focus of non-linear stability analyses  is on \emph{vacuum} (or electro-vac \cite{nz}) solutions to Einstein equations while in physical applications we are interested  in the radiation \emph{emitted by} compact astrophysical objects. More specifically, in the $\Lambda=0$ case the physical interest lies in retarded solutions in which there is no incoming radiation  --i.e., where $N_{ab}=0$ at $\scrimz$-- and, among solutions considered in the non-linear stability analysis, only Minkowski space meets this requirement. In the $\Lambda>0$ case there is a further twist.  The global, non-linear stability results for de Sitter space-time assume that the topology of $\scri^{\pm}$ is $\mathbb{S}^{3}$ and compactness of $\scri^{\pm}$ plays an important role in the analysis \cite{hf,hr}. As discussed in \cite{abk1}, for isolated systems such as black holes and oscillating stars, $\scri^{\pm}$ are \emph{non-compact}, with topology  $\mathbb{S}^{2}\times \mathbb{R}$, and the analysis becomes more complicated.  Together, these considerations bring out  the need to go beyond the conceptual setting and and mathematical tools provided by the non-linear stability analysis. 

Our goal is to carry out  this task. In this paper, we will formulate the ``no incoming radiation'' condition as the first step in the analysis of gravitational waves emitted by spatially compact sources, and discuss the associated geometrical structures and their physical content. Interestingly, the generalization of the Bondi et al framework requires us to combine physical concepts and mathematical techniques they introduced \cite{bondi,sachs,rp1} with those from the theory of quasi-local horizons developed \cite{afk,abl1,aabkrev} some 40 years later!

 In section \ref{s2} we introduce the appropriate past boundary on which the ``no incoming radiation'' boundary condition is to be imposed. We will refer to it as the \emph{`relevant scri-minus'} and denote it by $\scrimr$. In the $\Lambda=0$ case, explicit examples are useful in bringing out the motivation for various conditions imposed at $\scrimz$,  and understanding the physics and geometry of structures that emerge from them. In the same spirit, in section \ref{s3} we discuss two basic examples of isolated systems in presence of a positive $\Lambda$. (A third example is discussed in Appendix \ref{a1}).  In section \ref{s4} we discuss symmetry groups and in section \ref{s5} the associated charges that lead to definitions of total energy and angular momentum on $\scrimr$. In both cases we compare and contrast the structures with those at $\scrimz$ of the $\Lambda=0$ asymptotically flat space-times. In particular, we will find that, to begin with,  symmetry group at $\scrimr$ is infinite dimensional, with structure similar to that of the Bondi-Metzner-Sachs (BMS) group $\B$. However, addition of a physically motivated structure reduces it to a finite dimensional group, that then enables one to introduce the notion of energy and angular momentum.%
\footnote{A similar finite dimensional reduction of $\B$ occurs if one uses the no incoming radiation condition to introduce a family of `good cuts' on $\scrimz$ as additional structure; $\B$ reduces to the Poincar\'e group \cite{np,aa-radmodes}.\label{fn2}}
In section \ref{s6} we summarize our results and comment on {how the presence of a positive cosmological constant (or, more generally, continued accelerated expansion) forces us to change our intuition in several respects.  Appendix \ref{a2} collects results that are secondary to the main discussion of this paper but which may well  be useful for future work. }

Our conventions are as follows. Throughout we assume that the underlying space-time is 4-dimensional and the space-time metric has signature -,+,+,+.  Curvature tensors are defined via:  $2\nabla_{[a}\nabla_{b]} k_c = R_{abc}{}^d k_d$, $R_{ac} = R_{abc}{}^b$. Relation to the relevant Newman Penrose curvature components is presented in Appendix \ref{a2}. 


\section{$\scrimr$ and the no incoming radiation condition}
\label{s2}

In section \ref{s2.1} we recall from \cite{abk1,abk3} that, because of cosmological horizons, any given isolated system is visible only from a part of the full asymptotically de Sitter space-time. 
In terms of causal structure, then, this is the relevant region of space-time for the given isolated system. We will denote it by $\Mr$. The cosmological horizon that constitutes the past boundary of $\Mr$ is now the analog of  $\scrimz$ in the $\Lambda=0$ case. Therefore, we will refer to this horizon as the `relevant scri-minus' and denote it by $\scrimr$.  It is a \emph{null} 3-manifold  just as $\scrim$ is in the $\Lambda=0$ case (see Fig. \hskip-0.1cm \ref{ds-lin}).   We will see that the `no incoming radiation' condition can now be naturally imposed  by requiring that $\scrimr$ be a \emph{non-expanding horizon (NEH)}. In section \ref{s2.2}, we first recall the notion of a non-expanding horizon \cite{afk} and summarize its properties  that we will need. The older work on NEHs  (see, e.g., \cite{abl1,afk} ) was focused primarily on black holes. \emph{New issues arise} while exploring their role as  past boundaries $\scrimr$ of isolated systems in presence of a positive $\Lambda$. In subsequent sections we will find that now the relevant geometrical structures are closer to those at $\scrim$ in the $\Lambda=0$ case.  In section \ref{s2.3} we specify the class of space-times we consider in the rest of the paper.

\subsection{The Setting}
\label{s2.1}

\begin{figure}[]
  \begin{center}
  \vskip-0.4cm
    \includegraphics[width=2.0in,height=2.0in,angle=0]{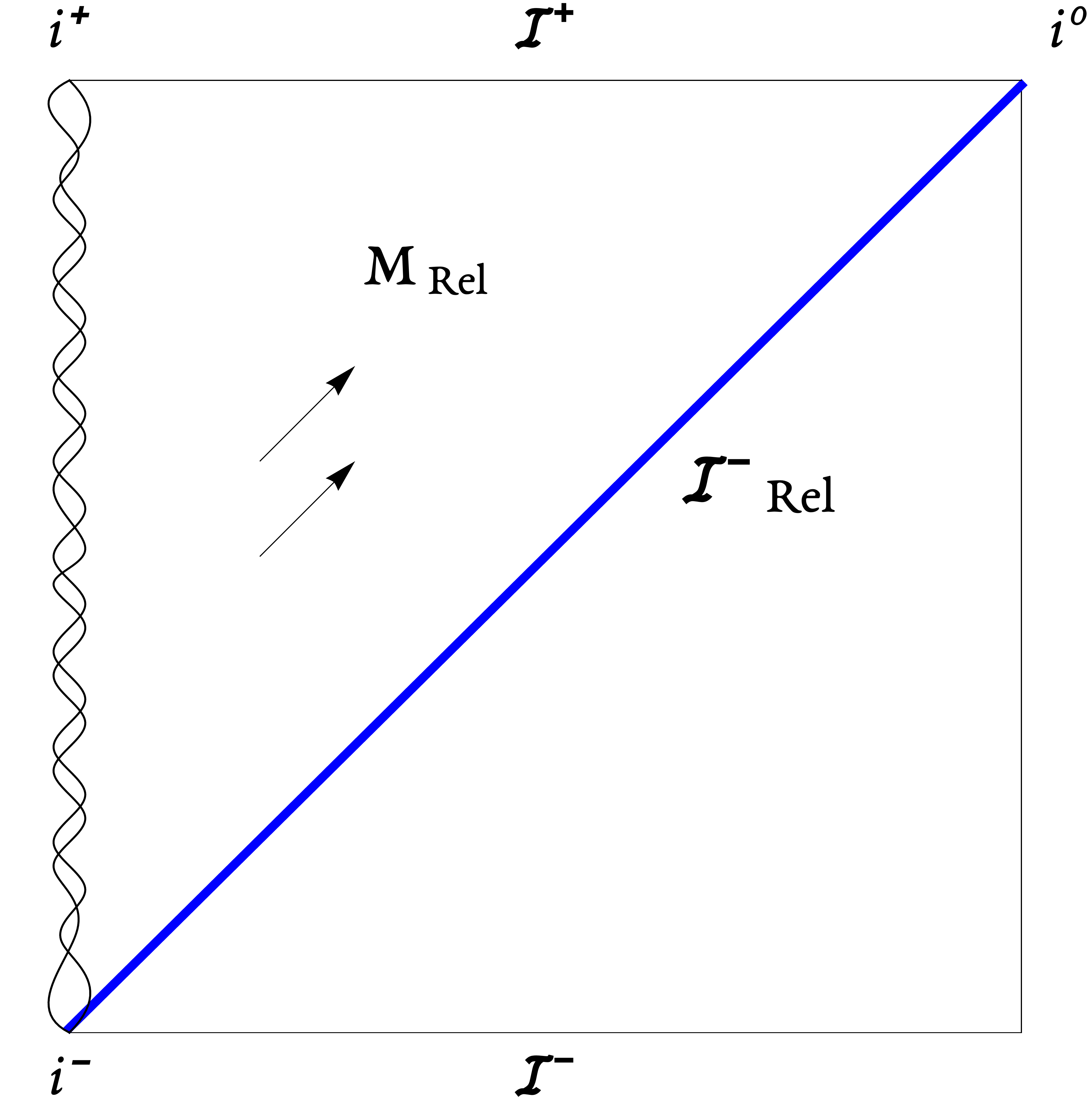} \hskip2cm   \includegraphics[width=2.2in,height=2.0in,angle=0]{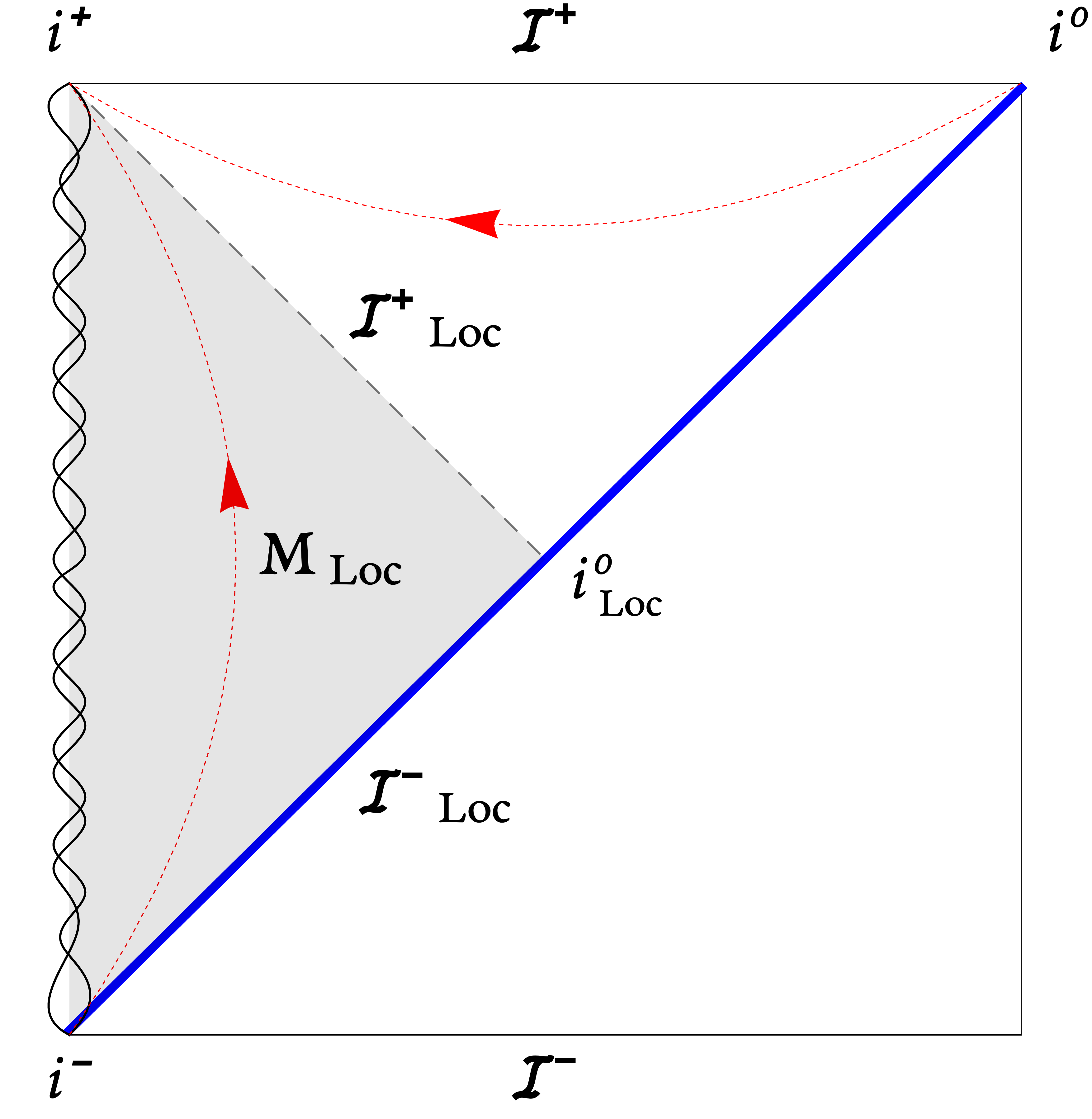}
\caption{{\footnotesize{A linearized compact binary on de Sitter background. The binary is depicted by intertwined lines on the left edge of the figure. It pierces space-like $\scri^{\pm}$ of de Sitter space-time at points $i^{\pm}$.  Solid (black) arrows denote the emitted radiation. \\ 
\emph{Left Panel:} The thick (blue) diagonal line represents the future event horizon $E^{+} (i^{-}) $ of $i^{-}$. Observers whose world-lines are confined to the portion of space-time to the past of $E^{+}(i^{-})$ --i.e. to the past Poincar\'e  patch-- cannot see the source, nor the radiation it emits. Therefore in the investigation of the isolated system, the relevant part $\Mr$  of space-time is only the future Poincar\'e patch. It's past boundary, denoted in the figure by $\scrimr$ serves as the \emph{relevant} $\scrim$. \\
\emph{Right Panel:}  For the future boundary, there are two choices: (i) space-like $\scrip$; or, ii) \emph{local} $\scrip$, the portion of the past event horizon $E^{-}(i^{+})$ of $i^{+}$ that lies in $\Mr$, denoted in the figure by $\scripl$. It intersects $\scrimr$ in a (bifurcation) 2-sphere, denoted  by $\izl$. The (red) dashed lines with arrows represent integral curves of a de Sitter `time-translation' Killing field adapted to the center of mass of the linearized source. It is time-like near the source but space-like near $\scrip$.}}}
\label{ds-lin}
\end{center}
\end{figure}

Let us begin with a linearized source (such as a star or a compact binary with time changing quadrupole) on de Sitter background, depicted in Fig. \hskip-0.1cm \ref{ds-lin}.  $\scri^{\pm}$  of de Sitter space-time are space-like 3-manifolds serving as future and past boundaries, and the world-tube of the spatially compact source intersects them in two points $i^{\pm}$, respectively. The future event horizon $E^{+} (i^{-}) $ of $i^{-}$ divides space-time into two parts, each of which serves as a ``Poincar\'e patch''. The causal domain of influence of the source is the future Poincar\'e patch. Therefore, in the investigation of properties of the radiation emitted by the given isolated system, only this portion of space-time is relevant. It is then natural to regard the past boundary $E^{+} (i^{-}) $ of this region as the  \emph{relevant} scri-minus. We will do so, and from now on denote it  by $\scrimr$.  Since we are interested only in the radiation emitted by the time-changing quadrupole moment of the source,  it is natural to impose the no incoming radiation  boundary condition at $\scrimr$ \cite{abk3,abklett}. (See the left panel of Fig. \hskip-0.1cm \ref{ds-lin}.)

This strategy is reenforced by energy considerations. The points $i^{\pm}$ naturally select a de Sitter `time-translation' Killing field $T^{a}$ whose trajectories are depicted (in the right panel of Fig. \hskip-0.1cm \ref{ds-lin}) by the (red) dashed lines with arrows.  The center of mass of the linearized source follows an integral curve of $T^{a}$. This Killing field is time-like near the source but becomes space-like in a neighborhood of $\scrip$. (Indeed, all Killing fields in de Sitter space-time have to be space-like in a neighborhood of $\scrip$ because $\scrip$ itself is space-like and every Killing field must be tangential to it.) As a consequence, in general the flux of energy $E_{(t)}$ associated with $T^{a}$ across $\scrip$ (or, a portion thereof)  can carry either sign. This is true for both gravitational and electromagnetic waves \cite{abk2}. Geometrically, one can pinpoint where positive and negative contributions come from. For definiteness, let us consider electromagnetic waves and consider the triangular region of  the right panel in Fig. \ref{ds-lin}, bounded by the space-like $\scrip$ to the future and two null boundaries to the past: (i)  the portion denoted by $\scripl$ (namely, the future half of the past event horizon $E^{-}(i^{+})$ of $i^{+}$ that intersects $\scrimr$ at a 2-sphere $\izl$), and, (ii)  the portion of $\scrimr$ that lies to the future of $\izl$. Conservation of stress-energy tensor implies that the energy flux across $\scrip$ equals the sum of energy fluxes across the two null boundaries in the past. Note however, that the Killing field $T^{a}$ is future directed and null on the boundary (i) (i.e. $\scripl$),  but \emph{past directed} on the boundary (ii). Therefore in any solution to Maxwell's equations, the energy flux across $\scripl$ is strictly non-negative while that across the other null boundary (ii) is strictly non-positive. Since the energy flux across $\scrip$ is the sum of these two contributions, in general it can be of either sign. However, if we are interested \emph{only} in the retarded solutions created by the source, then there is no incoming radiation across $\scrimr$. Hence for these solutions flux across the second null boundary (ii) vanishes identically, and that across $\scripl$ is positive, making the flux across $\scrip$ positive. Thus, while de Sitter space-time admits solutions to Maxwell's equations with negative energy, these do not result from a physical source \emph{if there is no incoming radiation at $\scrimr$}. (The situation is the same for gravitational waves but the argument requires symplectic geometric methods since we do not have a local, gauge invariant stress-energy tensor \cite{abk2,abk3}.) Thus,  in this example,  imposing no incoming radiation condition at $\scrimr$ has the desired physical consequence.

Explicit geometrical structures in this well-understood \cite{abk3}  example  motivate our general strategy.  Let us now consider isolated systems in the full, non-linear theory in presence of a positive $\Lambda$. These systems are naturally represented by  asymptotically de Sitter space-times where much of the structure we discussed is again available. Indeed, these space-times admit a conformal completion a la Penrose \cite{rp1} with space-like boundaries $\scri^{\pm}$. The spatially compact source would again intersect $\scri^{\pm}$ at points $i^{\pm}$ and, in the study of the isolated system, the relevant portion $\Mr$ of space-time will again lie to the future of $E^{+}(i^{-})$. Therefore, $E^{+}(i^{-})$ will again serve as the relevant $\scrim$ and we will denote it by $\scrimr$ also in the general context. In section \ref{s2.2} we will provide a precise formulation of the no-incoming radiation condition on $\scrimr$. Note that while the past boundary $\scrim$ of the full space-time $M$ is space-like, the past boundary $\scrimr$ of the relevant portion $\Mr$ of space-time is null, just as it is in the $\Lambda=0$ case.

While the focus of this paper will be on $\scrimr$, it is useful to note  structures that will provide the appropriate arena to  investigate properties of radiation emitted by the system.  Although this structure will be heavily used only in subsequent papers, we will discuss it here briefly because it plays a role in our present considerations as well.  In the discussion of outgoing radiation,  one possibility is to use the space-like future boundary $\scrip$ as the arena, as was done in the analysis that  generalized Einstein's quadrupole formula to include a positive $\Lambda$ \cite{abk3}. But there is also another possibility  \cite{aa-ropp}: use a more local, \emph{null} boundary, adapted to the cosmological horizon of the source, obtained as follows. Consider the past event horizon $E^{-}(i^{+})$ of $i^{+}$ and assume%
\footnote{\label{fn-hv} A priori, It is not clear whether in physically interesting radiating space-times $E^{-}(i^{+})$ will be `long enough' to intersect $\scrimr$. But non-linear stability results \cite{hv} for Kerr-de Sitter space-times suggest that there should be a large family of such space-times representing isolated systems in presence of a positive $\Lambda$. \label{fn3}}
that it is long enough to intersect $\scrimr$ in a 2-sphere that we will denote by $\izl$ (see the right panel of Fig. \hskip-0.1cm \ref{ds-lin}).  The intersection between the causal past and the causal future of the isolated system is the shaded triangular region $\Ml$ that is the `local neighborhood of the source' since it is bounded by the past and future event horizons of the world-tube of the source. Thus, $\Ml$ is the intersection of the causal future and the causal past of the isolated system; events in $\Ml$ can influence the system and can also be influenced by it. The required null boundary would  then be the future boundary of $\Ml$ --the portion of $E^{-}(i^{+})$ between $i^{+}$ and $\izl$. We will denote it by $\scripl$ and call it \emph{local scri-plus} (see Fig. \hskip-.1cm\ref{ds-lin}).  Note that $\Ml$ resembles the Penrose diagram of an asymptotically flat space-time containing an isolated system, with the 2-sphere $\izl$ playing the role of spatial infinity. Finally, if the source is spherically symmetric and static, then  the static Killing field $T^a$ is time-like everywhere in $\Ml$ except on the boundaries where it becomes null,  mimicking the behavior of the time-translation Killing fields in Minkowski  (and Schwarzschild) space-time. In section \ref{s3} we will examine the geometry of $\scrimr, \, \scriml,\, \izl,$ and $T^a$ in standard examples to gain further intuition.\\

\emph{Remark:} As mentioned in section \ref{s1},  in de Sitter space-time (without a linearized source), $\scri^{\pm}$ are spatially compact with topology $S^{3}$. This is also the case more generally in asymptotically de Sitter space-times that are usually considered in the geometric analysis literature in the cosmological context  \cite{hr}, because there is no isolated, (uniformly) spatially compact source that pierces $\scri^{\pm}$. Then there are no preferred points $i^{\pm}$ on $\scri^{\pm}$ and hence no $\scrimr, \scripl$ and $\izl$. The situation is then qualitatively different from the one of interest to this series of papers where the focus is on isolated systems in presence of a positive $\Lambda$.

 

\subsection{Non Expanding horizons and their properties}
\label{s2.2}

Since $\scrimr$ is a cosmological horizon, we can readily use the available results on quasi-local horizons  to impose the \emph{no-incoming radiation boundary condition} at $\scrimr$.  The appropriate notion turns out to be that of a non-expanding horizon (NEH). (For reviews on quasi-local horizons, see, e.g., \cite{aabkrev,ib,gj}.)

\textbf{Definition 1} \cite{abl1}:  A 3-dimensional sub-manifold  $\IH$  of space-time is said to be a \textit{non-expanding horizon} if \\
\noindent $i)$  $\IH$ is  diffeomorphic to the product $\tilde\IH \times \mathbb{R}$ where $\tilde\IH$ is a 2-sphere, and the fibers of the projection \, $ \ \tilde\IH \times \mathbb{R} \rightarrow \tilde\IH $\,\,\, are null curves  in $\IH$; \\
\noindent $ii)$ the expansion of any null normal $\l^{a}$ to $\IH$ vanishes; and, \\
\noindent $iii)$ Einstein's equations hold on $\IH$ and the stress-energy tensor $T_{ab}$ is such that $-T^a{}_b \l^b$ is causal and future-directed on $\IH$.\vskip0.1cm
\noindent Note that if these conditions hold for one choice of null normal, they hold for all. Condition $iii)$ is very mild; in particular, it is implied by the (much stronger) dominant energy condition satisfied by the Klein-Gordon, Maxwell, dilaton, Yang-Mills and Higgs fields as well as by perfect fluids. Finally, in view of the bundle structure, will refer to $\t\IH$ as the \emph{base space} and fields on it will carry a tilde. (In the literature on quasi-local horizons, one generally uses a `hat' rather than a `tilde' --we switched to a `tilde' because hats have been used to denote conformal completion in section \ref{s2.1}.)

\vskip0.2cm

Conditions in \emph{Definition 1} have a number of immediate consequences \cite{afk,abl1}. First, the space-time metric $g_{ab}$ induces a natural degenerate metric $q_{ab}$ of signature (0,+,+)  and an area 2-form $\epsilon_{ab}$ on $\IH$, satisfying $\Lie_{\l} q_{ab} =0, \, q_{ab}\l^{b} =0$ and $\Lie_{\l} \epsilon_{ab} =0, \, \epsilon_{ab}\l^{b} =0$ for all null normals $\l^{a}$. 
Thus $q_{ab}$ and $\epsilon_{ab}$ can be regarded as pull-backs to $\IH$ of the metric and the area 2-form on the base space $\t\IH$. In particular, then, the area of any 2-sphere cross-section of $\IH$ is the same. This is a reflection of the fact that there is no flux of energy --matter or radiation-- across $\IH$. Therefore if we ask that $\scrimr$ be an NEH, we would be guaranteed  that there is no incoming radiation into $\Mr$ from $\scrimr$. On a dynamical horizon, by contrast, there are fluxes of matter and/or radiation across the horizon and the area of cross-sections changes in response to these fluxes in a precise, quantitative fashion \cite{aabk1,aabk2}.

The second set of consequences  arises from fields associated with the space-time (torsion-free) connection $\nabla$ that is compatible with $g_{ab}$. The Raychaudhuri equation, together with conditions in \emph{Definition 1} implies that all null normals $\ell^{a}$ are also shear-free.  This property, together with condition $ii)$ implies that $\nabla$  induces a natural intrinsic, torsion-free derivative operator  $D$ on $\IH$ which is compatible with the induced metric $q_{ab}$  on $\IH$: $D_{a} q_{bc} =0$ on $\IH$. Furthermore, given any future-directed null normal $\l^{a}$, we have:
\be D_{a} \l^{b}  = \omega_{a} \l^{b}\, , \ee
for some 1-form $\omega_{a}$  on $\IH$ and, under the rescaling $\l^{a} \to \l^{\prime\,a} = f \l^{a}$ for any smooth positive function $f$ on $\IH$, we have:
\be \omega_{a}  \to  \omega_{a}^{\prime} = \omega_{a} + D_{a} \ln f.  \ee
(Thus, strictly, the 1-form $\omega_{a}$ should also carry a label $\l$ which we will omit  just for notational simplicity.) 
Following the Newman-Penrose notation, let us define $2\,{\Re} \Psi_{2} =  C_{abcd} \l^{a}n^{b}\l^{c} n^{d}$, and  $2\, \Im \Psi_{2} =  {}^{\star}C_{abcd} \l^{a}n^{b}\l^{c}n^{d}$, where $\ell^{a}$ is any null normal to $\IH$ and, given a null normal, $n^{a}$ is any null vector field that satisfies $\l^{a} n_{a} = -1$. (The NEH structure implies that the pull-back to $\IH$ of the space-time field $C_{abcd}\l^{d}$ vanishes, whence $\Psi_{2}$ is well-defined in spite of the freedom in choosing $n^{a}$.) The 1-form $\omega_{a}$ defined intrinsically on $\IH$ serves as a potential for the imaginary part $\Im\, \Psi_{2}$ of the Newman-Penrose component $\Psi_{2}$  of the 4-dimensional Weyl tensor evaluated on  $\IH$:
\be D_{[a} \omega_{b]}  = {\Im} \Psi_{2}\,\, \epsilon_{ab}.  \label{impsi1}\ee
 Since $\Im \Psi_{2} $ determines the angular momentum multipoles of the horizon \cite{aepv}, $\omega_{a}$ is called \emph{the rotational 1-form}. It's component $\kappa_{\l} := \omega_{a}\ell^{a}$ along $\ell^{a}$ is the \emph{surface gravity} associated with the null normal $\l^{a}$.  The real part $\Re\, \Psi_{2}$ of $\Psi_{2}$ determines mass multipole  \cite{aepv}.  Therefore, the field $\Psi_{2}$  plays a key role in characterizing the  geometry of WIHs and extracting their  physics \cite{aabkrev}.  We note  an important identity that relates  $\Re \Psi_{2}$  to the scalar curvature ${}^{2}\b{\R}$ of the metric $\b{q}_{ab}$ on any 2-sphere cross-section $C$ of an NEH:
\be {}^{2}\b{\R} = - 4 \,\Re\Psi_{2} + \f{2}{3} \Lambda +  {8\pi G}\, (2\l^{a}n^{b} T_{ab} + \f{1}{3} T)\, . \label{repsi1}\ee
where $\l^{a}$ is any null normal to the NEH, $n^{a}$ the other null normal to $C$ such that  $g_{ab}\l^{a}n^{b}=-1$, and  $T$ is the trace of the stress energy tensor. Eq. (\ref{repsi1}) is a special case of  a general geometric identity derived in Appendix \ref{a2},  now applied to $\IH$ on which  the shear and expansion vanish for any null normal $\l^{a}$.

Finally we note that surface gravity $\kappa_{\l}$ need not be constant on $\IH$ for a general choice of the null normal $\l^{a}$. However, given an NEH $\IH$,  one can exploit the freedom in the choice of null normals to restrict $\kappa_{\l}$. It turns out that  every NEH admits a sub-family of null normals  $\l^{a}$ such that $\Lie_{\l} \,\omega_{a} =0$ \cite{abl1}.  This condition says that not only is the intrinsic metric $q_{ab}$ of the NEH time-independent but a part of the connection $D$ on the NEH --namely the part that determines its action on these null normals $\l^{a}$-- is also time-independent.  Thanks to the identity 
\be \Lie_{\l} \,\omega_{a} =0\qquad \Leftrightarrow \qquad  D_{a}\kappa_{\l} =0 \ee
that holds on any NEH \cite{afk}, it follows that $\kappa_{\ell}$ is constant, i.e., the zeroth law of horizon dynamics holds for this sub-family. If an NEH $\IH$ is equipped with an equivalence class $[\l^{a}]$ of preferred null normals that  satisfy  $\Lie_{\l} \omega =0$, then the pair $(\IH, [\l^{a}])$ constitutes a \emph{weakly isolated horizon} (WIH); here two null normals are considered equivalent if they are related by  a rescaling with  a positive \emph{constant}.  While the `no  incoming radiation condition' introduced in section \ref{s2.3} refers only to the NEH structure, the WIH structure will play an important role in the subsequent discussion.

Note that if $\l^{a}$ is an affinely parametrized  geodesic vector field, $\kappa_{\l} =0$, and hence in particular a constant, whence $(\IH, [\l^{a}])$ is automatically a WIH.  These WIHs are said to be \emph{extremal}. If $\kappa_{\l} \not=0$, then $(\IH, [\l^{a}])$ is said to be \emph{non-extremal}.  WIHs are of special interest because they turn out to satisfy not only the zeroth law of horizon mechanics but also the first law. The WIH structure will play an important role in sections \ref{s4} and  \ref{s5}.
Specifically we will use three of their properties \cite{abl1}:\\
(i) Every NEH admits a \emph{canonical,  extremal WIH structure} $(\IH, [\lo^{a}])$. (On every extremal WIH, the rotational 1-form $\omega_{a}$ is the pull-back to $\IH$ of a 1-form $\t\omega_{a}$ on the `base-space' $\t\IH$, and on the canonical one, $\t\omega_{a}$ is divergence free on the base space $(\t\IH, \t{q}_{ab})$).\vskip0.1cm
\noindent(ii) An NEH does not admit a canonical non-extremal WIH structure. However, given a geodesically complete NEH, there is a 1-1 correspondence between \emph{non-extremal} WIH structures $(\IH, [\l^{a}])$ on it, and 2-sphere cross-sections $C_{[\l]}$ of $\IH$. The null normals $\l^{a} \in [\l^{a}]$ vanish on $C_{[\l]}$, are future directed to its past,  and past directed to its future.\vskip0.1cm
\noindent(iii) Every  \emph{non-extremal} horizon $(\IH, [\l^{a}])$ admits a \emph{canonical} foliation (such that the pull-back $\b\omega_{a}$ of the rotational 1-form $\omega_{a}$ on $\IH$ to the leaves of this foliation is divergence-free with respect to the 2-metric $\b{q}_{ab}$ on each leaf, pulled back from $\IH$.) In terms of the canonical extremal WIH structure $(\IH, [\lo^{a}])$ on the the underlying NEH, if we set the affine parameter $\vo$ along any $\lo^{a} \in [\lo]$ to a constant value on the preferred cross-section $C_{[\l]}$, then the leaves of the preferred foliation of  $(\IH, [\l^{a}])$ are precisely the $v_{o} ={\rm const}$ cross-sections of $\IH$.\\
 
 \emph{Remarks:}
 
1.  On any WIH $(\IH, [\l^{a}])$ we have $\big(\Lie_{\l} D_{a} - D_{a} \Lie_{\l}\big) \l^{b} =0$ and, as we remarked above, given any NEH,  one can always choose null normals $\l^{a}$ that satisfy this condition.  Thus, one can always pass from an NEH to a WIH simply by restricting oneself to a class of null normals. The restriction is analogous to the one often made on null infinity $\scripz$ where one restricts the null normal $\mathring{n}^{a}$ to be divergence-free to simplify the subsequent mathematical expressions. In both cases, the  restrictions are compatible with symmetries that a space-time may admit. Thus, in the $\Lambda>0$ case, if the space-time admits a Killing field whose restriction to $\IH$ is normal to it, then the normal automatically endows $\IH$ with the structure of a WIH.
 
 2.  It is tempting to strengthen the WIH condition and ask $\big(\Lie_{\l} D_{a} - D_{a} \Lie_{\l}\big)  t^{b} =0$ for all vector fields $t^{a}$ tangential to the $\IH$.  Then $(\IH, [\l^{a}])$ is called \emph{an isolated horizon}. However, an NEH $\IH$ \emph{need not} admit any null normal $\l^{a}$ satisfying this condition. Thus, while one can endow any NEH with a WIH structure `free of charge', one cannot in general endow it with the structure of a IH.  The notion of an IHs turns out to be well suited to describe \emph{black hole} horizons in equilibrium \cite{aabkrev, ib,gj}. By contrast, it  turns out to be too strong to describe  $\scrimr$ if $(\Mr, g_{ab})$  admits radiation. Therefore we have focused on NEHs and WIHs. This point will be discussed in detail in a forthcoming paper on  $\scripl$.
 
 3.  One may be tempted to ask: What about the actual universe we inhabit? Although it will be asymptotically de Sitter in the future (assuming the accelerated expansion continues indefinitely)  it  is not asymptotically de Sitter in the past as we assumed in Definition 
 2.  Note that we started with Penrose's \cite{rp1} conformal completion mainly to anchor the discussion in  familiar constructions. One could start with the physical space-time $(M,\, g_{ab})$ and consider sources whose spatial support is compact \emph{and} uniformly bounded and let $\Mr$ be the causal future of the world-tube of the source, and $\scrimr$ be its past boundary (see  Fig. \ref{ds-lin}).  $\Ml$ would then be the intersection of the causal future and causal past of the world-tube of the source. One could use Penrose's conformal completion just for $(\Mr,\, g_{ab})$, and introduce $\scrip$ and $i^{+}$, and use $i^{+}$ to  define $\scripl$. This construction will go through also for black holes formed by gravitational collapse, and enable us to define also the black hole horizon (see left panel of Fig. \ref{sds}). Thus all reference to $\scrim$ can be eliminated. Indeed, even when $\scrim$ exists, to investigate radiation \emph{emitted by} a given isolated system,  $\scrim$ is not the appropriate arena to specify the `no incoming radiation' condition; the appropriate arena is $\scrimr$.  For example, in the Penrose diagrams depicted in Fig. \ref{ds-lin}, there could be additional isolated sources in the past Poincar\'e patch --e.g., at the antipodal location depicted by the right vertical line-- in addition to the one of interest (depicted in the figure). In this case, even if we were to impose the `no incoming radiation condition' at $\scrim$,  radiation in the upper Poincar\'e patch would be an admixture of that emitted by the source of interest and that emitted by the other source that is not of interest. This problem is neatly bypassed by imposing the `no incoming radiation' condition at $\scrimr$, without having to know what is happening at $\scrim$.
  
 4.  Finally, note that the notion of an isolated system is an idealization that has been very useful in many areas of physics. In the $\Lambda=0$ case, space-time is just assumed to be asymptotically flat --one does not worry about the fact that real stars and black holes are produced at a finite time in the real universe.  Since there is now strong observational evidence that $\Lambda$ is positive, it is natural and meaningful to ask for a generalization of the $\Lambda=0$ framework to the $\Lambda>0$ case  --i.e. to use Einstein's equations $G_{ab} + \Lambda g_{ab} = 8\pi G_{N} \, T_{ab}$ with $\Lambda>0$-- while retaining the idealization of an isolated system, and therefore not worrying about the fact that real stars and black holes are produced at a finite time in the real universe. That is,  in this idealization $i^{-}$ denotes the birth of the star or the compact binary system, just as it does in the $\Lambda=0$ case.  Similarly, the `center of mass of the isolated system' is a loose physical term and we can just consider instead the world tube representing the system. 
  
 \subsection{ Past boundary conditions on the relevant part $\Mr$ of space-time}
 \label{s2.3}
 
 The strategy developed in section \ref{s2.1} and the structure available on non-expanding horizons summarized in section \ref{s2.2} now lead us to specify the class of space-times we will consider.  Let us first recall from \cite{abk1} the notion of asymptotically Schwarzschild-de Sitter space-times.\\
 \textbf{Definition 2}: A space-time is said to be \emph{asymptotically Schwarzschild-de Sitter} if 
there exists a manifold $\hat{M}$ with a future boundary $\scri^{+}$ and a past boundary $\scri^{-}$, equipped with a metric $\h{g}_{ab}$, and a diffeomorphism from $M$ onto the interior  $(\hat{M}\, \setminus\, \scri^{+} \cup \scri^{-})$ of $\hat{M}$ such that: 
\noindent  $(i)$\, there exists a smooth function $\Omega$ on $\hat{M}$ such that $\hat{g}_{ab}=\Omega^2 {g}_{ab}$ on ${M}$; $\Omega=0$ on $\scri^{\pm}$;\\ 
\indent\indent and $n_a := \nabla_a \Omega$ is nowhere vanishing on $\scri^{\pm}$;\, 

\noindent $(ii)$\,${g}_{ab}$ satisfies Einstein's equations with a positive cosmological constant,\\ \indent\indent i.e., ${R}_{ab} - \frac{1}{2} {R} {g}_{ab} + \Lambda {g}_{ab} = 8 \pi G \; {T}_{ab}$ with $\Lambda >0$; where $\Omega^{-1} {T}_{ab}$ has a smooth limit to \indent\indent  $\scri^{\pm}$; and,

\noindent $(iii)$ $\scri$ has topology $\mathbb{S}^2 \times \mathbb{R}$, and the vector field $n^{a}$ is complete in any divergence-free conformal frame (i.e., when the conformal factor  $\Omega$ is chosen to satisfy $\h{\nabla}^{a} \h{\nabla}_{a} \, \Omega =0$ at $\scri^{\pm}$).\vskip0.2cm

These conditions are appropriate for considering space-times representing isolated systems  in presence of a positive $\Lambda$ (assuming sources have spatially compact support that is uniformly bounded in time). Now,  since $\mathbb{S}^{2} \times \mathbb{R} = \mathbb{S}^{3}  \setminus  \{p_{1}, p_{2}\}$ (where $p_{1}, p_{2}$ are 2 points),  we can think of $\scri^{\pm}$ as being obtained from the de Sitter $\scri^{\pm}$ (with $\mathbb{S}^{3}$ topology) by removing points $i^{\pm}$ representing the future and past time-like infinity defined by the source, and  points $i^{o}$ that can be thought of spatial infinity (see Fig.\ref{ds-lin}). Discussion of  section \ref{s2.1}  leads to the next definition:\vskip0.1cm
\textbf{Definition 3}: The \emph{physically relevant portion}  $\Mr$ of the given space-time $(M, g_{ab})$ is that which lies to the future of the future horizon $E^{+}(i^{-})$ of the point representing the past time-like infinity $i^{-}$ of the isolated source.\vskip0.2cm

Being the past boundary of the physically relevant portion $\Mr$, \, $E^{+}(i^{-})$ can be taken as $\scrimr$, the ``relevant scri-minus.''  Finally, we impose the ``no incoming radiation'' boundary condition on $\scrimr$:\vskip0.1cm
\textbf{Definition 4}: We will say that the given space-time satisfies the \emph{no incoming radiation} condition if:\\
$(i)$ $\scrimr$ is a non-expanding horizon; and,\\
$(ii)$ It is geodesically complete. 
\vskip0.1cm
\noindent The geodesic completeness requirement can be rephrased as asking that the extremal null normals $\l^{a}$ (i.e., with $\kappa_{\l} =0$) are complete. If one extremal null normal is complete then they are all complete. This completeness requirement is completely analogous to the condition one imposes on $\scri^{-}$ of  the asymptotically Minkowski space-times in the $\Lambda=0$) case (see, e.g., \cite{aa-yau}). \vskip0.2cm 

In the rest of the paper we will work with asymptotically Schwarzschild-de Sitter space-times with no incoming radiation.  This is the class of space-times  representing  isolated  gravitational systems in presence of a positive cosmological constant and it will be denoted by $\CLi$.\\

\emph{Remark:} It is interesting to note the situation in the asymptotically flat, $\Lambda=0$ case. Let us again denote the physical space-time by $(M, g_{ab})$, the conformally completed space-time by $(\h{M},\h{g}_{ab})$, and work with a divergence-free conformal frame that is normally used to analyze structure at null infinity, which we will denote by  $\scri_{o}$.  Suppose the space-time is asymptotically Minkowskian \cite{aa-yau}. Then, interestingly, $\scri^{-}_{o}$ is null, geodesically complete and a non-expanding horizon in $(\h{M}, \h{g}_{ab})$; its structure closely resembles that of $\scrimr$ in the $\Lambda >0$.  However, there is a key difference: whereas $\scrimr$ is a sub-manifold of the \emph{physical} space-time, in the $\Lambda=0$ case $\scri^{-}_{o}$ is the boundary of the physical space-time (at which the physical metric $g_{ab}$ diverges). As a consequence, presence or absence of gravitational radiation is \emph{not} encoded in the NEH structure of $\scri^{-}_{o}$. To ensure that there is no radiation at $\scri^{-}$, one has to require, \emph{in addition}, that the Bondi news tensor $N_{ab}$ must vanish there. In the $\Lambda>0$ case,  by contrast, the NEH structure implies that the shear tensor $\sigma_{ab}$ of every null normal $\l^{a}$ vanishes on $\scrimr$ and this vanishing suffices to ensure that there is no flux of radiation across $\scrimr$.

\section{Examples}
\label{s3}

In this section we will examine the simplest examples of isolated systems in general relativity with a positive cosmological constant to illustrate the geometrical structures one can anticipate.  (Another example is discussed in Appendix \ref{a1}.) We will see explicitly that all conditions in our definitions are satisfied in these examples. Furthermore, the explicit form of the geometrical structures of these examples --such as Killing vectors, curvature quantities and their behavior in the $\Lambda \to 0$ limit-- will provide the much needed intuition in the discussion of the symmetry groups, physical fields and conserved charges at $\scrimr$ of general space-times. Although we have attempted to restrict ourselves to the most essential points, the discussion is rather long because the presence of a positive $\Lambda$ introduces certain unfamiliar structures that turn out to be important in the subsequent discussion. 

Throughout this paper, we use the symbol $\l$ in two different ways: $\l$ will stand for the cosmological radius $\sqrt{3/\Lambda}$, while $\l^{a}$ will denote null normals to $\scrimr$. 

\subsection{Linearized gravity with sources in the de Sitter space-time}
\label{s3.1}

In this subsection we will analyze the geometry and symmetries of the \emph{future Poincar\'e patch},  $\Mr$, of the space-time depicted in Fig. \hskip-0.2cm\ref{ds-lin}.  We begin by listing the five coordinate systems in which the background de Sitter metric is commonly displayed because they are useful to bring out various geometrical features that we will need.  The coordinates themselves are not important and will not play an essential role in the subsequent discussion; only the invariant structures they define in these examples will. These include existence of $\scrimr$ and the WIH structure thereon; $\scriml$, $\izl$ and $\scripl$; and the relation between physically interesting conserved quantities and these structures;  see the last part of the discussion in each example.

\subsubsection{Various forms of the metric}

(1) Standard cosmological coordinates $t, x, y, z$:\\
\be \rmd s^{2} = - \rmd t^{2} + a^{2}(t) \,( \rmd x^{2} + \rmd y^{2} +\rmd z^{2} ), \quad {\rm with} \quad a(t) = e^{t/\l}\, .  \label{cosmological}\ee 
$\scrimr$ corresponds to $r=\infty$ and $ t =-\infty$  (with $r^{2} := x^{2} +y^{2}+ z^{2}$), and $\scri^{+}$ to $t=\infty$. So the chart does not cover either. \\

(2) Conformal time $\eta$, and spherical coordinates $r,\theta, \varphi$, again on the cosmological slices:  \\
\be \rmd s^{2} = \t{a}^{2}(\eta)\, (\rmd \eta^{2} +  \rmd r^{2} + r^{2}\,  \dso^{2}), \quad {\rm with} \quad  \eta = - \l\, e^{-t/\l},\,\, \,  \t{a}(\eta) = - \f{\ell}{\eta} \equiv a(t)\, , \label{conformal}\ee
and where $\dso^{2}$ stands for the unit 2-sphere metric. In this chart, $\scrimr$ corresponds to $r= -\eta =\infty$, and $\scri^{+}$ to $\eta=0$.  Therefore, again $\scrimr$ and $\scrip$ are not covered by this chart.\\

(3) Static coordinates $T,R, \theta,\varphi$, in which the $\vec{T} \equiv T^a\partial_{a} := \partial/\partial T$ is manifestly a static Killing field and $R$ is the proper radius of 2-spheres of symmetry:\\
\be \rmd s^{2} = -f(R) \rmd T^{2} + \f{\rmd R^{2}}{f(R)} + R^{2}\,\dso^{2}, \quad {\rm with} \quad f(R) = 1-\f{R^{2}}{\ell^{2}}.  \label{static1}\ee
This chart covers the lower half of the Poincar\'e patch (i.e. the portion that lies to the past of $\scripl$) excluding the boundaries $\scrimr$ and $\scripl$ where $f(R)$ vanishes (see Fig. \hskip-0.1cm \ref{ds-lin}). In terms of $\eta, r$ of (2), we have:
\ba   & R = -\, \f{\l}{\eta} \, r    \qquad    & T = -\f{\l}{2}\, \ln \big(\f{\eta^{2} - r^{2}}{\l^{2}}\big)\nonumber\\
& r = \f{e^{-T/\l}\,\, R}{\sqrt{1- R^{2}/\l^{2}}};  \qquad   &\eta = - \f{e^{-T/\l}\,\, \l}{\sqrt{1- R^{2}/\l^{2}}}  \label{static2} \ea
This form is best suited for generalization to the  Schwarzschild-de Sitter metric.\\

(4) Eddington-Finkelstein coordinates $(v,R,\theta, \varphi)$ \\
\be \rmd s^{2} = -f(R) \rmd v^{2} + 2 \rmd v \rmd R + R^{2}\,\dso^{2} \quad {\rm with} \quad v = T + R_{\star} \equiv T + \f{\l}{2}\, \ln \f{(1+ R/\l)}{(1- R/\l)} . \label{ef} \ee
As with the static coordinates, this chart covers the lower half of the Poincar\'e patch, but now \emph{includes} the past boundary of this region --i.e., the lower half of $\scrimr$-- along which $v$ runs from $-\infty$ (at $i^{-}$) to $\infty$  (at $\izl$).\, It excludes $\scripl$ because $v=\infty$ there.\\

(5) Kruskal coordinates $(\U,\V,\theta,\varphi)$\\
\be  \rmd s^{2} =  \f{\l^{2}}{(1-\U\V)^{2}}\,\, \big(- 4 \rmd \U \rmd \V + (1 + \U\V)^{2} \, \dso^{2}\, \big). \label{kruskal1}\ee
This chart covers the entire Poincar\'e patch of interest to this paper, excluding $\scri^{+}$ (as well as the lower Poincar\'e patch that is not of interest to us). $\scrimr$ corresponds to $\U=0$ and is coordinatized by $\V$ which runs from $- \infty$ (at $i^{-}$) to $\infty$ (at $i^{o})$. (Thus $\partial_{\V}$ is future directed.) The 2-sphere $\izl$ (at which the Eddington $v$ diverges) corresponds to $\V=0$. $\scripl$ is the upper half of the $\V=0$ surface on which $\U$ runs from $0$ (at $\izl$) to  $\infty$ (at $i^{+}$).  The Kruskal coordinates are related to the double-null Eddington-Finkelstein coordinates $(u,v)$ via
\be \U =  e^{\f{u}{\l}}, \quad \V = - e^{ -\f{v}{\l}}, \qquad  {\rm where} \,\,\, u= T - R_{\star}\, ; \ee
to the static coordinates $(T,R)$ via:
\ba   &\U = \exp \, \Big(\f{T}{\l}\, -\, \f{1}{2}\, \ln \Big| \f{1 + \f{R} {\l}}{1 - \f{R} {\l}} \Big|\,\,\Big),\qquad & \V=  - \exp \, \Big(-\f{T}{\l}\, -\, \f{1}{2} \, \ln \Big| \f{1 + \f{R} {\l}}{1 - \f{R} {\l}} \Big|\,\,\Big),\nonumber\\
& \f{R}{\l} = \f{1+\U\V}{1-\U\V}  \qquad &  \f{T}{\l} = \f{1}{2} \, \ln \f{\U}{-\V} \label{kruskal2}\ea
and to the coordinates $(\eta, r,\theta,\varphi)$ adapted to the cosmological slices, via
\ba   &\U = \f{\l}{r-\eta} \qquad &  \V= \f{r +\eta}{\l} \nonumber \\
         & \eta = \f{\U\V-1}{2\U}\, \l \qquad & r = \f{\U\V+1}{2\U}\, \l  \label{kruskal3}\ea

\subsubsection{Symmetries}
 
 While (the global) de Sitter space-time carries 10 Killing fields, the Poincar\'e patch under consideration is left invariant only by 7 of them \cite{abk1}.  These Killing fields are manifest in the two cosmological charts (1) and (2) because they are adapted to spatially flat slices,
 but not in the other three.  We have three spatial-translations $S^{a}_{(i)}$ and three rotations $R^{a}_{(i)}$ associated with the spatial  cartesian coordinates $x,y,z$,  and a time-translation $T^{a}$ defined by 
 \be  \vec{T} := T^{a}\partial_{a}= - \frac{1}{\l}\,(\eta \partial\eta +x \partial_{x} +y\partial_{y} + z \partial_{z} ) \ee
 in the chart (2).  (Because of the form $T^{a}$ takes in these coordinates, it is sometimes referred to as a `dilation'.)  The commutation relations between $S^{a}_{(i)}$ and $R^{a}_{(i)}$ are the familiar ones and $T^{a}$ commutes with the three rotations $R^{a}_{(i)}$. These commutators do not refer to the cosmological constant $\Lambda$; they are the same as those in Minkowski space-time.  On the other hand, commutators between $T^{a}$ and space-translations are new and explicitly involve  $\ell = \sqrt{3/\Lambda}$: $[T, S_{i}]^{a} = \f{1}{\l} \, S_{i}^{a}$. Note that in the $\Lambda \to 0$ limit, $\ell \to \infty$ whence the commutators vanish, as in Minkowski space-time.
 
However, since the cosmological charts do not cover $\scrimr$ --and also fail to 
 extend to the Schwarzschild-de Sitter space-time-- to investigate the behavior of the Killing fields on $\scrimr$ we need to work with either the Eddington-Finkelstein or the Kruskal coordinates (charts (4) and (5)). In the Kruskal coordinates $(\U,\V,\theta,\varphi)$, the three rotations assume the familiar form:
 \be \vec{R}_{1} = -(\sin\varphi)\, \partial_{\theta} - (\cot\theta\, \cos\varphi)\,\partial_{\varphi};\quad 
 \vec{R}_{2} = (\cos\varphi)\, \partial_{\theta} - (\cot\theta\, \sin\varphi)\,\partial_{\varphi}; \quad
 \vec{R}_{2} = \partial_{\varphi}\, . \label{rot}\ee
 The time-translation becomes
 \be \vec{T} = \f{1}{\l} \big(\U\partial_{\U} - \V\partial_{\V} \big) \label{TT-kruskal} \ee
 while the form of the spatial-translations is more complicated because the chart is not tailored to
 spatially homogeneous slices: 
 \ba \vec{S}_{1} &=& \sin\theta\cos\varphi \big( -\U^{2}\partial_{\U} + \partial_{\V}\big) + \f{2\U}{1+\U\V} \big(\cos\theta\cos\varphi \partial_{\varphi} - \f{\sin\varphi}{\sin\theta} \partial_{\varphi}\big)\nonumber\\
 \vec{S}_{2} &=& \sin\theta\cos\varphi \big( -\U^{2}\partial_{\U} + \partial_{\V}\big) + \f{2\U}{1+\U\V} \big(\cos\theta\sin\varphi \partial_{\varphi} - \f{\cos\varphi}{\sin\theta} \partial_{\varphi}\big)\nonumber\\
 \vec{S}_{3} &=& \cos\theta \big(-\U^{2}\partial_{\U} + \partial_{\V}\big) - \f{2\U}{1+\U\V}  \sin\theta \partial_{\theta} \ea
 In this chart, $\scrimr$ corresponds to $\U=0$, whence it is manifest that all seven Killing fields are well behaved and tangential to $\scrimr$. Note in particular that the restriction to $\scrimr$ of the 4 translations is given by:
 \be \vec{S}_{1} = (\sin\theta\cos\varphi)\,\partial_{\V},\quad \vec{S}_{2} = (\sin\theta\sin\varphi)\,\partial_{\V}, \quad  \vec{S}_{3} = (\cos\theta)\, \partial_{\V}, \ee
 and
 \be \vec{T} = -  \f{\V}{\l}\, \partial_{\V}.  \ee
Since $\partial_{\V}$ is future directed on $\scrimr$,  and $\V$ is negative between $i^{-}$ and $i^{o}$, the vector field $\vec{T}$ is also future directed on $\scriml$. Recall that $\izl$ is coordinatized by  $\U=\V=0$. Therefore, the time-translation Killing field  $T^{a}$ vanishes at the local $\izl$ whence  $\izl$ is \emph{left invariant} under the action of the time-translation subgroup of the isometry group of $(\Mr,\, g_{ab})$.  Similarly, the three rotations are tangential to the 2-sphere $\izl$. Recall that $\scriml$ is the portion $\scrimr$ to the past of $\izl$. The time-translation $T^{a}$ and the three rotations $R^{a}_{i}$ leave $\scriml$ invariant. By contrast, none of the three space-translations $S^{a}_{(i)}$ vanish at $\izl$. Therefore $\scriml$ is \emph{not left invariant} by any of the space-translations. Thus, only 7 of the 10 de Sitter isometries leave the upper Poincar\'e patch $\Mr$ invariant, and only 4 leave the local cosmological region $\Ml$ around the source invariant.  We will see in section \ref{s4} that these features are reflected also in the structure of the symmetry group at $\scrimr$ of the class of space-times of interest to this paper.

 \subsubsection{Global structure and physical fields} 
 
The full space-time $(M, g_{ab})$ trivially satisfies Definition 2;  it is asymptotically Schwarz-de Sitter. The physically relevant portion $\Mr$ of this space-time is the upper half Poincar\'e patch. Since its past  boundary $\scrimr$ is given by $\U=0$ in the Kruskal coordinates, it follows immediately from (\ref{kruskal1}) that its topology is $\mathbb{S}^{2}\times \mathbb{R}$ and the expansion of any of its null normal vanishes. Furthermore, the stress-energy tensor $T_{ab}$ vanishes identically near $\scrimr$.  \emph{Therefore it meets all three conditions of Definition 1}; it is an NEH.  Finally,  using the form (\ref{kruskal1}) of the metric it is easy to verify that $\lo^{a}$ defined by $\mathring{\l}^{a}\partial_{a} =  \partial/\partial \V$ is a future pointing, affinely parametrized, null geodesic normal to $\scrimr$. (As the notation suggests,  $\lo^{a}$ is in fact the \emph{canonical} extremal null normal on $\scrimr$, discussed in section \ref{s2.2}.) Since $\V$ runs from $-\infty$ to $\infty$ on $\scrimr$, it is geodesically complete. Thus the space-time under consideration belongs to the class $\CLi$ of section \ref{s2.3}. In fact, this is the `simplest' example of a space-time in this class. It is analogous to Minkowski space with a linearized source  of compact spatial support in the class of all asymptotically flat space-times in the $\Lambda=0$ case. \vskip0.1cm

The time-translation Killing field $T^{a}$ is given by ${T}^{a}\partial_{a} = - (\V/\l)\,\partial_{\V} = \partial_{v}$ where, as before,  $v$ is the Eddington-Finkelstein null coordinate. Thus $T^{a}$ vanishes at $\izl$, and its affine parameter  $v$ runs from $-\infty$ (at $i^{-}$) to $\infty$ (at $\izl$). Therefore  this vector field is complete already on $\scriml$. Let us denote its restriction to $\scriml$ by $\l^{a}$. Since $\l^{a}$ is  a null normal to $\scriml$, it satisfies the geodesic equation $\l^{a} \nabla_{a}\l^{b} = \kappa_{\l}\, \l^{b}$ with $\kappa_{\l} = -1/\l$. Thus, while surface gravity of the Killing field $T^{a}$ evaluated on the cosmological horizon $\scriml$ is constant --as it must be since $\scriml$ is a Killing horizon-- in a stark contrast to the more familiar black hole horizons, \emph{it is negative}. The rotation 1-form associated with $\l^{a}$ is $\omega_{a} = -(1/\l)\, \partial_{a} v$, while that associated with the affinely parametrized geodesic vector field  $\lo^{a}$ vanishes identically.  Hence Eq. (\ref{impsi1}) implies that  $\Im\,\Psi_{2}$ must vanish on $\scrimr$. Similarly since the horizon is spherically symmetric with proper radius $\l$, Eq. (\ref{repsi1}) implies that $\Re\,\Psi_{2}$ must also vanish on $\scrimr$. Of course this also follows trivially from the fact that the de Sitter metic is conformally flat. \vskip0.2cm

Finally, let us consider the limit $\Lambda \to 0$. Interestingly, while the conformal coordinates (2) and the Kruskal coordinates (5) are well-suited to the study of different aspects of Killing vector fields, the forms (\ref{conformal}) and (\ref{kruskal1}) of the metric show that the differential structures they define (via $(\eta,r)$  and $(U,V)$, respectively) are  ill-suited to take the $\Lambda\to 0$ limit.%
\footnote{Note that in the limit $\Lambda \to 0$ (i.e., $\l \to \infty$) differential structures defined by coordinates in (1) to (5) are no longer equivalent even in sub-regions. For example, it follows from (\ref{static2}) that if we work in the differential structure given by $\eta,r$, then $T,R$ become ill-defined in the limit and vice versa. So, one has to first fix the differential structure and then take the limit. One cannot freely pass from one of the systems to another after taking the limit.}
Static and Eddington Finkelstein coordinates, on the other hand, are well-suited. In static coordinates (3), the limiting  metric 
\be  \lim_{\Lambda\to 0} \rmd s^{2} = \rmd s^{2}_{o} = - \rmd T^{2} + \rmd R^{2} + R^{2} \dso^{2} \ee
is the Minkowski metric in the spherical coordinates $T,R,\theta,\varphi$. Since $T$ ranges over $(-\infty, \infty)$ and 
$R$ over $(0, \l)$, the limiting space-time is the complete Minkowski space $(\Mo, g_{ab}^{o})$. It is manifest that the $T^{a}$ becomes the standard time-translation Killing field in Minkowski space, adapted to these coordinates. Note that since $\scriml$ and $\scripl$ are given by $R=\l$, and $\l \to \infty$ as $\Lambda \to 0$, it follows that in the limit $\scriml$ and $\scripl$ become, respectively  the past and future null infinity $\scri^{-}_{o}$ and $\scri^{+}_{o}$ of  $(\Mo, g_{ab}^{o})$. Thus, \emph{in the limit, the (shaded) triangular part $\Ml$ of the de Sitter space-time expands to fill out all of Minkowski space}.  However, since $T= \pm R_{\star}$ along $\scriml$ and $\scripl$, in the limit both $T$ and $R$  become ill-defined there.  As usual, one has to carry out a conformal completion to attach $\scri^{\pm}_{o}$ as future and past boundaries of $(\Mo, g_{ab}^{o})$. The limiting procedure and the final result is the same if we begin with the Eddington-Finkelstein coordinates (4).\\

\emph{Remark:}  One can also choose to work with the cosmological chart (1)  since in the limit $\Lambda \to 0$ the metric (\ref{cosmological}) remains well-defined:
\be  \lim_{\Lambda\to 0} \rmd s^{2} = \rmd s^{2}_{o} =  -\rmd t^{2} + \rmd x^{2} + \rmd y^{2} + \rmd z^{2}. \ee
Since each of the 4 coordinates range from $-\infty$ to $\infty$, the limiting space-time is again the complete Minkowski space $(\Mo, g^{o}_{ab}$).  But there is an interesting and important difference from the result we obtained above using the static  (3) or the Eddington-Finkelstein chart (4).  Let us introduce a chart $v_{o}, r,\theta, \varphi$ on the \emph{full Poincar\'e  patch}  $\Mr$, with $r^{2} = x^{2} + y^{2} + z^{2}$ as before, and $v_{o}= t+r$. In the limit, the Minkowski metric $g^{o}_{ab}$ is now expressed in the advanced null coordinates. So, by setting $\Omega= 1/r$,we can carry out a conformal completion and attach  a past null boundary $\scri^{-}_{o}$, coordinatized by $v_{o},\theta,\varphi$, to $\Mo$. Then as $v_{o}$ runs over $(-\infty, \infty)$, one goes from $i^{-}$ to $i^{o}$. Thus, \emph{ now, $\scri^{-}_{o}$ of the limiting Minkowski space corresponds to  the \emph{entire} $\scrimr$ --rather than its bottom half, $\scriml$!} (Furthermore, in this full Minkowski space, we can introduce another  chart $u_{o}, r,\theta, \varphi$, with $u_{o}= t-r$, and carry out a conformal completion to attach  a future null boundary $\scri^{+}_{o}$ to the resulting Minkowski space, coordinatized by $u_{o},\theta,\varphi$. In this completion, the entire $\scri^{+}$ of de Sitter space-times corresponds to `time-like infinity' $i^{+}_{o}$ of the conformally completed Minkowski space-time.)

This discussion brings out an important subtlety. Because of global issues, we do not have a canonical way to take the limit $\l \to \infty$.  Because we have to restrict ourselves to charts in which the limiting metric is well-defined, the freedom in the procedure used to take the limit is curtailed. However, even within the restricted freedom, \emph{global} aspects --such as which surface in the $\Lambda \not=0$ space-time goes over to $\scri^{\pm}_{o}$ in the limiting Minkowski space-time-- can depend on which admissible chart is used.  This is why we introduced both $\scrimr$ and $\scriml$ in the class $\CLi$ of metrics under consideration. 

\subsection{Schwarzschild-de Sitter space-time}
\label{s3.2}

Interestingly,  Schwarzschild anti-de Sitter space-times have drawn much more attention in the literature than Schwarzschild-de Sitter space-times which are physically more directly relevant. Even in the Schwarzschild-de Sitter literature, black hole horizons have been studied more extensively than  the cosmological horizon, probably because the latter do not exist in Schwarzschild anti-de Sitter space-times. Notable exceptions are Ref. \cite{gt}  where properties of cosmological horizons were explored from thermodynamical considerations, and Ref. \cite{cg} where the emphasis was on the ambiguity in the normalization of the time-translation vector field used in the mechanics of WIHs.
 We will complement that discussion with geometric considerations  that are brought to forefront by our strategy of using the cosmological horizon $E^{+}(i^{-})$ as $\scrimr$ (and its bottom half as $\scriml$).   Since the relevant  structures in this example are  very similar to those in Section \ref{s3.1}, we will primarily focus on new issues that arise due to the presence of the mass term in the metric.

Because of the mass term, space-time no longer admits a spatially homogenous  foliation. Therefore the first two charts used in section \ref{s3.1} no longer exist. However, the remaining three can be readily generalized. It is convenient to express the space-time metric in these three charts because, as in section \ref{s3.1}, different aspects of the structure  become transparent in different charts.\vskip0.2cm
(1) Static coordinates $T,R, \theta,\varphi$, in which the $T^a\partial_{a} := \partial/\partial T$ is manifestly a static Killing field and $R$ is the proper radius of 2-spheres of symmetry:\\
\be \rmd s^{2} = -f(R) \rmd T^{2} + \f{\rmd R^{2}}{f(R)} + R^{2}\,\dso^{2}, \quad {\rm with} \quad f(R) = 1- \f{2Gm}{R} -\f{R^{2}}{\ell^{2}}.  \label{ss-static}\ee
This chart covers only the region bounded by the black hole horizon and $\scripl$ to the future, and $\scriml$ to the past.\\

(2) Eddington-Finkelstein coordinates $(v,R,\theta, \varphi)$: \\
\be \label{sds-ef} \rmd s^{2} = -f(R) \rmd v^{2} + 2 \rmd v \rmd R + R^{2}\,\dso^{2} \quad {\rm with} \quad v = T + R_{\star} .  \ee
where as usual the Tortoise radial coordinate $R_{\star}$ is defined by $\rmd R = f(R)\, \rmd R_{\star}$. Its explicit form  is more complicated than in (\ref{ef}) because now $f(R)$ has 3 roots,  $R_{(b)}$ representing the radius of the black horizon, $R_{(c)}$ representing the radius of the cosmological horizon and a negative root $R_{o}$. For our purposes, it will suffice to note that $R_{\star}$ has the form
\be R_{\star} = \rho(R) -  \f{\l^{2}  R_{(c)}\,\, \ln \f{|R- R_{(c)}|}{\l}}{(R_{(c)} - R_{o})(R_{(c)} - R_{(b)})}\,
\equiv\,  \rho(R)  -  \f{1}{\alpha}\,\ln \f{|R- R_{(c)}|}{\l}\, ,  \label{Rstar}\ee
where $\rho(R)$ is a rather complicated function of $R$ that is well behaved on the cosmological horizon and the constant $\alpha$, given by
\be \label{alpha} \alpha = \f{(R_{(c)} - R_{o})(R_{(c)} - R_{(b)})}{\l^{2} R_{(c)}}\,,  \ee
is positive everywhere outside the black hole horizon (and has dimensions of inverse length). This chart contains the region covered by the static chart but now also includes $\scriml$ where $v$ ranges from $-\infty$ (at $i^{-}$) to $\infty$ (at $\izl$). But it excludes $\scripl$ because $v$ diverges there.\\

(3) Kruskal coordinates $(\U,\V,\theta,\varphi)$:  \\
\be \label{sds-kruskal} \rmd s^{2} = \, \f{4 e^{\alpha\rho(R)}}{\alpha^{2} \l R}\,\, (R-R_{o})(R-R_{(b)})\, \rmd \U\,\rmd \V + R^{2}\dso^{2} \, ,\ee
where the constant $\alpha$ is defined in (\ref{alpha}). These Kruskal coordinates are related to the past and future Eddington-Finkelstein null coordinates $v = T + R_{\star}$ and $u = T-R_{\star}$ via
\be  \U =   \, e^{\f{\alpha}{2} u} \qquad {\rm and} \qquad  \V = - \, e^{-\,\f{\alpha}{2} v}\, . \ee
Let us examine the range of coordinates and associated geometrical structures. As in section \ref{s3.1}, $\partial_{\V}$ is future directed on $\scrimr$: its affine parameter $\V$ assumes the value $-\infty$ at $i^{-}$, zero at $\izl$ and $\infty$ at $i^{+}$. Thus, as usual, the Kruskal coordinates extend the Eddington Finkelstein chart, in our case to $\scripl$ and its future all the way to space-like $\scri^{+}$ in the asymptotic region. As in section \ref{s3.1},  past boundary of the space-time covered by the Kruskal chart is the \emph{entire $\scrimr$, not just $\scriml$}. The future boundary is the union of the black hole horizon in the interior region and space-like $\scri^{+}$ in the asymptotic region.  (Note that, unlike in the $\Lambda=0$ case, the black hole horizon and the black hole region are excluded.)\vskip0.2cm

\begin{figure}[]
  \begin{center}
  \hskip0.2cm
    \includegraphics[width=1.6in,height=1.6in,angle=0]{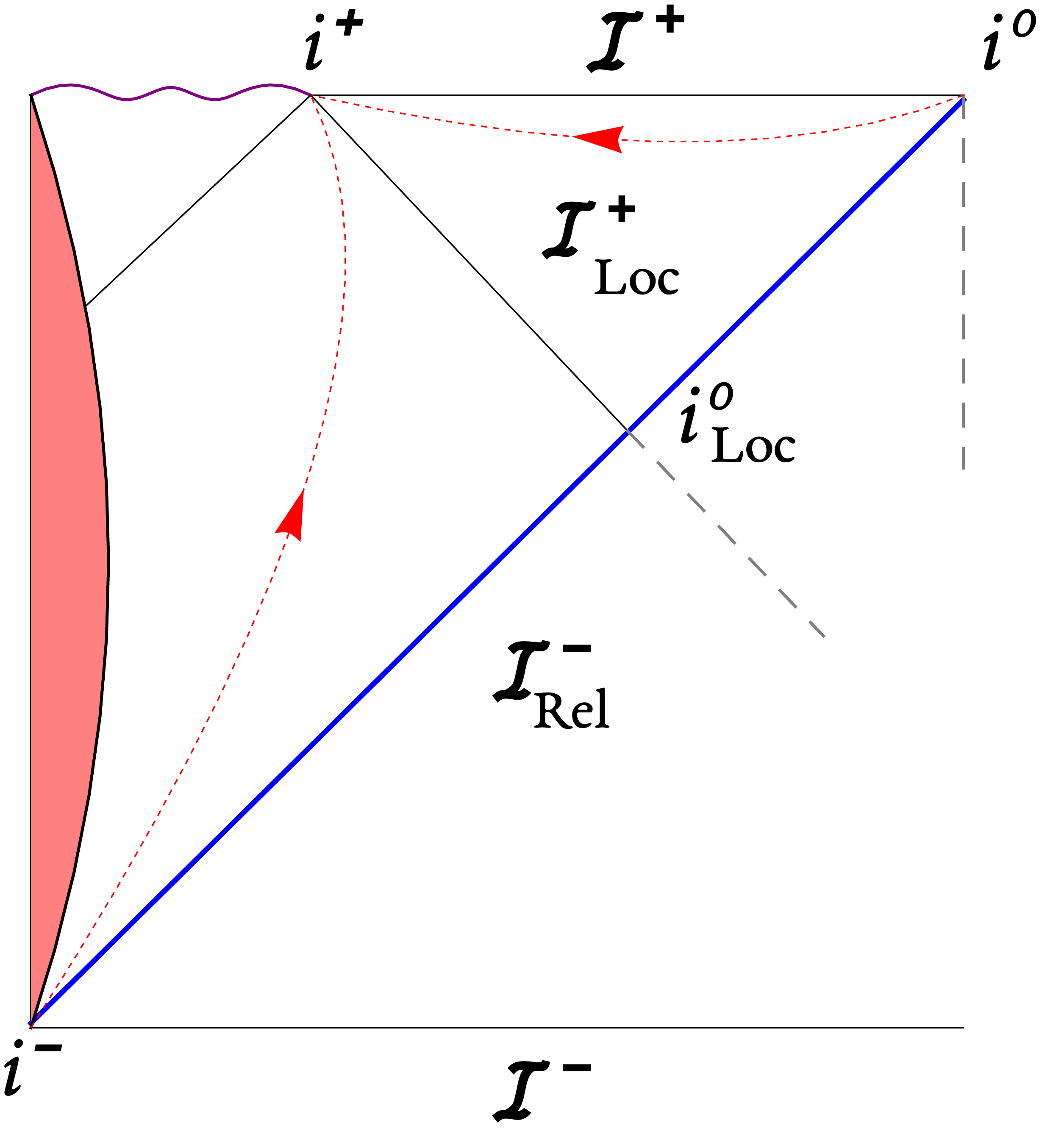}   \hskip1.1cm \includegraphics[width=4.2in,height=1.6in,angle=0]{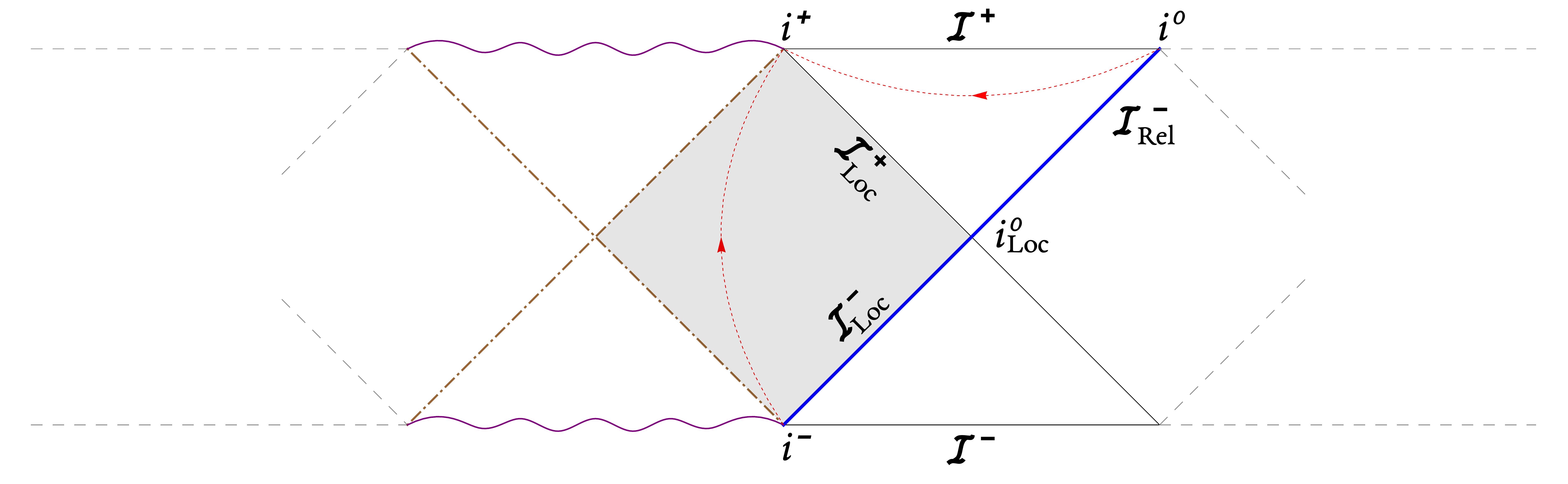}
\caption{{\footnotesize{\emph{Left panel:} Collapse of a spherical star in general relativity with a positive $\Lambda$. The collapse results in a space-like singularity in the future, denoted by the wiggly (magenta) line. The singularity is hidden from the exterior region by a black hole horizon, and we also have the future cosmological horizon of $i^{-}$ which serves as $\scrimr$, and (portion of) the past cosmological horizon of $i^{+}$ that serves as $\scripl$, and intersects $\scrimr$ in a 2-sphere cross-section $\izl$. The relevant space-time $\Mr$ is the portion to the causal future of $i^{-}$.  There is a static Killing field $T^{a}$ outside the star, whose integral curves are denoted by dashed (red) lines with arrows. It is time-like in the region bounded by the black hole horizon, $\scripl$ and $\scrimr$, but space-like near $\scrip$.\\
 \emph{Right panel:} Eternal spherically symmetric black hole in general relativity with a positive $\Lambda$. Because $\scri^{\pm}$ are space-like, the future (past) boundary of the maximally extended solution consists of an infinite sequence of singularities flanked by $\scrip$ (respectively, $\scrim$). Thus in contrast to the asymptotically flat, $\Lambda=0$ case, the space-time diagram continues ad-infinitum. However, following the strategy discussed in section \ref{s2}, for us the relevant part $\Mr$ of space-time is the causal future of $i^{-}$ which contains only  one future singularity and one $\scrip$. Situation with $\scripl$, $\izl$ and the static Killing field is the same as in the figure in the left panel. The shaded portion represents $\Ml$, the intersection of the causal future of $i^{-}$ with the causal past of $i^{+}$. }} }
\label{sds}
\end{center}
\end{figure}

Space-time has four Killing fields,  a time-translation $T^{a}$ and three rotations $R^{a}_{i}$, $i =1,2,3$. The rotations have the  same form (\ref{rot}) as in section \ref{s3.1}. In Kruskal coordinates, the static Killing field $T^{a}$ is given by
\be T^{a}\partial_{a}  = \f{\alpha\U}{2}\, \partial_{\U} \, - \, \f{\alpha\V}{2}\, \partial_{\V}     \ee
on entire $\Mr$. It is time-like in $\Ml$ and space-like in the asymptotic region near $\scrip$. 

Finally, the relevant global structures and physical fields can be summarized as follows.  First, $\scrimr$ is clearly a NEH since it is a Killing horizon. Second, $\lo^{a}$ defined by  ${\lo}^{a}\partial_{a} := \partial_{\V}$ is a future pointing,  affinely parametrized geodesic null normal and $\scrimr$ is complete because $\V$ runs from $-\infty$  (at $i^{-}$) to $+ \infty$ (at $i^{o}$). Thus, \emph{this space-time satisfies Definitions 2 and 4}  and therefore belongs to the class $\CLi$ under consideration. Next, let us consider the static Killing field $T^{a}$.  Its restriction $\l^{a}$ to $\scrimr$ is given by $ \l^{a} = - (\alpha/2)\, \V\, \lo^{a}$.  Since  $\lo^{a} $ is future pointing and  $\V$ is negative to the past of $\izl$ and positive to its future, it follows that $\l^{a}$ is \emph{future pointing on $\scriml$, vanishes at $\izl$ and past pointing to the future of $\izl$} (as in section \ref{s3.1}).  Surface gravity $\kappa_{\l}$ of this normal is given by $\kappa_{\l} = (1/2R_{(c)}) (1 - 3R_{(c)}^{2}/\l^{2})$.   
 The allowed range of $R^{2}_{(c)}$ is between $\l^{2}$ (when $m=0$)  and $\l^{2}/3$ which corresponds to the Nariai solution \cite{hn}.  For the entire class of these solutions, the surface gravity $\kappa$ of $T^{a}$ is \emph{negative} on $\scrimr$. Thus, while $(\scrimr,\, [\lo])$ is an extremal WIH, $(\scriml, \, [\l])$ is a non-extremal WIH.
 
 The rotation 1-form $\mathring\omega_{a}$ defined by $\lo^{a}$ again vanishes, and that associated with $\l^{a}$ is again exact, given by $\omega_{a} = \kappa_{\l} D_{a} v$. Therefore $\Im\, \Psi_{2} \,\epsilon_{ab}= D_{[a} \omega_{b]}$ vanishes on $\scrimr$, just as one would expect from the fact that since the space-time is spherically symmetric, all angular  momentum multipoles must vanish on $\scrimr$. The identity  (\ref{repsi1})  and spherical symmetry imply that  the real part $\Re \Psi_{2}$  on $\scrimr$ is given by 
\be \Re \Psi_{2} = \f{1}{2} \, \Big(\f{1}{R_{(c)}^{2} }- \f{1}{\l^{2}} \Big)\, . \ee
(In fact this relation between $\Psi_{2}$ and the area radius holds on any spherically symmetric cosmological or black horizon; see Appendix \ref{a2}.)  Next,  algebraic identities relate $\Psi_{2}$ to the parameter  $m$ in the expression of the space-time metric to $\Psi_{2} = - Gm/R^{3}$.  Therefore, in terms of  structures available at $\scrimr$ (or, $\scripl$), the parameter $m$ in the Schwarzschild-de Sitter geometry is given by the integral 
\be \label{mass1} m = - \f{1}{4\pi G} \oint_{C} R_{(c)}\, \Re \Psi_{2}\, \rmd^{2}V\ee 
evaluated on any 2-sphere cross-section $C$ of $\scrimr$ (or $\scripl$) Note that,  in the asymptotically flat context, in absence of incoming radiation, Bondi mass is given precisely by the limit to $\scrimr$ of  the 2-sphere integral on the right side.

What happens in the $\Lambda \to 0$ limit? As in section \ref{s3.1}, the Kruskal chart is ill-suited to take this limit. But we can take the limit using either the static or the Eddington-Finkelstein chart. In either case, the region bounded by the black hole horizon(s) and $\scri^{\pm}_{\rm Loc}$ expands out to give us the entire asymptotic region of the asymptotically flat Schwarzschild metric (representing a spherical collapse of a star as in the left panel of Fig. \hskip- 0.1cm \ref{sds} or eternal black hole as in the right panel). Thus, as in section \ref{s3.1}, in the limit $\scriml$ becomes $\scrimz$ and $\scripl$ becomes $\scripz$ of the asymptotically flat Schwarzschild metric.  The situation is completely analogous to that in section \ref{s3.1} for the case when the limit is taken using the static or the Eddington-Finkelstein chart.\\ 

\emph{Remarks:}

1. Given any solution with a time-translation isometry, energy $E_{t}$ is a linear map from the space of the time-translation Killing fields to real numbers. Thus, if we rescale the Killing field $t^{a} \to \b{t}^{a} = \lambda t^{a}$, the energy also rescales: $E_{\b{t}} = \lambda E_{t}$. This scaling is needed in the first law $ \delta E_{t} = \kappa_{t}\, \delta A$ of horizon mechanics. For,  surface gravity also rescales linearly while  area of the horizon is of course unaffected. Therefore, if the first law  holds for $t^{a}$, it holds also for $\b{t}^{a}$. For considerations of the horizon energy and the first law, then, we do not need to fix the rescaling freedom in the time-translation Killing field.

On the other hand,  only one of these energies $E_{t}$ can be regarded as mass $M$.  In the asymptotically flat case, it is $E_{t}$ associated with that time-translation Killing field $t^{a}$ which is unit at infinity. This method of fixing the rescaling freedom is not available in the $\Lambda >0$ case, because the norm of all time-translation Killing fields $T^{a}$ diverge at infinity in de Sitter space-time. However, we can take the limit $\Lambda \to 0$ and choose as preferred $T^{a}$ that vector field which, in the limit, goes to the \emph{unit} time-translation asymptotically as one approaches  $\scrimz$. The time-translation vector field $T^{a}$ used in this section is precisely this vector field. Can we characterize it intrinsically on $\scriml$, without reference to the limit? The answer is in the affirmative.  It turns out that the restriction $\ul\l^{a}$ of $T^{a}$ to $\scriml$  is the unique \emph{non-extremal} null-normal on $\scriml$ that satisfies two conditions:\\ 
(i) $\ul\l^{a}$  vanishes at $\izl$; and,\\
(ii) Its surface gravity is $\kappa_{\ul\l} = (1/2R_{(c)}) (1 - 3R_{(c)}^{2}/\l^{2})$. \\
This fact will serve as a guiding principle in section \ref{s5}. \vskip0.2cm

2. The left panel of Fig. \hskip-0.1cm\ref{sds}  depicts a spherical collapse while the right side depicts an eternal black hole. In both cases, the space-time continues ad infinitum --on the right side for the collapsing situation and on both sides for the eternal black hole.  This means in each case there is an infinite family of black and white holes. However, as is well-known, using the symmetries of the underlying space-time, for the eternal black hole one can carry out an identification so that we have only one black hole and only one white hole (see, e.g., section III.B in \cite{abk1}). 
But then the topology of space-time changes. Furthermore, {this is not possible for a collapsing star of the left panel without changing the physical system we are interested in.} From the perspective developed in sections  \ref{s1} and \ref{s2}, on the other hand, the situation is simpler. Since we ignore everything to the past of $\scrimr$ and impose the no incoming radiation condition there, we are led to consider only one collapsing star and only one black hole that results from the collapse; the relevant space-time $\Mr$ for us is precisely this region. 
\vskip0.2cm

3. For a linearized source in de Sitter space-time,  we found that there are two ways of taking the limit $\Lambda\to 0$. The first, discussed in the main text of section \ref{s3.1}, generalizes the one we used for Schwarzschild-de Sitter.  The second, discussed in the Remark at the end of section \ref{s3.1} exploited the presence of spatially homogeneous slicing in de Sitter space-time. If one uses the cosmological chart to take this limit, entire $\scrimr$ tends to $\scrimz$.  Can we not use a similar procedure here and obtain $\scrimz$ of the limiting  asymptotically flat Schwarzschild metric as the limit of full $\scrimr$? The procedure cannot be taken over directly because Schwarzschild-de Sitter space-time does not admit spatially homogeneous slices. Nonetheless,  one might imagine using the Eddington-Finkelstein chart, expressing the Schwarzschild-de sitter metric  as $\rmd s^{2}=\rmd s^{2}_{\rm de Sitter}  - \f{2Gm}{R} \rmd v^{2}$, and then using the cosmological chart $t,x,y,z$ for the de Sitter part of the metric. However, since the function $v$ diverges at $\izl$,  the extra term $\f{2mG}{R} \rmd v^{2}$ becomes ill-defined there. Therefore this strategy does not lead to a limit in which full $\scrimr$ of the Schwarzschild-de Sitter space-time  goes over to $\scrimz$ of the asymptotically flat Schwarzschild metric.  Thus, the second method of taking the $\Lambda \to 0$ limit in de Sitter space-time  exceptional and does not extend to more general space-times in our class. \\

Appendix \ref{a1} discusses the more complicated example of Kerr-de Sitter space-time. We again find that: (i)  the space-time admits $\scrimr$ and $\scriml$; (ii) $\scrimr$ is geodesically complete; (iii) there is a preferred time-translation Killing vector field $T^{a}$ that vanishes at $\izl$ and its restriction $\ell^{a}$ to $\scriml$ endows it with the structure of a non-extremal WIH. The rotational Killing field is tangential to $\izl$. Thus, $\izl$  is again left invariant by the isometries. We include this example to show that these structures are robust  in spite of important structural differences from the Schwarzschild-de Sitter case. But we chose to postpone it to the Appendix because expressions become long and the discussion is technically more complicated.

\section{Symmetries of $\scrimr$ and $\scriml$}
\label{s4}

Let us begin by recalling the symmetry groups in the asymptotically flat case. The past boundary of space-times representing isolated systems is $\scrimz$ which is endowed with certain universal structure --the geometric structure that is common to the past boundaries of \emph{all} asymptotically flat space-times. The symmetry group of $\scrimz$ is then the subgroup of ${\rm Diff}\, (\scrimz)$ that preserves this universal structure.  This is precisely the BMS group $\B$. (For a summary, see, e.g., \cite{aa-yau}.) Now,  if we are interested in isolated systems --as opposed to, say vacuum solutions to Einstein's equations-- then there is \emph{no incoming radiation} at $\scrimz$. This restriction can be used to introduce additional structure on $\scrimz$ : a 4-parameter family of preferred cross-sections --often called `good cuts'-- on which the shear of the ingoing null normal $n^{a}$ vanishes.  This family is left invariant by the BMS translations but not by the more general supertranslations. If one adds this family of `good-cuts'  to the universal structure,  symmetries would be only those elements of $\B$ that preserves this family. This is a 10 dimensional Poincar\'e group $\mathcal{P}$ of $\B$ \cite{np, aa-radmodes}. We will see that this situation in directly mirrored in the  $\Lambda>0$ case, now under consideration.\vskip0.1cm

In this case, the physically relevant portion $\Mr$ of space-time is the causal future of $i^{-}$.  As we saw in sections \ref{s2} and \ref{s3},  it is natural to impose the `no incoming radiation' condition at the past boundaries  $\scrimr$ of $\Mr$. Therefore,  symmetries of $\scrimr$  we are now seeking would be the analogs of the symmetries of the past null infinity, $\scrimz$. In section \ref{s4.1} we   examine the universal structure at $\scrimr$  --the structure that is shared by the past boundaries of the relevant portions $\Mr$ of \emph{all} space-times representing isolated systems in presence of a positive cosmological constant. The symmetry group $\G$ of $\scrimr$ would then be the subgroup of ${\rm Diff}\, (\scrimr)$ that preserves this universal structure. In section \ref{s4.2} we will analyze the structure of this symmetry group $\G$. We find that $\G$  \emph{is infinite dimensional, analogous in its structure to $\B$}, but with an interesting twist that captures the fact that we now have $\Lambda >0$. (These constructions were motivated by the analysis of \emph{non-extremal black hole horizons in the $\Lambda=0$ case} \cite{aank}  where the BMS group arises for different reasons.)

Motivated by considerations of sections \ref{s2.2} and \ref{s3},  in section \ref{s4.3} we introduce an additional structure --a preferred cross-section $\izl$  of $\scrimr$. ( For purposes of this section, it can be any cross-section; it need not be 
the intersection of the past event horizon $E^{-}(i^{+})$ of $i^{+}$ with $\scrimr$.)  $\scriml$ is the portion of $\scrimr$ that is to the past of $\izl$. We find that $\scriml$ is naturally foliated by a 1-parameter family of cross-sections. These are the analogs of `good cuts' in the $\Lambda=0$ case. The symmetry group  of $\scriml$ is therefore the subgroup of $\G$ that leaves this family invariant.  Addition of new structure always reduces the symmetry group. As in the asymptotically flat case, we find that the reduction is drastic: Infinite dimensional $\G$ is reduced to a seven dimensional group which we will denote by $\G_{7}$. 

We will use these symmetries to introduce conserved quantities in the next section.

\subsection{Universal structure of $\scrimr$}
\label{s4.1}

Recall that $\scrimr$ is an NEH that is geodesically complete. Thus, we are led to seek geometrical structures that are common to \emph{all} complete non-expanding horizons. Let us list these structures.

 First, every NEH is a 3-manifold that is topologically $\mathbb{S}^{2}\times \mathbb{R}$, ruled by the integral curves of null normals $\l^{a}$. Second, as noted in section \ref{s2}, each NEH comes equipped with a \emph{canonical} equivalence class $[\lo^{a}]$ of future directed null normals, where $\lo^{a} \approx \lo^{\,\prime\, a}$ if and only if $\lo^{\,\prime\, a} = c \lo^{a}$ for some positive \emph{constant} $c$. Each $\lo^{a}$ is a complete vector field on $\scrimr$.  

Next, each NEH is also equipped with a degenerate metric $q_{ab}$ of signature 0,+,+, satisfying $q_{ab} \lo^{b} =0$ and $\Lie_{\lo} q_{ab} =0$. However, the metric $q_{ab}$ itself is not universal. For example, on the de Schwarzschild-de Sitter $\scrimr$,  $q_{ab}$ is spherically symmetric, while on the Kerr-de Sitter $\scrimr$ it is only axi-symmetric.  More generally $q_{ab}$ may not admit any isometry. The scalar curvature ${}^{2}\R$ of $q_{ab}$ varies from one NEH to another. However, \emph{each NEH admits a unique 3-parameter family of unit round metrics $\qo_{ab}$ that are conformally related to its $q_{ab}$}. By construction, these round metrics $\qo_{ab}$ are themselves conformally related to one another. Furthermore, since  the $\qo_{ab}$ are all unit, round metrics, the relative conformal factors between them have a very specific form:
\be  \qo^{\,\prime}_{ab} = \alpha^{2} \qo_{ab}, \qquad {\rm where}\qquad  \alpha^{-1} (\theta,\varphi) =  \alpha_{0} + \alpha_{1} \sin\theta\cos\varphi + \alpha_{2} \sin\theta\sin\varphi + \alpha_{3} \cos\theta\, ,  \label{round}\ee
where  $\alpha_{0}, \alpha_{i}$ with $i =1,2,3$ are real constants satisfying, $-\alpha_{0}^{2} + |\vec\alpha|^{2} \equiv -\alpha_{0}^{2} + \alpha_{1}^{2} + \alpha_{2}^{2} + \alpha_{3}^{2} =-1$, and $\theta$ and $\varphi$ are a set of standard spherical coordinates associated with the metric $\qo_{ab}$. Thus, $\alpha^{-1}(\theta,\varphi)$  in (\ref{round}) is just a linear combination of the first four spherical harmonics of $\qo_{ab}$.  Note that the relative conformal factor $\alpha(\theta,\varphi)$ refers \emph{only} to the family $\{\qo_{ab}\}$ of round 2-sphere metrics; it has no memory of the physical metric $q_{ab}$ which varies from one space-time to another.

To summarize, the past boundary $\scrimr$ of every space-time in the class $\CLi$ under consideration, is equipped with the following three structures:\\
(1) $\scrimr$ is a 3-manifold, topologically $\mathbb{S}^{2}\times \mathbb{R}$;\\ 
(2) It carries with a preferred, equivalence class $[\lo^{a}]$ of complete vector fields $\lo^{a}$ where two are equivalent if they are related by a rescaling by a positive constant. Integral curves of these vector fields $\lo^{a}$ provide a fibration of $\scrimr$, endowing it with the structure of a fiber bundle over $\mathbb{S}^{2}$. \\
(3) $\scrimr$ carries an equivalence class of unit round 2-sphere metrics $\qo_{ab}$, related to each other by a conformal transformation of the type (\ref{round}), such that $\qo_{ab} \lo^{b} =0$ and $\Lie_{\lo} \qo_{ab} =0$ for every $\qo_{ab}$ in this family.\\
This is the universal structure at $\scriml$.\vskip0.2cm

Overall, the situation is analogous to that at null infinity, $\scrimz$, of  asymptotically flat space-times. If we were to restrict ourselves to Bondi conformal frames --as is often done-- then the universal structure at $\scrimz$ consists of pairs  $(\qo_{ab}, \lo^{a})$ of fields on $\scrimz$, where  $\qo_{ab}$ is a unit, round, 2-sphere metric,  and $\lo^{a}$ a null normal, such that any two pairs are related by conformal rescalings of the type $(\qo^{\,\prime}_{ab},\, \lo^{\,\prime\,a}) = (\alpha^{2} (\theta,\varphi) \qo_{ab}, \alpha^{-1}(\theta,\varphi) \lo^{a})$ with $\alpha(\theta,\varphi)$ again is given by (\ref{round}).  (See, e.g., \cite{aa-yau}.) There is however one difference: while $\scrimr$ admits a \emph{canonical} $[\lo^{a}]$ that is not tied to the 3-parameter family of unit round metrics $\qo_{ab}$, in the asymptotically flat case, $\scrimz$ admits a 3-parameter family of null normals $\lo^{a}$, each tied to a round metric $\qo_{ab}$, undergoing a rescaling by $\alpha^{-1}(\theta,\varphi)$ when $\qo_{ab}$ is rescaled by $\alpha^{2}(\theta,\varphi)$. This difference will play a key role: It lies at the heart of the subtle but important difference between the BMS group $\B$ at $\scrimz$ and the symmetry group $\G$ of $\scrimr$, discussed in section  \ref{s4.2}.

Note that the universal structure on $\scrimr$  refers neither to the physical metric $q_{ab}$, nor to the intrinsic derivative operator $D_{a}$, nor to the rotation 1-form $\omega_{a}$, as these structures vary from the $\scrimr$ of one physical space-time to another.  These constitute \emph{physical fields} on $\scrimr$  that capture physical information --such as mass, angular momentum and multipole moments--  contained in the gravitational field of the specific space-time under consideration.  (In particular,  in the universal structure, $[\lo^{a}]$ are only complete vector fields; not geodesic vector fields since the notion of geodesics requires a connection.) This situation is completely analogous to that in the asymptotically flat case. There, the connections $D_{a}$ on $\scrimz$ whose curvature defines the news tensor  $N_{ab}$ \cite{aa-radmodes},  and the Newman Penrose components \cite{rp1,aa-yau} $\Psi_{0}^{o},\, \ldots, \Psi_{4}^{o}\,$ of the (appropriately rescaled) Weyl curvature of the of the conformally rescaled metric $\h{g}_{ab}$  are \emph{physical fields}  on $\scrimz$ that vary from one space-time to another.

\subsection{Symmetries of  $\scrimr$} 
\label{s4.2}

As in the asymptotically flat case,  discussion of asymptotic symmetries is most transparent if one first introduces an abstract 3-manifold, $\scriab$, that is not tied to any specific space-time, but is endowed with the universal structure of $\scrimr$. Thus, $\scriab$ will be: 
\\ (1) a 3-manifold, topologically $\mathbb{S}^{2} \times \mathbb{R}$; \\
\indent \hskip0.2cm equipped with: \\
 (2) a class $ [\lo^{a}] $  of complete vector fields $\lo^{a}$, related to each other by a rescaling by a positive constant; 
 and, \\
 (3)  a class $\{\qo_{ab}\}$ of conformally related, unit, round 2-sphere (degenerate) metrics   such that $\qo_{ab}\lo^{b} =0$ and $\Lie_{\lo}\, \qo_{ab} = 0$. (Note that the fact that the $\qo_{ab}$ are unit, round metrics that are conformally related implies that the relative conformal factor $\alpha$ must of the type (\ref{round}).)\\
\noindent The space of integral curves of $\lo^{a}$ is topologically $\mathbb{S}^{2}$ and we will denote it by $\scriabt$.
 \vskip0.2cm
 
 The symmetry group $\G$ is then the subgroup $\Diff (\scriab)$ that preserves this structure. Given any concrete space-time in our class $\CLi$, there exist diffeomorphisms from the concrete $\scrimr$ to $\scriab$ that send $[\lo^{a}]$ and $\{\qo_{ab}\}$ on $\scrimr$ to $[\lo^{a}]$ and $\{\qo_{ab}\}$ on $\scriab$. However, these diffeomorphisms are not unique. Any two are related by an element of the symmetry group $\G$ (since elements of $\G$ are the diffeomorphisms from $\scriab$ to itself that preserve the universal structure on $\scriab$). 
 
Let us  examine the structure of $\G$. As in the asymptotically flat case, it is simplest to first discuss the structure of its Lie algebra $\LG$. Since $\G$ is a subgroup of  ${\rm{Diff}}({\scriab})$, every element of  $\LG$ is represented by a vector field $\xi^{a}$ that generates  a 1-parameter family of diffeomorphisms preserving the universal structure. Elements of $[\lo^{a}]$ differ from each other only by a rescaling $\lo^{a} \to c \lo^{a}$ where $c$ is a positive constant, and elements of  $\{\qo_{ab}\}$ are related by $\qo_{ab} \to \alpha^{2}(\theta,\varphi) \qo_{ab}$ where the conformal factor $\alpha (\theta,\varphi)$ is specified in (\ref{round}). Therefore, under a 1-parameter family $d(\lambda)$ of diffeomorphisms generated by  a symmetry vector field $\xi^{a}$, we must have:
\be \lo^{a} \to c(\lambda)\, \lo^{a} \qquad {\rm and} \qquad \qo_{ab} \to \alpha^{2}(\lambda)   \qo_{ab} \ee
where for each $\lambda$,  $c(\lambda)$ is a constant,  and $\alpha(\lambda)$ is a function of $\theta,\varphi$ of the form given in Eq. (\ref{round}), with $c|_{\lambda=0} = 1$ and $\alpha|_{\lambda =0} (\theta,\varphi)=1$. Furthermore,  Eq. (\ref{round}) implies that the four constants $\alpha_{0}(\lambda)  \ldots \alpha_{3}(\lambda)$ in the expression of the conformal factor  $\alpha(\lambda) $ must satisfy $\alpha_{0}^{2}(\lambda) - |\vec{\alpha}(\lambda)|^{2} =1$ for each $\lambda$. Now, to obtain infinitesimal action of the Lie algebra element, we just need to take the derivative with respect to $\lambda$ and evaluate the result at $\lambda =0$.  Thus, to qualify as an infinitesimal symmetry,  the vector field $\xi^{a}$ on $\scriab$ must satisfy 
\be  \Lie_{\xi}\, \lo^{a}  = - \kappa \,\lo^{a} \qquad {\rm and} \qquad \Lie_{\xi}\, \qo_{ab} = 2\phi (\theta,\varphi)\, \qo_{ab},  \qquad \forall\,  \lo^{a}\,\, {\rm and} \,\,  \forall\,\qo_{ab}  \label{infsym}\, .\ee  
where $- \kappa = (d c(\lambda)/d\lambda)|_{\lambda=0}$ and $\phi = (d\alpha(\lambda)/d\lambda)|_{\lambda=0}$. Here the constant  $\kappa \in \mathbb{R}$ depends on $\xi^{a}$ but is independent of the choice of $\lo^{a} \in [\lo^{a}]$ (and the minus sign in front of $\kappa$ is introduced in (\ref{infsym}) for later convenience). The function $\phi$, on the other hand, varies from one round metric $\qo_{ab}$ to another. Restrictions on $\alpha(\lambda)$ imply that $\phi$ is a linear combination of the first three spherical harmonics defined by $\qo_{ab}$ and in particular satisfies $\Lie_{\lo} \phi =0$. Thus, it projects down to a function $\t\phi$ on the space $\scriabt$ of integral curves of  $\lo^{a}$ (and satisfies $ \Dot_{a} \Dot_{b} \,\t\phi = \t\phi\, \qot_{ab}$, where $\Dot_{a}$ is the derivative operator defined by $\qot_{ab}$ on $\scriabt$). 

Since the conditions (\ref{infsym}) that characterize infinitesimal symmetries $\xi^{a}$ are so simple, it is rather straightforward to analyze the structure of the Lie algebra $\LG$.  

Let us first consider the space $\LVer$ of symmetry vector fields $\xi^{a}$ that are vertical, i.e. proportional to $\lo^{a}$. These would be analogous to supertranslations on $\scrimz$. Let us  fix a fiducial  $\lo^{a}$ in $[\lo^{a}]$  and set $\xi^{a} = \xi \lo^{a}$. (Thus, given a $\xi^{a}$ there is an ambiguity $\xi \to c^{-1}\xi$ in the choice of function $\xi$.)  Since\, $\Lie_{\lo} \qo_{ab} =0$\, and\, $\qo_{ab} \lo^{b} =0$,\, it follows immediately that $\Lie_{\xi} \qo_{ab} =0$ for all $\qo_{ab}$ in our universal structure. Therefore, the second of Eq.(\ref{infsym}) is automatically satisfied (with $\phi(\theta,\varphi) =0$). The first condition on the other hand is a genuine restriction on the function $\xi$. To write the solution explicitly, let us introduce a cross-section $C$ of $\scriab$ and denote the affine parameter of $\lo^{a}$ that vanishes on this $C$ by $\vo$. Let us also introduce spherical coordinates $(\theta,\varphi)$ on $C$ and extend them to all of $\scriab$ by demanding that they be constant along fibers (i.e. integral curves of $\lo^{a}$). Then, the general solution to the first of Eqs (\ref{infsym}) is simply
\be \xi = \kappa \vo + f (\theta,\varphi)\, \qquad {\hbox{\rm so that}}\qquad  \xi^{a} = \big(\kappa \vo + f (\theta,\varphi)\big) \,\lo^{a}  \label{ver}\ee
 Thus $\xi^{a} \in \LVer$ if and only if it has the form (\ref{ver}), whence every vertical symmetry vector field $\xi^{a}$ can be labelled  by a pair $(\kappa, f)$ where  $\kappa$ is a real number and $f$ a function on $\scriab$ satisfying $\Lie_{\lo} f = 0\,$.%
 \footnote{ However, this labeling depends on our choice of $\lo^{a}$ and the choice of an affine parameter $\vo$ (or a cross-section $C$ of $\scriab$). Under the most general change of these fiducial  choices, we have $\lo \to \lo^{\,\prime\, a} = c\lo^{a}$ and $\vo \to \vo^{\prime} = (1/c)\, \big(v_{o} + a(\theta,\varphi)\big)$, with $c>0$, we have: $ \kappa^{\prime} = \kappa$ and $c f^{\prime}(\theta,\varphi)  =  f(\theta,\varphi) - a(\theta,\varphi)\kappa$.}
 Now, given a vertical vector field $\xi^{a}_{1} = \xi_{1} \lo^{a}$ in $\LVer$ and a \emph{general} infinitesimal   symmetry $\xi_{2}^{a}$, by first of the Eq. (\ref{infsym}), the commutator is given by
\be  [\xi_{2}, \, \xi_{1}]^{a} =  \big( (\Lie_{\xi_{2}} \, - \kappa_{2}) \xi_{1} \big) \,\,\lo^{a}\, .  \label{normal1}\ee
 Since the right side is again vertical, it follows that  the space $\LVer$ of vertical symmetry fields constitutes a Lie-ideal of $\LG$. 
 
 Let us therefore take a quotient $\LG/\LVer$. Each element of $\LG/\LVer$ is an equivalence class $\{\xi^{a}\}$ of symmetry vector fields $\xi^{a}$, where two are equivalent if they differ by a vertical symmetry vector field. Since, furthermore, $\Lie_{\lo} \xi^{a}$ is again vertical for any $\xi^{a} \in \LG$, it follows that every symmetry vector field $\xi^{a}$ can be projected to a vector field $\t\xi^{a}$ on the base space $\scriabt$ of $\scriab$ unambiguously and, furthermore, \emph{all vector fields in a given equivalence class  $\{\xi^{a}\}$ have the same projection} $\t{\xi}^{a}$. Therefore elements of the quotient $\LG/\LVer$ are in 1-1 correspondence with the projected vector fields $\t\xi^{a}$ on $\scriabt$. Now,  the second of Eq (\ref{infsym}) implies that each $\t\xi^{a}$ is a conformal Killing field for every round metric $\qot_{ab}$ on $\scriabt$ in our universal structure. (If it is a conformal Killing field for one $\qot_{ab}$, it is also a conformal Killing vector field for every other because all our round metrics are conformally related.)  Thus, the quotient $\LG/\LVer$ is isomorphic with the Lie algebra of conformal Killing vectors on our family of unit, round metrics $\qot_{ab}$ on the 2-sphere $\scriabt$. But it is well known that this Lie algebra is isomorphic to the Lie algebra $\LLor$ of the Lorentz group $\Lor$ in 4 space-time dimensions. Thus, we conclude that \emph{the quotient $\LG/\LVer$ of the symmetry Lie algebra by the sub-algebra of vertical symmetry fields is isomorphic with the Lorentz Lie algebra $\LLor$.} \\
  
Returning to the group $\G$, we have shown that the vertical diffeomorphisms in $\G$ constitute a normal subgroup $\Ver$, and the quotient $\G/\Ver$ is isomorphic with the Lorentz group. \emph{Thus, $\G$ is a semi-direct product of the Lorentz group $\Lor$ with $\Ver$: \, $\G = \Ver \rtimes \Lor$.}  Recall that, in the asymptotically flat case, the BMS group $\B$ has similar structure: It is a semi-direct product $\B = \S \rtimes \Lor$ where $\S$ is the group of supertranslations.  However, there are two key differences that can be traced back directly to the differences in the universal structure in the two cases:\vskip0.1cm
\noindent (1) The normal subgroup $\Ver$ of $\G$ is generated by vertical vector fields of the form $\xi^{a} = ( \kappa v_{o} + f (\theta, \varphi))\lo^{a}$ where $\kappa \in \mathbb{R}$ and  $f (\theta,\varphi)$ is a smooth function on the base space $\scriabt$. 
In the case of the BMS group, the supertranslation subgroup $\S$ is generated by vector fields on $\scri_{o}$ of the type $\xi^{a} = f(\theta, \varphi) \lo^{a}$ (where $\lo^{a}$ is again a fiducial null normal, now representing a `pure' time-translation in a fiducial Bondi conformal frame). Thus heuristically, $\Ver$ has `one more' generator than $\S$. Furthermore, while the supertranslation subgroup $\S$ is Abelian, $\V$ is not.  \vskip0.1cm
\noindent (2) Another --and more important-- difference is that the semi-direct product structure is quite different. In the  parametrization introduced above,  a general element $\xi^{a}$ of $\LG$ can be represented as
\be \xi^{a} = \big[ \kappa \vo + f(\theta,\varphi) \big] \lo^{a} + \bar{K}^{a}  \label{xi}\ee
where $\bar{K}^{a}$ \, has the following properties: (i) $\Lie_{\lo} \b{K}^{a} =0$; and,  (ii) $\bar{K}^{a}$ is tangential to each $\vo= {\rm const}$ cross-sections of $\scriab$ and a conformal Killing field of the metric $\mathring{\bar{q}}_{ab}$, obtained by pulling back to the cross-section any round, unit 2-sphere metric $\qo_{ab}$ in the universal structure. Thus, for any given choice of the affine parameter $\vo$ of $\lo^{a}$, we have a decomposition of $\xi^{a}$ into a vertical vector field, proportional to $\lo^{a}$ and a horizontal vector field $\bar{K}^{a}$, that is tangential to \emph{all} $\vo ={\rm const}$ cross-sections. The six horizontal $\bar{K}^{a}$ are generators of a Lorentz subgroup $\Lor$ of $\G$.%
\footnote{Since $\mathring{\bar{q}}_{ab}$  is a unit 2-sphere metric, the six conformal Killing vectors $K^{a}$ have the form  $K^{a} = \mathring{\bar{q}}^{ab} D_{b} \phi (\theta,\varphi) +  \mathring{\bar{\epsilon}}^{ab} D_{b} \psi(\theta,\varphi)$. Here $\mathring{\bar{\epsilon}}^{ab}$ is the alternating tensor defined by $\mathring{\bar{q}}^{ab}$, and  $\phi (\theta,\varphi)$ and $\psi (\theta,\varphi)$ are linear combinations of the three $Y_{1,m} (\theta,\varphi)$, i.e., solutions to  $\mathring{\bar{D}}^{2} \phi = -2\phi$ and $\mathring{\bar{D}}^{2} \psi = -2 \psi$ on each $\vo= {\rm const}$ cross-section. {Finally, if we change the vector $\lo^{a} \to   \lo^{\prime\,a} = c \lo^{a}$  in $[\lo^{a}]$ but retain the cross-section $C$ as the origin of the affine parameter --so that $\vo^{\prime} = \vo = 0$ on $C$-- then $\kappa^{\prime} = \kappa;\,  f^{\prime} (\theta,\varphi) = (1/c) f(\theta,\phi);\, K^{\prime\,a} = K^{a}$.}}
The situation in the BMS group is different. There, \emph{none} of the Lorentz subgroups leave invariant an entire family of cross-sections, $\vo = const$, of $\scri_{o}$. Indeed, every Lorentz subgroup $\Lor$ of $\B$ leaves invariant precisely one cross-section.\smallskip
 
There is another way to display the structure of $\G$ that brings out a different aspect of its relation to the BMS group.
It is clear from Eq. (\ref{ver}) that \emph{supertranslations} --i.e. vertical vector fields of the type $\xi^{a} = f(\theta, \varphi)\, \lo^{a}$-- form an Abelian sub-Lie algebra of $\LG$. Furthermore, if $\xi_{1}^{a} = f_{1} (\theta,\varphi)\, \lo^{a}$ is a supertranslation, and $\xi_{2}^{a}$\, is a \emph{general element} of $\LG$, then Eq. (\ref{normal1}) implies:
\be [\xi_{2},\, \xi_{1}]^{a}\, =\,  \big( (\Lie_{\xi_{2}} - \kappa_{2}) f_{1}(\theta, \varphi) \big)\, \lo^{a}  \, = F(\theta,\varphi)\, \lo^{a} \ee
for some $F(\theta,\varphi)$, since  it is easy to verify that $\Lie_{\lo} \big( (\Lie_{\xi_{2}} - \kappa_{2}) f_{1}(\theta, \varphi) \big) = 0$. \emph{Thus, the subgroup $\S$ of supertranslations is also a normal subgroup of the symmetry group $\G$}.  As one would expect from the fact that $\LVer$ can be thought of as `the Lie algebra $\LS$ of supertranslations, augmented with one extra element', the quotient is a 7 dimensional group $\G_{7}$. \emph{Thus, $\G$ can also be expressed as another  semi-direct product where the normal sub-group is the group $\S$ of supertranslations: $\G = \S  \rtimes \G_{7}$.} Recall that for the BMS group $\B$ we have $\B = \S \rtimes \Lor$ where $\Lor$ is the 6-dimensional Lorentz group.

To explore the  structure of $\G_{7}$, let us work with Lie algebras. An element of the Lie algebra $\LG_{7}$ is an equivalence class  $\{\xi^{a}\}$ of elements of $\LG$ where two are equivalent if they differ by a super-translation. It is immediate from the form (\ref{xi}) of a general infinitesimal symmetry $\xi^{a}$ that a general element of $\LG_{7}$ can be written as an equivalence class $\{\kappa \vo \, \lo^{a} + \bar{K}^{a} \}$ of elements of $\LG$.  Each equivalence class is labelled by a real number $\kappa$ and a conformal Killing field $\bar{K}^{a}$ on a unit, round 2-sphere. Now, since the vector space of conformal Killing fields is 6 dimensional, it follows that $\LG_{7}$ is a 7 dimensional Lie algebra. It is easy to verify that the element $\{ \kappa\vo\, \lo^{a} \}$ commutes with every element of $\LG_{7}$. Therefore, at the level of groups, $\G_{7}$ is just a direct product $\G_{7} = \mathbb{R} \times \Lor$. Put differently, \emph{ $\G_{7}$ is a central extension of the Lorentz group,} albeit a trivial one.\medskip

Let us summarize. The symmetry group $\G$ of $\scriab$ is infinite dimensional. Its structure is similar to that of the BMS group $\B$, in that it is a semi-direct product of the abelian group $\S$ of supertranslations with a finite dimensional group. However, while $\B = \S \rtimes \Lor$, where $\Lor$ is the Lorentz group, $\G = \S \rtimes \G_{7}$ where $\G_{7}$ is a trivial central extension of the Lorentz group: $\G_{7} = \mathbb{R} \times \Lor$.  In presence of a positive $\Lambda$, this extension has a crucial role.  As we saw in the examples considered in section \ref{s3}, the time-translation isometry (singled out by the source) is \emph{precisely} the `extra' element $\kappa\vo\,\lo^{a}$ in $\G_{7}$ added to the Lorentz group; it is missing in the BMS group $\B$. In the Schwarzschild-de Sitter space-time, we could take the limit $\Lambda \to 0$. In this limit, Killing vector $T^{a}$ of the Schwarzschild-de Sitter space-time on $\Ml$ tends to the time-translation Killing field of the Schwarzschild space-time in the asymptotic region outside the horizon. The restriction of this $T^{a}$ to $\scrimr$ is precisely the `extra' element $\kappa\vo\,\lo^{a}$ in $\G_{7}$.\\
\goodbreak

\emph{Remarks:} 

1. Recall that in the case of the BMS group $\B$, since the Lorentz group $\Lor$ arises as the quotient $\Lor = \B/\S$,  there are `as many' Lorentz subgroups of $\B$ as there are supertranslations: the group $\S$ of supertranslations acts simply and transitively on the space $S\mathcal{L}$ of all Lorentz  subgroups of $\B$.  But $\S$ also acts simply and transitively on the space  $S\mathcal{C}$ of all cross-sections of $\scrimz$.  Therefore the two spaces, $S\mathcal{C}$ and $S\mathcal{L}$, are isomorphic. In fact there is a \emph{natural} isomorphism: Each cross-section $C$ in $S\mathcal{C}$ is left invariant by one and only one Lorentz subgroup $\Lor$ of $\B$. \\
 What is the situation at $\scrimr$? Here,  $\G_{7}$ arose as the quotient $\G_{7} = \G/\S$ of the full symmetry group $\G$  by its subgroup $\S$ of supertranslations, whence  $\S$ acts simply and transitively on the space $S\G_{7}$ of all $\G_{7}$ subgroups of $\G$ that  are isomorphic with $\G/\S$ under the projection. But we also know that $\S$ acts simply and transitively on the space $S\mathcal{C}$ of all cross-sections of $\scrimr$  (just as it does on $\scrimz$ in the asymptotically flat case).  And again there is a \emph{natural} isomorphism between the two spaces, $S\G_{7}$ and $S\mathcal{C}$, on each of which the supertranslation group $\S$ acts simply and transitively: \emph{Each cross-section $C$ in $S\mathcal{C}$ is left invariant precisely by one $\G_{7}$ sub-group in $S\G_{7}$.}
\smallskip

2. In section \ref{s3.1}, we considered a linearized source on a de Sitter background. In this case, we found that $\scrimr$  (as well as the upper Poincar\'e patch $\Mr$) is left invariant by a seven dimensional subgroup of the de Sitter group.  Let us call it $G_{7}$. In this section we encountered another seven dimensional group $\G_{7}$. In terms of the full symmetry group $\G$ on $\scrimr$, the two groups have the following roles.  $G_{7}$ is the subgroup of $\G$ that arises in de Sitter space-time and is generated there by:  (i) the time-translation $\vo \lo^{a}$;\, (ii) three space-translations $Y_{1m}\lo^{a}$; and, \,(iii)  three rotations.  While the  time-translation and the three rotations leave one cross section, $\izl$, of $\scrimr$ invariant, the three space-translations do not leave \emph{any} cross-section of $\scrimr$ invariant!  By contrast, in general space-times in the class $\CLi$ under consideration,  the group $\G_{7}$ arises as the \emph{quotient} $\G_{7} = \G/\S$;\,  it  is not a canonical subgroup of $\G$.  As we noted above in Remark 1, given a cross-section $C$ of $\scrimr$ we \emph{can} naturally embed $\G_{7}$  into $\G$; but by construction that subgroup leaves the chosen cross section invariant while $G_{7}$ leaves no cross-section of $\scrimr$ invariant! \smallskip

3. There is a discussion of symmetry groups also in the literature on quasi-local horizons \cite{abl2,jltp,aabkrev} where these groups were found to be \emph{finite dimensional}. However, that analysis referred to symmetries of \emph{specific} WIHs. The vector fields were required to preserve not just the universal structure but certain physical fields, in particular the physical (degenerate) metric $q_{ab}$ and the rotational 1-form on the given WIH. In the present paper, on the other hand,  the focus is on symmetries of $\scrimr$ of \emph{all} space-times in our collection $\CLi$. Therefore, we were led to introduce the universal structure shared by \emph{all}  (geodesically complete) NEHs and consider as infinitesimal symmetries all vector fields on $\scrimr$ that leave this universal structure invariant. Since this is a \emph{much} weaker  requirement, the Lie algebra of symmetry vector fields turned out to be \emph{infinite dimensional}.  If a specific space-time in our class $\CLi$ were to admit a Killing vector, not only would it belong to the infinite dimensional $\LG$ but its action would also leave the geometrical fields on $\scrimr$ invariant. Therefore, it would also be a symmetry in the stronger sense considered in the quasi-local horizon literature.

\subsection{$\izl$ and symmetry reduction: Symmetries of $\scriml$}
\label{s4.3}

As we saw in section \ref{s4.2},  given a cross-section $C$ of $\scrimr$, we can set the affine parameter $\vo$ of $\lo^{a}$ to be zero on $C$ and then obtain a natural foliation of $\scrimr$ by the $\vo={\rm const}$ surfaces. This foliation refers to the entire class $[\lo^{a}]$ of vector fields that $\scrimr$ is endowed with. Under $\lo^{a} \to \lo^{\prime} = c \lo^{a}$,  the labeling of the leaves of the foliation changes  via $\vo^{\prime} = (1/c) \vo$, but the leaves of the foliation remain the same. Let us use this foliation in the decomposition (\ref{xi}) of $\xi^{a}$ into vertical and horizontal parts. Then, each part is individually left unchanged under $\lo^{a} \to \lo^{\prime} = c \lo^{a}$. In particular, we have $K^{\prime\, a} = K^{a}$.  As we vary symmetry vector fields $\xi^{a}$, we obtain different horizontal vector fields $K^{a}$ and together, they constitute a Lorentz subalgebra $\LLor$ of $\LG$. Thus, the foliation $\vo = {\rm const}$ selects a specific Lorentz-subgroup $\Lor$ of the symmetry group $\G$.

What happens under a change of the initial cross-section that serves as the `origin' of the affine parameter? If $C \to C^{\prime}$ then we have $\vo \to \vo^{\prime} = v_{o} + a(\theta,\varphi)$  where $a$ is any function on $\scrimr$ satisfying $\Lie_{\lo} a =0$.  Now the new foliation $\vo^{\prime} ={\rm const}$ is distinct, related to the original one $\vo = {\rm const}$ by a supertranslation. Therefore the decomposition of symmetry vector fields $\xi^{a}$ changes:
\be  { \bar{K}^{a} \to \bar{K}^{\prime\,a} = \bar{K}^{a} - (\Lie_{\bar{K}}\, a(\theta,\varphi))\lo^{a}}\, ; \ee
The vector fields $K^{\prime\,a}$ that are tangential to the $\vo^{\prime} = {\rm const}$ cross-sections constitute  another Lorentz Lie-algebra $\LLor^{\prime}$ of $\LG$. 

Now suppose \emph{we add to our universal structure the cross-section $\izl$} at which $\scripl$ intersects $\scrimr$. Then we acquire a preferred foliation of $\scrimr$ and hence a preferred Lorentz subgroup. Furthermore, as we recalled in section \ref{s2.2},  there is a 1-1 correspondence between cross-sections of any complete NEH and \emph{non-extremal} WIH structures thereon: Given any $C$, the corresponding non-extremal null normals $[\ell^{a}]$ vanish on that $C$. What is the canonical non-extremal WIH structure $[\l^{a}]$ induced on $\scrimr$ by the cross-section $\izl$? It is given by $[\l^{a}] = [c\vo\,\lo^{a}]$.  Thus, each of these preferred non-extremal normals is indeed a symmetry vector field in $\LG$.  Note that these symmetries also leave invariant $\izl$, and the foliation $\vo = {\rm const}$. An inspection of the form (\ref{xi}) of general symmetry vector fields $\xi^{a}$ shows that the symmetry vector fields that leave $\izl$ --and the associated family of cross-sections $\vo = {\rm const}$-- invariant are precisely linear combinations of the preferred non-extremal null normals $\l^{a}\in [\l^{a}]$, and the horizontal vector fields $K^{a}$ that are tangential to the preferred foliation selected by $[\l^{a}]$:
\be \xi^{a}_{\rm loc}  = \kappa v_{o}\, \lo^{a} + K^{a}  \equiv \l^{a} + K^{a} \qquad \hbox{\rm for some}\,\, \l^{a} \in [\l^{a}] \equiv [\vo\,\lo^{a}] \label{xi2}\ee 
They constitute a seven dimensional sub-Lie-algebra, isomorphic to $\LG_{7} = \LG/\LS$. That is, because we fixed a cross-section $\izl$ of $\scrimr$, we are able to find a canonical lift of the quotient $\LG/\LS$ into $\LG$. Recall that $\G_{7}$ is the trivial central extension of the Lorentz group:  $\G_{7} = \mathbb{R} \times \Lor$. Motivated by examples discussed in section \ref{s3} and Appendix \ref{a1}, we will refer to the $\mathbb{R}$ part,  `the 1-dimensional group of time-translations', and label it by $\T_{1}$. Its induced action on a suitable phase space will lead us to a Hamiltonian that we will identify with energy. Similarly the induced action of $\Lor$ will lead us to the notion of angular momentum.
\smallskip

To summarize, the addition of the 2-sphere cross-section $\izl$ to the universal structure reduces the infinite dimensional symmetry group $\G$ of $\scrimr$ to a seven dimensional subgroup $\G_{7}$. Its generators are given by (\ref{xi2}).  This reduction has several interesting features that \emph{bring out a  non-trivial confluence of ideas and structures} from: (i) the theory of WIHs,\,  (ii) specific examples we discussed in detail in section \ref{s3} and Appendix \ref{a1}; and, (iii) our strategy of using $\scrimr$ and $\scriml$ as the appropriate analogs of $\scrimz$ in the asymptotically flat case. These features can be summarized as follows.
\vskip0.2cm
1. The reduced symmetry group $\G_{7}$ preserves $\scriml$ because the reduction gets rid of the supertranslation subgroup $\S$  ---precisely the elements of the full symmetry group $\G$ that fail to leave $\izl$ --and hence $\scriml$-- invariant.\vskip0.1cm
2.  The affine parameter $v$ of every preferred \emph{non-extremal} null-normal $\l^{a} \in [\l^{a}]$ selected by $\izl$ runs from $-\infty$ (at $i^{-}$) to $\infty$ (at $\izl$).  Thus, each of these non-extremal null normals is a future directed and complete vector field on $\scriml$. (It is also a complete vector field on the complement $\scrimr\, \setminus \,\scriml$ of $\scriml$ but there it is past-directed.) In examples discussed in section \ref{s3} and Appendix \ref{a1}, the preferred non-extremal null normals $\l^{a}$ are all restrictions to $\scrimr$ of a `time-translation' Killing vector field, which are future directed and time-like in a neighborhood of $\scriml$ in $\Ml$. The affine parameter of $\vo$ of $\lo^{a}$ corresponds to the Kruskal coordinate $V$, while the affine parameter $v$ of $\l^{a}$ corresponds to the Eddington-Finkelstein coordinate, which was also denoted by $v$. \vskip0.1cm
3. Recall from section \ref{s2.2} that each non-extremal horizon admits a canonical foliation on which the pull-back $\bar\omega_{a}$ of the rotational 1-form $\omega$ is divergence-free. Therefore, it would appear that we have two preferred foliations of $\scriml$: one provided by the non-extremal null normals $[\l^{a}]$, and another provided by the $\vo= {\rm const}$ cross-sections, which serve as affine parameters of the \emph{extremal} null normals $[\lo^{a}]$, with $\vo =0$ at $\izl$. However, the first family actually coincides with the second! This can be seen as follows. The pull-back $\mathring{\bar\omega}_{a}$ to the $v_{o} = {\rm const}$ cross-sections of the rotation 1-form $\mathring{\omega}_{a}$ of $\lo^{a}$ is divergence-free,  by the very definition of the canonical $[\lo^{a}]$. Now,  since $\l^{a} = \kappa \vo\, \lo^{a}$, and the rotation 1-form $\omega_{a}$ of the non-extremal $\l^{a}$ is given by $\omega_{a} = \mathring{\omega}_{a} + D_{a} \ln \kappa \vo$. Since the pull-back of the second term to the $v_{o} = {\rm const}$ 2-spheres vanishes, it follows that $\bar\omega_{a} = \mathring{\bar\omega}_{a}$. Hence the canonical foliation on non-extremal WIHs determined by the condition that $\bar\omega_{a}$ be divergence-free on each leaf of the foliation is satisfied by the $\vo = {\rm const}$ foliation. Thus, there is a pleasing coming together of: \, (i) the  canonical extremal null normals $[\lo^{a}]$, \,the canonical foliation associated with the \emph{non}-extremal null normals $[\l^{a}]$ selected by any cross-section of $\scrimr$,\; and, (iii)  the symmetry vector fields  $\xi^{a}$ in $\LG_{7}$. \vskip0.1cm
4. Since $v= \ln \vo$ is an affine parameter for $\l^{a} = \vo\, \lo^{a}$ it follows that the canonical foliation can be labeled either by $\vo = \rm{const}$ or by $v= {\rm const}$.  This foliation provides  us with a family of `good-cuts'  of $\scriml$. As we saw, the reduction from the infinite dimensional $\G$ to its 7-dimensional sub-group $\G_{7}$ occurs if we add this 1-parameter family to our universal structure. \vskip0.1cm 
5. It is instructive to compare the situation at the past null infinity $\scrimz$ of asymptotically flat space-times. There, 
if we work with Bondi conformal frames, we obtain a preferred 4-parameter family of null-normals. Motions along these  null normals generate the 4-dimensional subgroup $\mathcal{T}$ of BMS translations of  $\B$. Therefore, if we are given a cross-section $C$ of $\scrimz$ one obtains a \emph{4-parameter family} of cross-sections, related to the initial $C$ by elements of $\T$. On $\scrimr$, by contrast we have a 1-parameter family of preferred null normals $[\lo^{a}]$. Therefore, if we fix a cross-section  $C$ on $\scrimr$, we obtain   a \emph{1-parameter family} of cross-sections. In particular, then, by fixing $\izl$, we obtain a \emph{rest frame} on $\scriml$. Such a frame is not available at $\scrimz$ of asymptotically flat space-times.\vskip0.1cm
6. Symmetry reduction from $\G$ to  $\G_{7}$  is analogous to what happens at $\scrimz$ of asymptotically flat space-times of isolated systems. To begin with, the symmetry group of $\scrimz$ is the infinite dimensional BMS group $\B$. However,  because the Bondi-news tensor vanishes on $\scrimz$, we obtain a canonical 4-parameter family of `good-cuts' \cite{rp1,aa-radmodes,aa-yau}. As we noted in the beginning of section \ref{s4}, if we add this family to the universal structure of $\scrimz$, then $\B$ reduces to a 10 dimensional Poincar\'e sub-group $\mathcal{P}$ thereof. Note that, while in the Poincar\'e group there is no preferred time-translation, $\G_{7}$ admits a preferred 1-parameter family --in fact this is the only time-translation subgroup in $\G_{7}$. This difference is directly related to the presence of a canonical rest frame on $\scriml$. \medskip

We conclude by noting that the main considerations of this section hold if we add to the universal structure \emph{any} cross-section $C$ (which then provides, via our canonical extremal null normals $[\lo^{a}]$, a 1-parameter family of cross-sections).  Our use of $\izl$ for $C$ was motivated by the special role it plays in examples. More generally, if  the past horizon $E^{-}(i^{+})$ is long-enough to intersect $\scrimr$, we have available three notions --\,\,$\scripl$, $\izl$ and $\scriml$-- that are analogous to $\scripz$, $i^{o}$ and $\scrimz$ in the $\Lambda=0$ case  (see footnote \ref{fn3}). 

\section{Physical fields and conserved Charges}
\label{s5}

This section is divided into two parts. In the first we collect the `leading order' physical fields that are available at $\scrimr$ and $\scriml$ of any space-time representing an isolated system in our class $\CLi$. In the second, we use these fields together with symmetries at $\scriml$  to introduce the notion of total mass and angular momentum of the system from the perspective of the local space-time $\Ml$. It would be interesting  to investigate how our notions of symmetries  and conserved quantities are related to those introduced in \cite{cfp} on general null boundaries.

\subsection{Physical fields at the past boundary}
\label{s5.1}

Fields on $\scrimr$ are of two types:  (i) the `universal ones' that are common to all space-times in our class $\CLi$ that were discussed in section \ref{s4.1}; and, (ii)  fields that vary from space-time to space-time, some of which were mentioned in  section \ref{s2}. In this sub-section, we will gather the geometric structures and fields on $\scrimr$ from sections \ref{s2} - \ref{s4}. This succinct list will help us streamline the discussion of conserved quantities in the next subsection.  (For proofs and derivations, see \cite{afk,abl1}. ) \vskip0.1cm

Let us fix a space-time $(\Mr,\,g_{ab})$ in our class and examine the structures that are induced on $\scrimr$ by the space-time metric $g_{ab}$.  First, the past boundary $\scrimr$ of $\Mr$ comes with a preferred equivalence class  $[\lo^{a}]$ of null normals, which are complete, affinely parametrized geodesics w.r.t. $g_{ab}$.  Here $\lo^{a} \in  [\lo^{a}]$ and $\lo^{\prime\,a} \in [\lo^{a}]$ if and only if $\lo^{\prime\, a}= c \lo^{a}$ for some positive constant $c$. These null geodesics provide a ruling of $\scrimr$ and the quotient,  $\t{\scri}^{\,-}_{\rm Rel}$, is topologically $\mathbb{S}^{2}$.  However, because surface gravity $\kappa_{\lo}$, expansion $\Theta_{\lo}$ and shear $\sigma_{\lo}$ all vanish for \emph{every} $\lo \in [\lo]$, there is no simple way to remove the rescaling freedom in $c$ and extract a preferred null normal $\lo^{a}$ in the equivalence class.  As we will see in Appendix \ref{a2}, this fact has an important consequence.  
The second field is $q_{ab}$, the pull-back of $g_{ab}$ to $\scrimr$. $q_{ab}$ is  a degenerate metric of signature (0, +, +),  satisfying $q_{ab} \lo^{a} =0$ and $\Lie_{\lo} q_{ab} =0$. Thus, $q_{ab}$ is the pull-back to $\scrimr$ of a metric  $\t{q}_{ab}$ on the base space $\t{\scri}^{-}_{\rm Rel}$.  The third field on  $\scrimr$ an area 2-form $\epsilon_{ab}$, the pullback to $\scrimr$ of the area 2-form $\t\epsilon_{ab}$ compatible with the metric $\t{q}_{ab}$ on $\t{\scri}^{-}_{\rm Rel}$.  These are the \emph{`zeroth order' fields} on $\scrimr$, in the sense that they are directly induced by the space-time metric itself. 

The \emph{`first order' field}  is a (torsion-free) intrinsic derivative operator $D$ on $\scrimr$, induced by the (torsion-free) space-time derivative operator $\nabla$ on $\Mr$ compatible with $g_{ab}$.  Since $D$ is the pull-back of $\nabla$, it follows immediately that it satisfies $D_{a} q_{bc} =0$ and $D_{a} \epsilon_{bc} =0$. Next, given \emph{any} null normal $\l^{a}$ to $\scrimr$, we acquire a 1-form $\omega_{a}$ on $\scrimr$ through $D_{a}\l^{b} = \omega_{a}\l^{b}$ (since $D_{a}\l^{b}$ is necessarily proportional to $\l^{b}$ in any NEH).  As we explained in section \ref{s2.2}, the 1-form $\omega_{a}$ is tied to the null normal $\l^a$ and under $\l^{a} \to \l^{\prime\,a}= f \l^{a}$, we have $\omega_{a} \to 
\omega^{\prime}_{a} = \omega_{a} + D_{a} \ln f$.  But for notational simplicity we will not attach a label $\l^{a}$ to $\omega_{a}$. The 1-form $\omega_{a}$, in turn leads to several interesting structures that will play an important role for us: \smallskip

\noindent 1. The component $\kappa_{\l} = \omega_{a} \l^{a}$ of the 1-form  $\omega_{a}$ is the \emph{surface gravity} of $\l^{a}$. If $\kappa_{\l} =0$, the null normal $\l^a$ to $\scrimr$ is said to be \emph{extremal}; if $\kappa_{\l}\not= 0$, it is said to be \emph{non-extremal}. The natural null normals $\lo^{a} \in [\lo^{a}]$ on $ \scrimr$ are all extremal.\\
2. Given a \emph{non}-extremal WIH structure $[\l^{a}]$ on $\scrimr$, one acquires a \emph{unique} foliation of $\scrimr$ by a 1-parameter family of 2-spheres \cite{abl1}. The defining property of this foliation is that the pull-back $\b{\omega}_{a}$ of $\omega_{a}$ is divergence-free on each leaf of the foliation: $\b{q}^{ab} \b{D}_{a} \b\omega_{b} =0$, where $\b{q}^{ab}$ is the natural metric on the leaves of the foliation and $\bar{D}$ the derivative operator compatible with it. The 1-parameter family of diffeomorphisms generated by any $\l^{a} \in [\l^{a}]$ leaves this foliation invariant. In particular, for each $\l^{a}$ one obtains a \emph{unique} affine parameter $v$, up to the shift of origin,\, i.e., up to $v\to v +{\rm const}$.\\
3. Given a cross-section $C$ of $\scrimr$, there is a \emph{unique} non-extremal equivalence class  $[\l^{a}]$ that endows $\scrimr$ with the structure of a WIH.  Each $\l^{a}$ in $[\l^{a}]$ vanishes on the cross-section $C$ (and nowhere else on $\scrimr$). The converse is also true:\, every non-extremal WIH structure $[\l^{a}]$ on $\scrimr$ determines a unique cross section $C$ on which each $\l^{a} \in [\l^{a}]$ vanishes.\\
4. \emph{If} the past event horizon $E^{-}(i^{+})$ is long enough so as to intersect $\scrimr$ in a cross-section (which we labelled by $\izl$), then because of property 3, we acquire a preferred non-extremal WIH structure $[\l^{a}]$ on $\scrimr$. In this case, the portion of $\scrimr$ joining $i^{-}$ to $\izl$ defines $\scriml$ --which can be regarded as \emph{local} $\scrim$-- and the portion of $E^{-}(i^{+})$ joining $i^{+}$ to $\izl$ defined $\scripl$ --which can be  regarded as \emph{local} $\scrip$.\\
5. The null normals $\l^{a}$ on $\scriml$ generate the 1-dimensional time-translation subgroup $\T_{1}$  of the symmetry group $\G_{7}$. Because of property 2, $\scriml$ is equipped with a preferred foliation, defining a `rest frame'. \\
6. Under a constant rescaling of a null normal, $\l^{a} \to \l^{\prime\, a} = c \l^{a}$, we have $\kappa_{\l^{\prime}} = c \kappa$. Therefore,  given a \emph{non-extremal} WIH structure $[\l^{a}]$,  one can select a preferred null normal $\l^{a}$ in the equivalence class $[\l^{a}]$ by specifying a (non-zero) value of surface gravity. This is in stark contrast with the preferred family $[\lo^{a}]$ of  null normals on $\scrimr$ which are \emph{extremal}. \smallskip

These properties, together with the interplay between physics and geometry in de Sitter, Schwarzschild-de Sitter and Kerr-de Sitter space-times  (discussed in section \ref{s2} and Appendix \ref{a1}), will lead us to a natural strategy to define the mass of a general space-time in our class $\CLi$.  In these examples, $E^{-}(i^{+})$ is indeed long enough to provide us with $\izl$. Furthermore,  the WIH structure provided by the resulting $[\l^{a}]$ is induced on $\scrimr$ by a Killing vector field $t^{a}$ which  is null on $\scrimr$ and vanishes on $\izl$ (and nowhere else).%
\footnote{In section \ref{s2}, this Killing field is denoted by $T^{a}$ and in Appendix \ref{a1}, by $\ub{t}^{a}$.} 
This Killing field ${t}^{a}$ is a `time-translation' in the sense that it is time-like in a (large) neighborhood of $\scri^{\pm}_{\rm Loc}$, with orbits that are topologically $\mathbb{R}$. Furthermore, in these space-times  $t^{a}$ is the \emph{unique} Killing field, up to a constant rescaling,  with these properties. Now, since in the limit  $\Lambda \to 0$, the $\scriml$ of these space-times becomes the $\scrimz$ of Minkowski, Schwarzschild and Kerr space-times, we can fix this rescaling freedom in  ${t}^{a}$ by requiring that it approach the \emph{unit} time-translation Killing field  of these space-times in a neighborhood of their $\scrimz$. Interestingly,  this normalization can be directly transferred to general space-times of interest using property 5 above.  More precisely,  in all examples, the correctly normalized  `time-translation' Killing field $\ub{t}^{a}$ has the property that its restriction $\ul{\l}^{a}$ to $\scriml$ has a specific surface gravity: $\kappa_{\ul{\l}} = (1/2R_{(c)})( 1- 3 (R_{(c)}^{2}/\l^{2}))$.  This will lead us to associate mass $M$ of a general space-time with that null normal  $\ul{\l}^{a}$ in the equivalence class  $[\l^{a}]$, selected by $\izl$, which has surface gravity $\kappa_{\ul{\l}}$.

This concludes our discussion of the `first order structure' at $\scrimr$ made available by the derivative operator $D$. \medskip

The \emph{`second order' structure} at $\scrimr$ is induced by space-time curvature.  The fact that $\scrimr$ is an NEH immediately leads to constraints on the Ricci tensor $R_{ab}$ of the space-time metric $g_{ab}$, evaluated on $\scrimr$:
\be R_{ab}\ell^{a} X^{b} =0 \,\,\,\, \forall \,X^{a} \,{\hbox {tangential to\,\,}} \scrimr, \quad{\hbox {\rm which in particular implies}}\,\,\, R_{ab}\l^{a}\l^{b} =0\, . \ee
In the Newman-Penrose notation these conditions translate to the vanishing of  4 components,  $\Phi_{00}$ and $ \Phi_{01}$ of the Ricci tensor. For the Weyl tensor,  we have
\be C_{abcd}\,  X^{a}_{1}X^{b}_{2} X^{c}_{3} \,\l^{d}=0, \quad \forall \, X^{a}_{1}, \,X^{a}_{2}, \, X^{a}_{3}\,\,\,\,  {\hbox{\rm tangential to\,\,} }\scrimr \,\ee
which, in the Newman-Penrose notation implies that 4 components, $\Psi_{0}$ and $\Psi_{1}$, of the Weyl tensor  must also vanish on $\scrimr$. (Recall that in the $\Lambda=0$ case, $\Psi_{0}^{o}$ is the radiation field on $\scrimz$ and both $\Psi_{0}^{o}$ and $\Psi_{1}^{o}$ vanish if the Bondi news vanishes on $\scrimz$.) 

As we already remarked in section \ref{s2.2}, the one-form $\omega_{a}$ serves as a potential to $\Im\,\Psi_{2}$  on $\scrimr$:
\be  2\, \Im\, \Psi_{2}  = \epsilon^{ab}\, D_{[a} \omega_{b]}\, . \label{impsi2}  \ee
We will see that the angular momentum at $\scrimr$  --which represents the total angular momentum of space-time-- is determined by $\Im\, \Psi_{2}$. The mass, on the other hand, is encoded in $\Re\Psi_{2}$  which is related by Eq. (\ref{repsi1}) to the scalar curvature of the 2-metric $\bar{q}_{ab}$ on any cross-section $C$ of $\scrimr$:
\be {}^{2}\b{\R} = - 4 \,\Re\Psi_{2} + \f{2}{3} \Lambda +  {8\pi G}\, (2\l^{a}n^{b} T_{ab} +\f{1}{3} T)\, . \label{repsi2}\ee
where $\l^{a}$ is any null normal to the NEH, $n^{a}$ the other null normal to $C$ such that  $g_{ab}\l^{a}n^{b}=-1$, and  $T$ is the trace of the stress energy tensor. (For a proof, see  Appendix \ref{a2}.)

\subsection{Conserved charges}
\label{s5.2}

This sub-section is divided into three parts. In the first, we introduce the notion of mass $M$ using a physical thought experiment involving tidal acceleration. In the second, we obtain an expression for energy as the Hamiltonian generating time-translations $\T_{t}\in \G_{7}$ and discuss its relation to  $M$.  In the third, we discuss angular momentum as the Hamiltonian generating Lorentz transformations $\Lor \in \G_{7}$. 

\subsubsection{Mass at $\scriml$}
\label{s5.2.1}

Already in the $\Lambda=0$ case, we had to develop intuition as to what constitutes mass in general relativity.  The early analysis by Arnowitt Deser and Misner and others \cite{adm} of the structure of the gravitational field at spatial infinity and by by Bondi, Sachs, Penrose and others \cite{bondi,sachs,rp1} at null infinity led us to precise notions of mass in the two regimes. As a result, we habitually identify the parameter $m$ in the Kerr family as the ADM or the Bondi mass of the space-time. But as Appendix \ref{a1} shows,  this identification is no longer tenable for Kerr-de Sitter metrics: the notion of mass is more complicated even for this special, explicitly known  family. Therefore, to define mass at  $\scriml$ for general space-times, we need further guidance. In this section we will begin by introducing some physical considerations as motivation, then define mass on $\scriml$, and finally discuss properties of this notion of mass.\bigskip

\centerline{\it Motivation}\bigskip

Since $\scriml$ is analogous to $\scrimz$ of isolated systems in the $\Lambda=0$ case, let us begin by recalling the notion of the Bondi mass in that case.  Fix an asymptotically flat space-time $(\Mo, g^{o}_{ab})$. We will work in a neighborhood of $\scrimz$ where $T_{ab}=0$. Let us introduce Bondi coordinates $(v, r, \theta, \phi)$ such that $\partial_{v}$ is the asymptotic  time-translation in the asymptotic rest frame of the system. In these coordinates, the   3-surfaces $v= {\rm const}$ are portions of ingoing null cones. Let us fix one, say $v=v_{o}$, and foliate it by a family of  2-spheres $R=const$, where $R$ is the area-radius of the 2-spheres. Denote these 2-spheres by $C_{R}$. As $R$ increases, the 2-spheres $C_{R}$ approach a cross-section $C$ of $\scrimz$. Next, introduce a Newman-Penrose null tetrad adapted to this foliation of the $v=v_{o}$ surface:  Let $n^{a}$ be  a (future directed)  null normal to the $v=v_{o}$ 3-surface, and let $\l^{a}$ be the other (future pointing) null normal to the $R=R_{o}, v= v_{o}$ 2-spheres, with $n^{a}\l_{a} = -1$. Then, using the no incoming radiation condition on $\scrimz$, the Bondi mass can be defined by the following limiting procedure:
\be M_{\rm Bondi}  = - \f{1}{4\pi G} \lim_{C_{R} \to C}\,\, \oint_{C_{R}}  R\,\, \Re\, \Psi_{2}\, \rmd^{2}V\, . \label{bondiM}\ee
Here  $\Re\,\Psi_{2}= \f{1}{2} {C}_{abcd} {n}^{a} {\l}^{b} {n}^{c} {\l}^{d}$ is the  component  of the Weyl tensor that falls off as $1/r^{3}$ in asymptotically flat space-times \cite{rp1}, capturing the `Coulombic aspect' of the asymptotic gravitational field. (Note that $\Psi_{2}$ is insensitive to the rescaling of the initial choice of $n^{a}$.) Because of the `no-incoming radiation' condition,  $M_{\rm Bondi}$ equals the ADM mass and thus represents the total mass of the system.

While in the post-Newtonian limit one finds an explanation of how mass can be identified, e.g. using geodesics of test particles in standard textbooks,  somewhat surprisingly 
it appears that a similar \emph{physical} `justification'  as to why the right side of  (\ref{bondiM})  should represent the mass at $\scrimz$  does not exist in the literature. Therefore we will first present such a justification and then use it to motivate the definition of mass at $\scriml$ in the $\Lambda >0$ case. Let us begin with an isolated system in \emph{Newtonian gravity}. So the matter density has compact spatial support and the Newtonian potential is given by $\Phi = - GM/r + O(1/r^{2})$. In full general relativity, it is the tidal force  $\nabla_{a} \nabla_{b}  \Phi$ that has a clean counterpart in terms of curvature. So, let us express mass $M$ in terms of the tidal force. For this, we can consider a large 2-sphere of radius $r$ surrounding the matter source, and a nearby concentric 2-sphere of radius $r-\delta$. Let us now consider a shell of (massive) test particles at rest on each of these two 2-spheres. Let us drop them at $t=0$. Then, to the leading order, the 2-spheres will continue to remain 2-spheres but their separation will increase because of tidal effects associated with the inhomogeneity of the field because particles on the inner 2-sphere will experience a slightly greater acceleration  than those on the outer 2-sphere, whence $\delta$ will increase in time. To the leading order, we have: 
\be \ddot\delta = \f{2GM}{r^{3}}\, \delta \,\,  = \,\, \delta\,\h{r}^{a}\, \h{r}^{b} D_{a} D_{b} \Phi 
\label{tidal}\ee
where $D_{a}$ is just the 3-dimensional derivative operator of the Euclidean space. This equation leads to an expression of mass of the isolated system in terms of the tidal acceleration, as a limit of a 2-sphere integral
\be \label{newtonianM} M = \f{1}{8\pi G} \, \lim_{{r_{o}\to \infty}} \,\oint_{r=r_{o}}  r\, \h{r}^{a}\, \h{r}^{b} D_{a} D_{b} \Phi\,\, \rmd^{2}V  \ee

We can now carry over this physical idea to general relativity by replacing the Newtonian tidal acceleration with the appropriate component of curvature that features in the geodesic deviation equation. Let us consider an isolated system in general relativity (with $\Lambda =0$) represented by an asymptotically flat space-time as in our discussion that led to Eq. (\ref{bondiM}). We can consider two concentric spheres  $R=R_{o}$ and $R= R_{o}-\delta$  on the $v=v_{o}$ surface in the asymptotic region, where $R$ denotes the area radius. Using the null vector fields $\l^{a}$ and $n^{a}$, let us define a unit time like vector field $\h{t}^{a}$ and a space-like (radial) vector field $\h{r}^{a}$, both orthogonal to the family $C_{R}$ of 2-spheres: $ \h{t}^{a} = (1/\sqrt{2})\,(\l^{a} + n^{a})$ and $\h{r}^{a} = (1/\sqrt{2})\,(n^{a} - \l^{a})$.  We again consider a shell of (massive) test particles on each of the two shells with 4-velocities aligned with $\h{t}^{a}$ at the initial time and let them freely fall, i.e., follow geodesic orbits. Then,  by the standard geodesic deviation equation, at the initial time  we have: 
 \ba \h{r}_{a} \,(\delta\, \h{r}^{a})^{\cdot\cdot} &=& -\, (\delta)\,\, \h{r}^{a}\, \h{r}^{c} (\h{t}^{b}\h{t}^{d}\, R_{abcd})  \nonumber\\
 &=& - (\delta)\,\, n^{a}\, n^{c} \l^{b} \l^{d}\, C_{abcd}  
 \,\,\,= - 2(\delta)\, \Re\, \Psi_{2}  \ea
where in the second equality  we have used the fact that the test particles are all in the asymptotic, source-free region where the Ricci tensor of $g_{ab}^{o}$ vanishes, and in the third step we have used the definition of the component $\Re\,\Psi_{2}$ of the Weyl tensor.  The thought experiment suggested by Newtonian considerations leads us to replace the newtonian tidal acceleration $\h{r}^{a} \h{r}^{b} D_{a} D_{b}\Phi$  by $-2 \Re \Psi_{2}$ and think of the resulting integral 
\be  - \f{1}{4\pi G}\, \oint_{C_{R}} R\, \Re\, \Psi_{2}^{o}\, \rmd^{2}V\ee
as the `mass contained in the 2-sphere $C_{R}$'. Now, in general relativity gravity itself gravitates. Therefore, even if the matter source is confined to some spatially compact region, to obtain the \emph{total} mass we have to take a limit as $R\to \infty$ i.e. the family of 2-spheres $C_{R}$ tend to the cross-section  $C$ of $\scrimz$. When this is done, we recover precisely the expression (\ref{bondiM}).  \smallskip

\emph{Remark:}  Since the displacement vector $\delta r^{a}$ is initially orthogonal to $\h{t}^{a}$, it continues to remain orthogonal since $\Lie_{t} \h{r}^{a} =0$,\, $\h{t}^a \nabla_{a} \h{t}_{b} =0$, and $\h{t}^{a} \hat{t}_{a} = -1$. Therefore $(\delta\, \h{r}^{a})^{\cdot\cdot}$ has components only along $\h{r}^{a}$ and angular directions $\hat{m}^{a}$ and we have: $(\delta\, \h{r}^{a})^{\cdot\cdot} = - 2(\delta)\, [\Re\,\Psi_{2}\,\, \h{r}^{a}  +\Re\, \Psi_{1} \,\, \h{m}^{a}]$ in the Newman-Penrose notation (see Appendix \ref{a2}). However,  because of the `no incoming radiation' condition at $\scrimz$, the contribution from $\Psi_{1}$ vanishes in the limit  and only the term  $\Re \Psi_{2}$ survives; i.e., in the limit the vector $(\delta\, \h{r}^{a})^{\cdot\cdot} $ becomes just $\ddot\delta \, \h{r}^{a}$ and the analogy with the Newtonian expression becomes even closer.
\bigskip

\centerline{\it $\Lambda >0$: Definition of mass and its properties}
\bigskip

The strategy is to carry over this physical idea to the $\Lambda >0$ case.  However, there is a new conceptual subtlety: now the Ricci tensor is non-zero outside matter sources, given by $R_{ab} = \Lambda g_{ab}$, and this part of the curvature also contributes to the geodesic deviation. In particular, while there is no geodesic deviation in Minkowski space-time, there \emph{is} non-trivial geodesic deviation  in de Sitter space just due to cosmic expansion. Thus,  there is a part of geodesic deviation that has nothing to do with the presence of physical mass in the space-time and we have to subtract it out to obtain the mass of the isolated system under consideration. Fortunately this can be done rather easily because the Riemann tensor neatly decomposes into the Weyl and the Ricci parts.

Let us then consider the same thought experiment, replacing the asymptotically flat space-time $(\Mo, g^{o}_{ab})$ by a space-time $(M, g_{ab})$ in our class. Furthermore, since $\scriml$ is now `at a finite distance' we need not consider a limiting procedure, but  simply start by considering a shell of (massive) test particles of area radius $R=R_{(c)}$ that lies on $\scriml$, and another shell of radius $R = R_{(c)} - \delta$. Then, the geodesic deviation equation for (massive) test particles on these two 2-spheres now yields;
\ba (\delta\, \h{r}^{a})^{\cdot\cdot} &=& -\, (\delta)\,\, \, \h{r}^{c} (\h{t}^{b}\h{t}^{d}\, R^{a}{}_{bcd})   \nonumber\\
&=& -\,(\delta)\,\, \h{r}^{c} \h{t}^{b}\h{t}^{d} \big[C^{a}{}_{bcd} + \f{2}{\ell^{2}}\, \delta^{a}{}_{[c} \,  g_{d]b}\big]  
\nonumber\\  
 &=& \,(\delta)\,\, [\f{1}{\l^{2}} \h{r}^{a} - C^{a}{}_{bcd} \h{t}^{b} n^{c}\l^{d} ]   =(\delta)\,  [\f{1}{\l^{2}} - 2 \Re\Psi_{2}\big] \h{r}^{a}\ea
where in the last step we have used the fact (noted above in the Remark) that $(\delta\, \h{r}^{a})^{\cdot\cdot} $ is orthogonal to $\h{t}^{a}$ and $\Psi_{1}$ vanishes on $\scriml$. 

The first term in the last step can be directly identified as the contribution to the geodesic deviation due to the cosmological constant. It is non-zero already in de Sitter space, where the separation $\delta$ between the shells will increase just because of the accelerated expansion of the universe, even though there is no physical mass in the space-time. The second term vanishes in de Sitter space-time and represents the geodesic deviation over and above the contribution due to the cosmic accelerated expansion. It is then natural to attribute this part to the presence of the mass within the 2-sphere $C_{R}$. These considerations lead us to define the mass at $\scriml$ as:
\be M = - \f{1}{4\pi G}\, \oint_{C} R_{(c)}\, \Re\, \Psi_{2}\,\, \rmd^{2}V\, ,   \label{mass}
\ee
where $C$ is any cross-section of $\scriml$ and $R_{(c)}$ is the area-radius of $\scriml$. Thus, the mass $M$ is completely determined by the following physical fields on $\scrimr$:  the area radius $R$, the component $\Psi_{2}$ of the Weyl tensor, and the area 2-form $\epsilon_{ab}$ on $\scrimr$ that defines the volume element $\rmd^{2}V$ on $C$. Note that in the right side of (\ref{mass}) we can use any cross-section $C$  of $\scrimr$ and any (non-vanishing) null normals $\l^{a}, n^{a}$ to $C$. Because $\Psi_{0} = \Psi_{1} =0$ on the entire $\scrimr$, it follows that $\Psi_{2}$ is insensitive to these choices. Finally, 
$M$ is conserved because  $\Lie_{l} R_{(c)}=0$ and $\Lie_{\l} (\Re\, \Psi_{2}) =0$ on $\scriml$. \smallskip

This notion of mass has several interesting properties. \smallskip

1. The identity (\ref{repsi2}) relates $\Re\,\Psi_{2}$ with the scalar curvature ${}^{2}\R$ of the 2-metric $q_{ab}$ on the (base space $\t\scri^{\,-}_{\rm Loc}$ of ) $\scriml$, the cosmological constant $\Lambda$ and the trace $T$ of the stress energy tensor of matter fields at $\scriml$.  For simplicity, let us suppose that the stress-energy tensor of matter at $\scriml$, if any,  is tracefree (e.g. a Maxwell or Yang-Mills field).  Then,  since by the Gauss theorem $\oint_{C} {}^{2}\R \,\rmd^{2}V = 8\pi$, Eq. (\ref{repsi2}) implies 
\be  M =  \f{R_{(c)}}{2 G} \Big(1 - \f{R_{(c)}^{2}}{\l^{2}} \Big)\, .  \label{M} \ee
The right side vanishes if and only if $R_{(c)} = \l$, which is achieved in de Sitter space-time. Physically, one might  expect that `due to the attractive nature of gravity', the cosmological horizon is, so to say, `pulled in' by the presence of a mass in the interior. This expectation is borne out in the Schwarzschild-de Sitter family, where $R_{(c)} \le \l$ and equals $\l$ \emph{only} for the de Sitter solution.  Therefore the  right side is always positive, and reaches its maximum $M_{\rm sup} = \f{1}{3\sqrt{3}}\, \l$  in the Nariai solution, when $R_{(c)} = \l/\sqrt{3}$. For the Schwarzschild-de Sitter family, then,   $M$ equals the parameter $m$ that enters the solution.  

2. This is no longer the case for the Kerr-de Sitter family. Nonetheless,  $M$ is again positive and numerical calculations of $R_{(c)}$  show that the maximum value of $M$ is again $M_{\rm sup} = \f{1}{3\sqrt{3}}\, \l$. For this 3-parameter family, we can focus on a neighborhood  $N$ of $\scriml \cup \scripl$ within $\Ml$ and take the limit $\Lambda \to 0$, keeping  $m$ and $a$ fixed. In the limit, the space-time geometry in $N$ tends to  the space-time geometry of a neighborhood of $\scrimz \cup \scripz$ of the Kerr family and $M$ of Eq. (\ref{mass}) \emph{tends to the Bondi mass} on $\scri^{\pm}_{0}$ of the $\Lambda=0$  null infinity. We expect that this will be the case for all space-times in our class $\CLi$ for which there is a physically well-motivated procedure to take the $\Lambda \to 0 $ limit.  An example of such a procedure would be to consider a double null-surface framework to solve source free Einstein's equations in a neighborhood of $\scriml \cup \scripl$ within $\Ml$, using $\scriml$  for one of the two null surfaces. One could use a power series expansion of the solution away from $\scriml$ as in \cite{bk,jlcl}, and then take the limit $\Lambda \to 0$ in that expansion.

3. In this discussion of mass, we focused on $\scriml$ because for $\Lambda >0$, it is the natural analog of $\scrimz$ in the asymptotically flat case. However, from the strict $\Lambda >0$ perspective, we could have worked with $\scrimr$ as well, and used  an extremal null normal $\lo^{a}$ to $\scri$ (in place of the non-extremal null normal $\l^{a}$ adapted to $\scriml$) in the expressions  (\ref{mass}) of $M$ without changing the result. Since $\lo^{a}$ is nowhere vanishing on the entire $\scrimr$, the other null-normal $\no^{a}$ to any cross-section $C$ of $\scrimr$ is also well-defined, and we can use any 2-sphere cross section of $\scrimr$ to evaluate the integral. Thus the mass $M$ is really associated with the entire $\scrimr$, not just with $\scriml$. 

4. Since $M$ is conserved, we can also think of it as being associated with the point $i^{o}$ at spatial infinity, or the point $i^{-}$ at past time-like infinity, of $\Mr$ (see, e.g., Fig. \!\ref{ds-lin}). We have other definitions of mass at both these points. The one at $i^{o}$ uses space-like surfaces (such as the cosmological slices in the de Sitter space-time) that extend  to $i^{o}$ (see, e.g., \cite{abbott,km}). The one at $i^{-}$  is obtained by working with (the space-like)  $\scrim$, and imposing the `no incoming radiation condition'  by requiring that the magnetic part of the (appropriately conformally rescaled) Weyl tensor vanishes there (see, e.g. \cite{abk1}). It is likely that these definitions agree with (\ref{M}) under appropriate conditions. However, to establish these results one would need to understand the precise relation between limits of various physical fields as one approaches $i^{o}$ along $\scrimr$ and along space-like surfaces, and $i^{-}$ along $\scrim$ and along $\scrimr$.

\subsubsection{The Hamiltonian framework, energy and the first law}
\label{s5.2.2}

In section \ref{s5.2.1}  we arrived at a definition of the total mass of the space-time $(\Ml, g_{ab})$ using physical considerations involving the motion of appropriately chosen test particles near $\scriml$.  In the $\Lambda=0$ case, these considerations do yield the correct definition of mass at $\scrimz$ \cite{bondi,sachs,rp1} as well at spatial infinity $i^{o}_{o}$ \cite{adm,aarh} . However, in that case we also have conserved charges  that arise as Hamiltonians generating the action of asymptotic symmetries on a suitably defined phase space. In particular, using the 4-dimensional group of asymptotic translations, one can define the ADM and the Bondi 4-momentum of the system.
For $\Lambda>0$, the symmetry group of $\scriml$ is $\G_{7}$, and elements of the Lie algebra $\LG_{7}$ have the form: \, $\xi^{a}_{\rm loc} = \kappa\vo\lo^{a} + K^{a} \equiv \l^{a} + K^{a}$ (see Eq. (\ref{xi2})). The vertical vector fields $\l^{a}$ generate the 1-dimensional time-translation subgroup $\T_{1}$ of $\G_{7}$. Therefore, we are led to ask:\\
(i) Are there charges $Q_{\l}$ associated with  the generators $\l^{a}$ of the $\T_{1}$ ?  If so,\\ 
(ii) What is the relation between those charges and the mass $M$ defined in (\ref{M})?\\
(iii) Do these charges serve as Hamiltonians generating canonical transformations corresponding to these symmetries 
on an appropriate phase space, tailored to $\scrimr$ or $\scriml$?  and, \\
(iv) Since, in addition to being the analog of $\scrimz$ in asymptotically flat space-times, $\scriml$ is also a non-extremal WIH, do the charges associated with the time-translation symmetry of $\scriml$ satisfy a first law of horizon mechanics?\\
In this subsection we will show that the answer to these questions is in the affirmative.  Charges associated with the Lorentz generators $K^{a}$ will be discussed in the next sub-section.  Together, they provide charges associated with the symmetry Lie algebra $\LG_{7}$ on $\scriml$. On full $\scrimr$, we also have supertranslations $\xi^{a} = f(\theta,\phi) \lo^{a}$  that belong to $\LG$. The corresponding charges will be discussed in Appendix \ref{a2}.  \smallskip

Let us then begin with the 1-dimensional time-translation sub-group $\T_{1}$.  The existing literature on WIHs \cite{afk,abl2,aabkrev} spells out a procedure to construct a covariant phase space $\Gcov$ from solutions to Einstein's equations admitting a WIH boundary.   We can apply that procedure because our solutions  $g_{ab}$ on $\Ml$ do admit a WIH horizon --namely $\scriml$-- as a boundary.  Given any null normal, $\l^{a}$, generating $\T_{1}$ on $\scriml$, we can extend it to a neighborhood of $\scriml$ within $\Ml$ by a time-like vector field $t^{a}$ and consider the 1-parameter family of transformations induced on our $\Gcov$ by the diffeomorphisms generated by this $t^{a}$. It turns out that this induced action is Hamiltonian if and only if the first law holds, i.e., if and only if there is a function $E_{t}$ on $\Gcov$ such that \cite{afk}
\be \delta E_{t} = \kappa_{\l}\, \delta A  \label{1law}\ee
for any vector field $\delta$ on $\Gcov$,  where $A$ is the area of any cross section of $\scrimr$ and $\kappa_{\l}$ is the surface gravity of the null normal $\l$ we began with. If this condition is satisfied, then $E_{t}$ is the Hamiltonian function on $\Gcov$ generating the canonical transformation. Note that this condition refers only to the boundary value $\l^{a}$ of $t^{a}$ and not to the details of our extension of $\l^{a}$ away from $\scriml$ (whence we could have used the symbol $E_{\l}$ in place of $E_{t}$). 

Now, $\scriml$ is endowed with a canonical null normal $\ul\l^{a}$ with surface gravity $\kappa_{\ul\l} = (1/2R_{(c)})\, (1-3 R_{(c)}^{2}/\l^{2})$, where, as before $R_{(c)}$ is the area radius of any cross-section of  the cosmological horizon $\scriml$. Every $\l^{a}$ that generates a time-translation symmetry is proportional to $\ul\l^{a}$:\, $\l^{a} = k \ul\l^{a}$ where $k$ is a positive constant. Therefore,  one can easily integrate (\ref{1law}) on $\Gcov$ to obtain 
\be E_{t} \,=\, k \, \f{R_{(c)}}{2G}\,  \Big(1 - \f{R_{(c)}^{2}}{\l^{2}} \Big)\,\, \equiv\,\, k M \, ,\label{energy} \ee
where in the last step we have used Eq. (\ref{M}).  Thanks to the expression  (\ref{mass}) of $M$,  this function $E_{t}$ 
on $\Gcov$ provides  an explicit linear map from the space of time-translations $\l^{a}$ on $\scriml$ to $\mathbb{R}$ via
\be \l^{a}\,\,\,   \to \,\,\, E_{t}  =  - \, \f{1}{8\pi G}\, \oint_{C}  R_{(c)} \,C_{abpq} \,\ul{l}^{a}\ub{n}^{b}\, \l^{p} \ub{n}^{q} , \rmd^{2}V \,\, 
\equiv    - \, \f{k}{4\pi G}\, \oint_{C}  R_{(c)}  \Re\, \Psi_{2}\,  \rmd^{2}V  \label{Et} \ee
where $\ub{n}_{b}$ is any future pointing null vector field on $\scriml$ satisfying $\ul{\l}^{a} \ub{n}_{a} = -1$. Thus, the numerical value of the Hamiltonian $E_{t}$ generating the time-translation $t^{a}$ is $M$ precisely if $t^{a}|_{\scriml} = \ul{l}^{a}$, the preferred null normal. As discussed in the Appendix \ref{a1},  in the Kerr-de Sitter space-time, $\ul\l^{a}$ is the restriction to $\scriml$ of the \emph{unique} Killing field $\ub{t}^{a}$ with the following key properties: (i) It is time-like in a neighborhood of $\scriml \cup \scripl$ within $\Ml$; and (ii) is so normalized that in the limit $\Lambda \to 0$, its norm with respect to the physical metric tends to $-1$ as one approaches $\scrim_{0} \cup \scrip_{0}$.  Thus, near $\scriml \cup \scripl$, the Killing field  $\ub{t}^{a}$ in Kerr-de Sitter space-time is the precise analog of the properly normalized time-translation Killing vector field in Kerr space-time near $\scrim_{0} \cup \scrip_{0}$.  That fact led us to a `correctly normalized' generator $\ul{\l}^{a}$ of time-translations in $\T_{1}$  for all space-times in $\Gcov$.
We have now found that the energy associated with these time-translations by Hamiltonian considerations is precisely the mass $M$ we obtained from the `tidal acceleration' considerations in section \ref{s5.2.1}.\vskip0.2cm

To summarize, there \emph{is} a Hamiltonian framework that lets us define energy $E_{\l}$ for each generator $\l^{a}$ of the time-translation subgroup $\T_{1}$ of the symmetry group $\G_{7}$. If the generator is so normalized as to correspond to the unit time-translation ${t}^{a}$ in the Kerr family, then the energy equals mass: $E_{\ul\l} = M$. For the Kerr family the energy is positive and heuristics motivated by the `attractive nature of gravity' suggest that the energy and (hence also the mass) on $\scriml$ should be positive in general. Note that, in contrast to the asymptotically flat case, we do not have a notion of 3-momentum (or, alternatively, the 3-momentum vanishes) because the available structure naturally leads us to a preferred rest frame, reflected in the fact that the translation subgroup is one dimensional and, furthermore,  there is only 1-parameter family of `good cuts' of $\scriml$. By contrast, in  the asymptotically flat case we have a 4-dimensional translation subgroup on $\scrimz$ and absence of radiation leads us to a 4-parameter family of `good cuts'. Different 1-parameter sub-families define different rest frames, whence we are led to the (Bondi) 4-momentum. \smallskip

\subsubsection{Angular momentum}
\label{s5.2.3}

Recall from section \ref{s4.3} that  $\scriml$ admits a natural foliation, and its symmetry group $\G_{7}$ admits a canonical Lorentz subgroup $\Lor$ whose action leaves each leaf of this foliation invariant. As before let us denote the vector fields generating $\Lor$ by $K^{a}$; these are the `horizontal' vector fields in $\LG_{7}$. The WIH framework provides a natural strategy to define charges $Q_{K}$ associated with each of these vector fields. Let us extend these vector fields in a neighborhood of $\scriml \cup \scripl$ within $\Ml$ and denote the extension also by $K^{a}$. Then diffeomorphisms generated by any one $K^{a}$ induce a 1-parameter family of  canonical transformations on $\Gcov$ and $Q_{K}$ are precisely the corresponding Hamiltonians \cite{abl2,aabkrev}.  As we noted in section \ref{s2}, on any WIH these   angular momentum  charges can be expressed using the `rotational 1-form' $\omega_{a}$ defined by $D_{a} \l^{b} = \omega_{a} \l^{b}$:
\be Q_{K} = - \f{1}{8\pi G} \oint_{C} \omega_{a}\, K^{a}\, \rmd^{2}V   \ee
where $C$ is a leaf of the preferred foliation on the non-extremal WIH $\scriml$, with $v=\vo$ for some constant $\vo$.
Since $K^{a}$ is tangential to these 2-spheres, it can be expanded as
\be K^{a} = \b\epsilon^{ab} \b{D}_{b} f\, +\, \b{q}^{ab} \b{D}_{b} g  \label{decomp}\ee
where $\b{q}_{ab}$ and $\b\epsilon_{ab}$ are the pull-backs to the leaves of the foliation of the physical metric $q_{ab}$ and the area 2-form $\epsilon_{ab}$ on $\scriml$.  Recall that the defining property of the preferred foliation $v=v_{o}$ is that the pull-back $\b\omega_{a}$ of $\omega_{a}$ is divergence-free on each leaf with respect to the induced physical metric $q_{ab}$: \, $\b{q}^{ab} \b{D}_{a} \omega_{b} =0$. Therefore we can simplify the expression of $Q_{K}$:
\ba Q_{K}  &=& - \f{1}{8\pi G} \oint_{C} \omega_{a}\, \big(\b\epsilon^{ab} \b{D}_{b} f\, +\, \b{q}^{ab} \b{D}_{b} g\big)\, \rmd^{2}V \nonumber \\
&=&  \f{1}{8\pi G} \oint_{C} f\, \b\epsilon^{ab} \b{D}_{b} \b{\omega}_{a} \, \rmd^{2}V \nonumber \\
&=& - \f{1}{4\pi G} \oint_{C} f\, \Im\, \Psi_{2} \, \rmd^{2}V \, \, , \label{QK}\ea
where in the second step we have carried out an integration by parts and in the third step used (\ref{impsi2}). There are some noteworthy aspects of the final expression.\smallskip
 
\noindent 1. While energy $E_{t}$ is determined by $\Re\Psi_{2}$, (see  Eq. (\ref{Et})),  the angular momentum charges are governed by $\Im\Psi_{2}$. This in line with the fact that while $2\, \Re \Psi_{2} = K_{abcd} \l^{a}n^{b}\l^{c}n^{d}$ is a scalar, $2\, \Im \Psi_{2} = {}^{\star}K_{abcd} \l^{a}n^{b}\l^{c}n^{d}$ is a pseudo-scalar. In the asymptotically flat case, the situation at $i^{o}$ is completely analogous: The ADM energy is defined using $\Re\Psi_{2}$ while angular momentum is contained in $\Im\,\Psi_{2}$ \cite{aarh}.\\
2.  If $f=0$, i.e., $K^{a} = \b{q}^{ab} \b{D}_{b} g$, we have $Q_{K} =0$. So in place of the `relativistic angular momentum' associated with the full 6-dimensional Lorentz-group, we have an angular momentum 3-vector associated with a $SO(3)$ subgroup of $\Lor$.  This is in line with the fact that, whereas $\scrimz$ in the $\Lambda=0$ case is endowed with a \emph{4-parameter family} of (relatively boosted) `good cuts,'   $\scriml$ is endowed with a \emph{1-parameter family} of `good cuts'.  The angular momentum `3-vector' refers to the rest frame selected by $[\l^{a}]$ and the `center of mass world-line' selected by the 1-parameter family of good cuts.%
\footnote{ Recall from special relativity that the angular momentum tensor $M_{ab}$ of a system/field in Minkowski space-time refers to a Lorentz group, selected by choosing an origin. If we are also given a rest frame, i.e., a preferred time translation Killing field $t^{a}$, one can further decompose the Lorentz Lie algebra into a rotation part and a boost part. One can always select a world-line  passing through the given origin --called the center of mass world-line-- along which the boost angular momentum vanishes --i.e. $M_{ab}t^{a} =0$.  On $\scriml$ the choice of a cross section is analogous to the choice of an origin in Minkowski space,  the canonical time translation provides a rest frame, and the 1-parameter family of preferred cross sections is the analog of the center of mass world line.}

3. Note, however, that the decomposition (\ref{decomp})  of $K^{a}$ into a `rotation part'  $\b\epsilon^{ab} \b{D}_{b} f$ and a `boost part' $\b{q}^{ab} \b{D}_{b} g$ depends on the physical metric $q_{ab}$ on $\scriml$, and thus varies from one space-time to another in the covariant phase space $\Gcov$. Therefore, as we move from one space-time to another, the $SO(3)$ subgroup of $\Lor$ that defines the angular momentum 3-vector changes. As a consequence, given \emph{any} $K^{a}$ in the Lie algebra of $\Lor$, there is a space-time in $\Gcov$ for which $Q_{K}$ is non-zero. Therefore, from the Hamiltonian perspective, none of these diffeomorphisms corresponds to gauge transformation in $\Gcov$; they are all physical symmetries.

4. Suppose the space-time admits a rotational Killing field $\varphi^{a}$. Then we can also calculate the Komar integral associated with $\varphi^{a}$. Even though we now have $R_{ab} = \Lambda g_{ab} \not=0$ outside sources, the Komar integral is conserved in the following sense:  It's values evaluated on 2-spheres $S_{1}$  and $S_{2}$ in the source-free region agree if there is a 3-surface $\Sigma$  --with $S_{1}$ and $S_{2}$ as boundaries--  to which $\varphi^{a}$ is everywhere tangential. Therefore, the Komar integral is an interesting quantity. When correctly normalized, its value agrees with the component of the angular momentum $Q_{\varphi}$ obtained by setting $K^{a} = \varphi^{a}$ \cite{abl2}. This provides an additional support for the definition of $Q_{K}$.\\
Note, incidentally, that if we have a space-time that admits a `time-translation' Killing field $t^{a}$, the corresponding Komar integral is not as interesting if $\Lambda \not=0$ because it is generically not possible to find a 3-manifold $\Sigma$ that joins a 2-sphere $S_{1}$ in the interior (but still outside sources) and $S_{2}$ on $\scriml$ or $\scrim$ and, in addition,  $t^{a}$ is tangential to it.\\

\section{Discussion}
\label{s6}

Although Einstein \cite{ae} showed that general relativity  admits gravitational waves in the linear approximation around Minkowski space already in 1916, there was much confusion about the reality of gravitational waves in full general relativity for several subsequent decades \cite{Kennefick:book}.  {Strange as this state of affairs may seem,  especially in light of the recent discoveries by the LIGO-Virgo collaboration, the confusion was not due to some trivial misunderstanding.} Rather, it was rooted in the fact that, when space-time geometry is itself dynamical,  it is quite subtle to separate gravitational radiation from coordinate effects in the \emph{full, non-linear} theory. The issue was fully resolved only in the early 1960s by the careful work by Bondi et al.%
\footnote{  `Wave propagation'  was discussed already in the 1952 seminal work on the Cauchy problem by Choquet-Bruhat \cite{ycb}. However,  those considerations were local, and one cannot decide locally if there is radiation carrying energy, momentum, etc. For example, the  c-metric \cite{lc} --an exact solution discovered by Levi-Civita in 1918-- admits a Killing vector that is time-like in large patches. Therefore it was often thought that the solution has no gravitational radiation. It is only in 1981 that a detailed analysis of its asymptotic structure at $\scripz$ became available in full general relativity and established that it \emph{does} carry gravitational radiation \cite{aatd} (emitted by two eternally accelerated black holes). This could not have been dome using local considerations; the Bondi, Sachs, Penrose et al framework was essential.}
As was natural at the time, the work assumed that the cosmological constant $\Lambda$ is zero, and therefore modeled isolated systems by asymptotically flat space-times. In this case space-time curvature decays as we move away from sources, giving rise to several simplifications. In presence of a positive $\Lambda$ on the other had, space-time curvature does not decay \emph{no matter how far you move away from sources.} Therefore, it is now much more difficult to distinguish ripples in space-time geometry representing genuine gravitational waves from gauge artifacts.  To capture the notion of an isolated system, on the other hand, one needs to provide a gauge invariant criterion to ensure that  there are \emph{no} physical gravitational radiation incident on the system from infinity. In this paper, we have addressed this problem by introducing the notions of $\scrimr$ and $\scriml$.

Recall that already in the first discussions of conformal completions of space-times, Penrose \cite{rp1} considered the possibility of a cosmological constant and showed that for $\Lambda >0$, the boundaries $\scri^{\pm}$ of the conformal completion are space-like. In the asymptotically flat case, one specifies the `no incoming radiation' condition by requiring that the gauge invariant Bondi news tensor $N_{ab}$ should vanish on the past boundary $\scrimz$. Why did we not simply repeat that strategy at the space-like past boundary $\scrim$ in the $\Lambda>0$ case? As we pointed out in section \ref{s1},  we do not yet have the analog of $N_{ab}$ on $\scri^{\pm}$ in the $\Lambda>0$ case. Why not take recourse to the notion of the radiation field $\Psi_{4}^{o}$ on $\scripz$ (and $\Psi_{0}^{o}$ on $\scrimz$) that is routinely used in numerical simulations of binary black hole simulations to calculate the wave forms? In the $\Lambda=0$ case,  $\scri^{\pm}_{o}$ are null and their normals provide the null vector that is needed to define $\Psi_{4}^{o}$ (and $\Psi_{0}^{o}$). For $\Lambda >0$, $\scri^{\pm}$ are space-like and we no longer have a canonical null direction to extract $\Psi_{4}^{o}$ on $\scrip$ (or $\Psi_{0}^{o}$ on $\scrim$) in a gauge invariant manner \cite{kp,rp2}. There are two further conceptual obstacles associated with $\scrim$ that are naturally overcome if one uses $\scrimr$  and $\scriml$ instead. First, consider  gravitational collapse of a star depicted in the left panel of Fig. \ref{sds}. If we use the boundaries $\scri^{\pm}$, then, in contrast to the $\Lambda=0$ case, the space-time diagram continues to the right because the analytical continuation of the Schwarzschild de Sitter metric goes on ad infinitum. 
If on the other hand we focus just on the relevant part $\Mr$ of space-time, this problem disappears since the part of space-time to the right of $\scrimr$ is simply not  relevant. Second, already in the Kerr-de Sitter space-time, in the limit $\Lambda \to 0$ the part of space-time near $\scrim$ disappears. Therefore, if we imposed the no incoming radiation condition on $\scrim$ and extracted physical information from fields thereon, it would not be directly related to the physical information extracted from structures at $\scrimz$ of the $\Lambda=0$ theory. \vskip0.2cm

Our strategy of using $\scrimr$ or $\scriml$ as the past boundary in place of $\scrim$ led to a rich structure. First, we saw that in the standard examples discussed in section \ref{s3} and Appendix \ref{a1}, $\scrimr$ does have all the structure we introduced to impose the no incoming radiation condition,  to discuss symmetry groups and to define conserved charges. In particular we saw that, in these examples: \\ (i) $\scrimr$ is geodesically complete;\\ (ii)  $\scripl$ is `long enough' to intersect  $\scrimr$ in a 2-sphere $\izl$;\\ (iii) via a general construction, the 2-sphere $\izl$ endows $\scriml$ with a specific Weakly Isolated Horizon (WIH) structure. This structure is also the natural one from the perspective of individual examples and their isometries. For example, in the Schwarzschild de Sitter space-time, this is precisely the WIH structure induced on the cosmological horizon by the standard `static' Killing field $T^{a}$;\\(iv) in  the region $\Ml$ of the Kerr-de Sitter space-time, the Killing field selected by the WIH null normals $[\l^{a}]$ is very similar in its structure  to the standard stationary Killing field $t^{a}$ in the asymptotic region of Kerr space-time.\\  The Vaidya solution depicting `evaporation' of a Schwarzschild de Sitter black hole to de Sitter space-time also provides support for our framework. It is somewhat more interesting because it is dynamical \cite{abk1} but we chose not to discuss it in detail because the discussion of examples is already quite long. These examples together with the results on linearized gravitational waves on de Sitter background \cite{abk2,abk3} provide some 
concrete evidence in favor of the boundary conditions introduced in section \ref{s2.3}. It is interesting to note in retrospect that in the $\Lambda=0$ case concrete evidence in favor of the boundary conditions was the same when Bondi, Sachs, Penrose and others  \cite{bondi,sachs,rp1}  first introduced them.

However, since then there have been significant advances in approximation methods, numerical simulations and geometric analysis. They can all be used to create additional evidence for or against the conditions introduced in section \ref{s2}. For example, one can use approximation methods to analyze  radiating solutions  `near' Kerr-de Sitter by making an order by order expansion along the lines of \cite{bk,jlcl}, but now in a neighborhood of the cosmological horizon $\scrimr$, rather than the black hole horizon.  On the numerical side, the framework is supported  by simulations of collapse of gravitational waves \cite{num1},  and head-on collisions of black holes \cite{num3}. Finally, on the geometric analysis side, as we pointed out in footnote \ref{fn-hv}, there are interesting results  \cite{hv} on non-linear perturbations of the Schwarzschild-de Sitter solution (which allow the angular momentum to change). Those results suggest  --but do not establish--  that there is a large class of examples with gravitational radiation in which $\scripl$ is `sufficiently long'. These solutions asymptotically approach Kerr-de Sitter geometry near $i^{+}$ in the shaded region of the right panel of Fig. \ref{sds}. Similarly,  there are results \cite{schlue} suggesting that there exists a non-linear neighborhood of the Schwarzschild-de Sitter space-time in which $\scripl$ and $\scrimr \setminus \scriml$ are `sufficiently long', and $\scrip$ admits radiation. Interestingly, in these space-times, the asymptotic geometry near $\scrip$ will  \emph{not} be that of Kerr-de Sitter because the magnetic part of the Weyl tensor will not vanish there \cite{abk1,schlue}. Thus, the first steps needed to establish that the class $\CLi$ of space-times  introduced in section \ref{s2.3} admits an infinite dimensional family of radiating solutions have been taken. It would be very helpful to use the techniques already developed to solve the characteristic initial value problem to establish global existence (for small data) \emph{in the future light cone of $i^{-}$} of the right panel of Fig. \ref{sds}. The characteristic initial data would be specified on the null boundary of this region, such that it is \emph{trivial} on $\scrimr$, and non-trivial on the rest of the boundary (that consists of the white and black hole horizons). Triviality on $\scrimr$ will ensure that $\scrimr$ will continue to serve as the `relevant scri-minus' also for the radiating solution, and the non-trivial data on the rest of the null boundary will mimic the radiation that would be emitted by a more realistic compact binary in the shaded portion of the right panel of Fig. \ref{sds}. To summarize, the setup introduced in section \ref{s2.3} suggests generalizations of  analytical approximation methods along the lines of \cite{bk,jlcl},  more numerical simulations along the lines of \cite{num1,num3}, and geometric analysis investigations to extend results of \cite{hv,schlue}. \vskip0.2cm

In section \ref{s4} we found that the symmetry group $\G$ of $\scrimr$ is analogous to the BMS group $\B$  at $\scrimz$ in asymptotically flat space-times: both are semi-direct products of an Abelian group $\S$ of supertranslations with  a finite dimensional group. However,  there is also an interesting twist that captures an essential signature of a positive $\Lambda$. While on $\scrimz$ the finite dimensional group is just the 6 dimensional Lorentz group $\Lor$, on  $\scrimr$ the finite dimensional group is the 7 dimensional $\G_{7}$, which is a  (trivial) central extension of $\Lor$:\, $\G_{7} = \T_{1} \otimes \Lor$. The `extra' one dimensional sub-group $\T_{1}$ of $\G_{7}$ is the time-translation group selected by the \emph{canonical non-extremal WIH structure} on $\scriml$, which has no analog on $\scrimz$. We compared and contrasted in detail the structures of $\scrimz$ and $\scrimr$, and of the BMS group $\B$ and the symmetry group $\G$. The `no incoming radiation' condition endows $\scrimz$ with a 4 parameter family of preferred cross sections, called the `good cuts'. By contrast, $\scrimr$ is endowed with a 1-parameter family of `good cuts'. Thus in contrast with $\scrimz$, we have a preferred `rest frame' on $\scriml$ (which extends to $\scrimr$). Finally,  while $\G_{7}$ initially arises as the quotient,  $ \G_{7} = \G/\S$,  there is a \emph{canonical} embedding of $\G_{7}$ into $\G$ that leaves every good cut of $\scriml$ invariant.  By contrast, in the $\Lambda=0$ case, there is no Lorentz subgroup $\Lor$ of the BMS group $\B$ that leaves any 1-parameter family of good cuts on $\scrimz$ invariant. \vskip0.1cm

Subsequently, in the main text we focused on $\scriml$ and this canonical $\G_{7}$ subgroup of $\G$, leaving the further discussion of the supertranslation subgroup $\S$ to Appendix \ref{a2}. In section \ref{s5} we discussed the notion of mass $M$ of $\scrimr$ and of charges $E_{t}$ and $Q_{K}$ associated with the time-translation subgroup $\T_{1}$ and the Lorentz subgroup $\Lor$ of $G_{7}$. The definition of mass was motivated by a thought experiment that extracted $M$ from the (tidal acceleration or) geodesic deviation of a suitable set of test particles. The definition of charges was arrived at using a covariant phase space $\Gcov$ tailored to $\scriml$. Specifically the charges are the Hamiltonians that generate canonical transformations on $\Gcov$, induced by the action of time-translation and Lorentz vector fields  in the Lie algebra $\LG_{7}$ of $\G_{7}$.  Thus, the mass $M$ and the charge $E_{t}$ associated with the time-translation symmetry group $\T_{1}$ were arrived at from entirely different considerations, whence their initial expressions appear completely unrelated: $M$ arises as the integral of a component of the Weyl curvature over a 2-sphere cross section of $\scriml$, while $E_{t}$ arises as a function of the area-radius of this cross-section. Yet, because of a differential geometric identity, and the WIH structure of $\scriml$, the two seemingly unrelated expressions are equal to each other!  In Kerr-de Sitter space-times, not only is the mass $M$ positive, but it is also bounded above. From general physical considerations, we would expect that $M$ would be positive for all space-times under consideration. One approach to establishing positivity in the case when the only past boundary of $\Ml$ is $\scriml$ would be to use a  spinorial argument a la Witten \cite{ew}.   \vskip0.1cm

Perhaps the most striking difference from the past boundary $\scrimz$ in the $\Lambda=0$ case is the dual role played by $\scriml$ (and $\scrimr$).  On the one hand $\scriml$ is  analogous to $\scrimz$ and in fact goes over to $\scrimz$ as $\Lambda \to 0$ in examples where there is a clear-cut limiting procedure. On the other hand it is also a \emph{non-extremal WIH}. 
\footnote{By contrast, $\scrimz$ is an \emph{extremal} WIH,  and that too from the perspective of the  \emph{conformally completed} space-time,  not the physical one.}
 Thanks to this dual role of $\scriml$, we could go back and forth between the two seemingly different sets of structures. For example, the symmetry group $\G$ resulted by examining the structure of $\scrimr$ from the perspective of $\scrimz$, while the preferred foliation and the symmetry group $\G_{7}$ on $\scriml$ arose from, the structure $\scriml$ inherits from being a non-extremal WIH. Similarly, we  treated $\scriml$ as the analog of $\scrimz$ to fix the normalization of the `time-translation' in $\T_{1}$ and also to introduce the definition (\ref{mass}) of the mass $M$ in terms of $\Re\Psi_{2}$. On the other hand, we used the fact that it is a non-extremal WIH to define horizon charges --energy $E_{t}$ and angular momentum $Q_{K}$-- and express $M$ and $E_{t}$ using the area radius $R_{(c)}$ through Eqs (\ref{M}) and (\ref{energy}); these expressions are simply not available at $\scrimz$.
 
  Indeed, because $\scrimr$ and $\scriml$ lie, so to say, `in the middle of space-time'  rather than at an infinite separation from sources, a priori it was not clear that it would have any of the structures that are  needed  to extract physics of the isolated system in a gauge invariant fashion. The fact that this is possible can be traced back directly to the fact that $\scrimr$ and $\scriml$ have the structure of non-expanding horizons.  Finally, let us consider $\scripl$.  First results reported in \cite{aa-ropp} indicate that $\scripl$ will also have a dual structure. To describe properties of gravitational radiation across $\scripl$, one can emphasize its similarity with $\scripz$, while to speak of symmetries and corresponding charges, one can endow it with a `fiducial' structure of a WIH that is `dragged' from $\izl$.  Thus, constructions introduced in this paper and the results that they led to serve as points of departure to  obtain a gauge invariant characterization of gravitational waves at $\scripl$ (and/or $\scrip$), and to study their properties. 

\section*{Acknowledgements}
We would like to thank Beatrice Bonga, Alex Corichi,   Aruna Kesavan, Neev Khera, Jerzy Lewandowski, and Volker Schlue for discussions. Several results in  this paper were presented by AA in conferences, particularly at the Tsinghua Sanya Mathematics Forum, Sanya Island, China in 2016, and at the GRAV17 conference, Cordoba, Argentina in 2017. He thanks participants for valuable comments, particularly on the stability of the Kerr-de Sitter cosmological horizons. This work was supported in part by the NSF grant PHY-1806356, grant UN2017-9945 from the Urania Stott Fund of Pittsburgh Foundation and the Eberly research funds of Penn State.

\begin{appendix}
\section{The Kerr-de Sitter space-time}
\label{a1}

In this Appendix we will summarize the relevant geometrical structures of the Kerr-de Sitter space-time and their relation to our discussion of symmetries and conserved quantities in sections \ref{s4} and \ref{s5}. We will find that the geometry is much more intricate than in the two examples discussed in section \ref{s3}. In particular,  
there is an unforeseen complication: it is no longer transparent which of the 2-parameter family of Killing fields should be identified as the time-translation symmetry --the analog of $T^{a}$ in  the Schwarzschild-de Sitter space-time.  

\begin{figure}[]
  \begin{center}
  \vskip-0.4cm
    \includegraphics[width=3.5in,height=2.5in,angle=0]{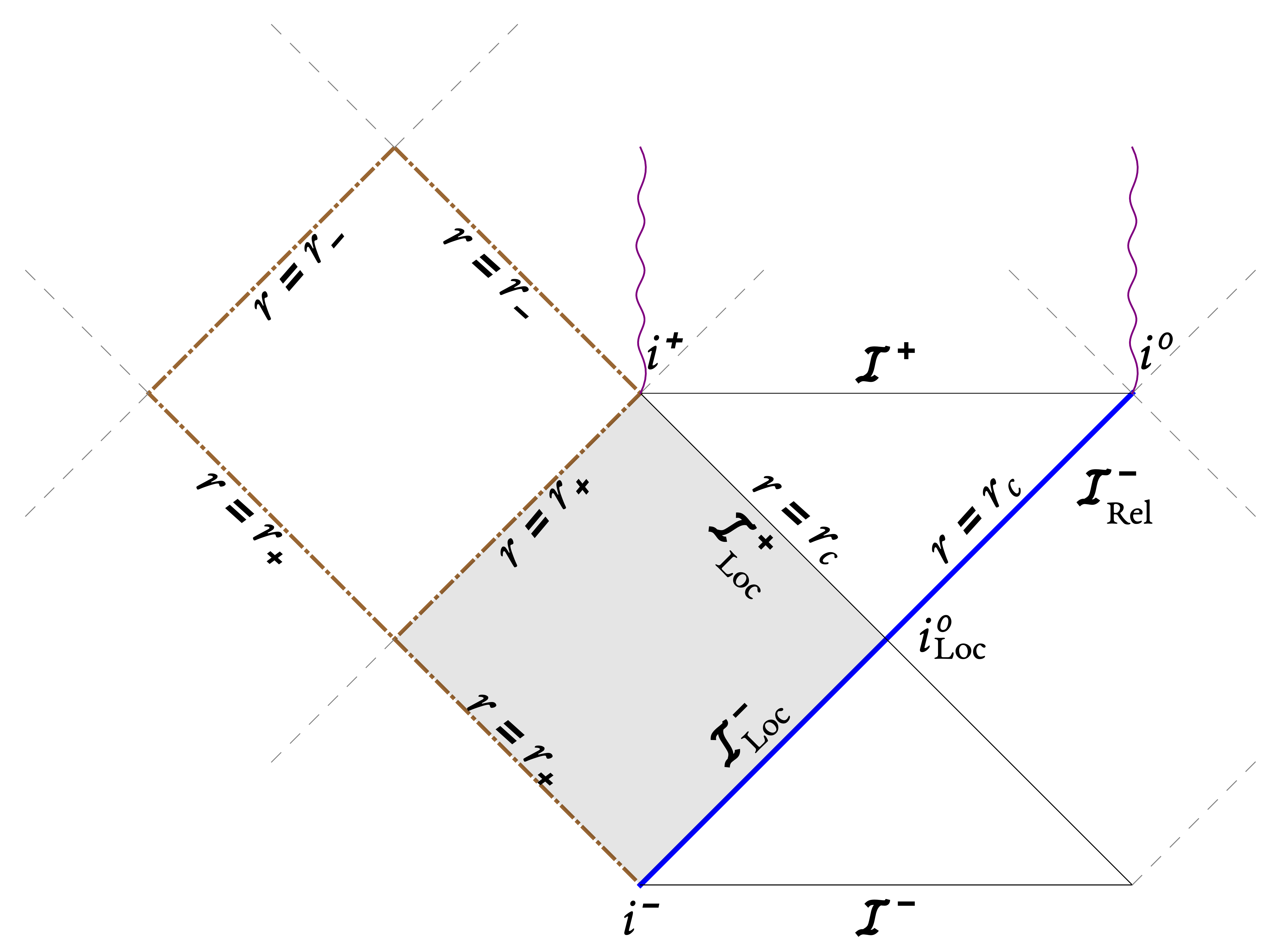}    
    \caption{{\footnotesize{\emph{Kerr-de Sitter space-time.}  The future and past boundaries, $\scri^{\pm}$, of the asymptotic region are space-like  because we have a positive $\Lambda$. The future event horizon $\scrimr$ of $i^{-}$ intersects the past  cosmological horizon of $i^{+}$ in a 2-sphere $\izl$ just as in Figs 1 and 2. The vertical wiggly lines depict the singularities. We now have \emph{three} horizons separating the singularity from the asymptotic regions near $\scri^{+}$: the inner black hole horizon $r=r_{-}$,  the outer black hole horizon $r=r_{+}$ and the cosmological $r=r_{c}$. As in Figures 1 and 2,  black hole and cosmological horizons serve as the past and the future boundaries of the (shaded) local space-time region $\Ml$,  the intersection of the causal future of $i^{-}$ with the causal past of $i^{+}$. The full past boundary $\scrimr$  is the extension of $\scriml$ all the way to spatial infinity $i^{o}$, the `right end' of $\scrip$.}} }
\label{kerr-ds}
\end{center}
\end{figure}

The Kerr-de Sitter metric is generally written in the Boyer-Lindquist coordinates as \cite{bc,am}:
\ba
 \label{kds}
&& g_{ab}\rmd x^{a} \rmd x^{b} \equiv \rmd s^{2} = \Big[a^2 \sin^2{\theta} \big(1+ \frac{a^2}{\ell^2} \cos^2{\theta}\big) - \Delta(r)\Big]\,\, \frac{\rmd t^2}{(r^2+a^2 \cos^2{\theta}) \Big(1+ \frac{a^2}{\ell^2}\Big)^2} \nonumber\\
&& +2 \Big[\Delta(r)- (r^2+a^2) \big(1+\frac{a^2}{\ell^2} \cos^2{\theta}\big)\Big]\,\,\frac{a \sin^2{\theta} \ \rmd t \rmd \varphi}{(r^2+a^2 \cos^2{\theta}) \Big(1+ \frac{a^2}{\ell^2}\Big)^2} \nonumber\\
&& + \frac{r^2+ a^2 \cos^2{\theta}}{\Delta(r)} dr^2 + \frac{r^2 + a^2 \cos^2{\theta}}{1+ \frac{a^2}{\ell^2} \cos^2{\theta}} \rmd\theta ^2 \nonumber\\
&& + \Big[(r^2+a^2)^2 \big(1+ \frac{a^2}{\ell^2} \cos^2{\theta}\big)- a^2 \sin^2{\theta} \Delta(r)\Big]\,\, \frac{\sin^2{\theta} \ \rmd \varphi^2}{(r^2+ a^2 \cos^2{\theta}) \Big(1+ \frac{a^2}{\ell^2}\Big)^2} \,  ,
\ea
where 
\be\label{deltar} 
\Delta(r) = -\frac{r^4}{\ell^2} + \Big(1-\frac{a^2}{\ell^2}\Big) r^2 - 2 Gm r + a^2.
\ee
Since $\Delta(r) $ is a polynomial of order 4, it has four roots.  The Boyer-Lindquist chart fails at the three positive roots 
$r_{\mp}, r_{c}$ (with $r_{-} \le r_{+}\le r_{c}$) of $\Delta(r) $ (the fourth root is negative). These correspond, respectively,  to  the inner black hole (or, the Cauchy) horizon,  the outer black hole (or, the event) horizon, and the cosmological horizon, shown in Fig. \!\ref{kerr-ds}.  Since the cosmological constant $\Lambda$ is positive,  space-time boundaries $\scri^{\pm}$ in the Penrose conformal completion are of course space-like.  $\scrimr$, shown as a (blue) bold-faced line, is the future event horizon of $i^{-}$ that connects $i^{-}$ on $\scri^{-}$ with $i^{o}$ on $\scrip$ (exactly as in figures 1 and 2). It is again a non-expanding horizon, ruled by complete null geodesics. For the single, rotating  black hole under consideration, the \emph{relevant} portion $\Mr$ of space-time is the causal future of $i^{-}$. 
This structure implies that Kerr-de Sitter space-time belongs to the class $\CLi$ of space-times introduced in section \ref{s2.3}. The local region $\Ml$ of space-time is intersection of the causal future of $i^{-}$ with the causal past of $i^{+}$, depicted in the figure by the shaded region, bounded by $\scriml$ and $r=r_{+}$ in the past and $r=r_{+}$ and $\scripl$ in the future.
 
In the $\Lambda =0$ case,  Kerr black holes are characterized just by the two parameters $m,\,a$ with $m >0$ and $|a|\le Gm$; and we have the extremal Kerr solution at $Gm=|a|$ for which the inner and outer black hole horizons coincide and the surface gravity vanishes. With $\Lambda >0$ the situation is much more complicated because now the solution carries \emph{three} parameters, $m,\,a,\l$ and \emph{three} horizons. All three coincide if $Gm = 4\l\,\,\big((2/\sqrt{3}) -1\big)^{3/2} \approx 0.24 \l$ and $a= (2-\sqrt{3}) \l \approx 0.27\l$. Note that in this case $Gm < a$! Next, we have the possibility that only two of the three horizons coincide: the two black hole horizons can coincide (as in the extremal Kerr for $\Lambda=0$), the cosmological horizon  remaining distinct, lying outside the common black hole horizon;  or, the outer black hole horizon can coincide with the cosmological horizon, leaving the inner horizon distinct. The parameter values at which these possibilities are realized involve rather complicated relations between $m,\, a,\, \l$.
\footnote{For example,  for there to be three distinct horizons, we must have $m_{-} < m < m_{+}$ where $m_{\mp}$ are functions of $a,\,\l$, given by $m_{\mp} = \frac{\sqrt{A_{\pm}}}{ 8 \ell^2} (A_{\pm} + 4 B_{\pm}) $ with $A_{\pm} = ({8 a ^2 \ell^2})/\big({\ell^2-a^2 \pm \sqrt{(\ell^2-a^2)^2- 12 a^2 \ell^2}}\big)$, and $B_{\pm} = \frac{\ell^2- a^2 \pm \sqrt{(\ell^2-a^2)^2- 12 a^2 \ell^2}}{2}$.}

As in the $\Lambda=0$ case,  it is clear from inspection of Eq. (\ref{kds}) that the space-time admits two commuting  Killing fields $t^{a}$ and $\varphi^{a}$. They lead us to the physical notions of mass and angular momentum. From the perspective of $\scriml$ developed in sections \ref{s3} - \ref{s5},  the relevant symmetry to define the mass at $\scriml$ (or $\scrimr$)  is generated by the Killing field that is: (i) a null normal $\ul\l^{a}$  to $\scriml$, \,(ii) vanishes at $\izl$, and,  (iii)  normalized such that the surface gravity $\kappa_{\ul{\l}}$ is given by $\kappa_{\ul{\l}} = (1/2R_{(c)}) (1- 3R_{(c)}^{2}/\ell^{2})$. (See Remark 1 at the end of section \ref{s3.2}.)  Here $R_{(c)}$ is the area-radius of the cosmological horizon $\scrimr$:
\be R^{2}_{(c)} = \f{a^{2} + r_{(c)}^{2}}{1 + \f{a^{2}}{\l^{2}}} \, . \ee
Therefore,  we are led to seek the linear combination of the two Killing fields that coincides with $\ul\l^{a}$ on $\scrimr$. Now, motions generated by both Killing fields leave  the local region $\Ml$ of the Kerr space-time invariant, whence they leave $\scripl, \, \scriml$ and $\izl$ invariant. Hence condition (ii) is satisfied by \emph{every} linear combination of $t^{a}$ and $\varphi^{a}$.  In the Schwarzschild-de Sitter case,  the restriction of $t^{a}$ to $\scriml$ is null, whence it satisfies condition (i) and we only had to rescale it so it has the desired surface gravity on $\scrimr$ to define mass.  However,  if $a\not=0$, the vector field ${t}^{a}$  is \emph{space-like} on $\scrimr$. The vector field which \emph{is} proportional to $\ul\l^{a}$ is given by the following linear combination of the two Killing fields
\be \label{ubt} \ub{t}^{a} = K \big(t^{a}  + \Omega_{c}\, \varphi^{a} \big)  \qquad {\rm with} \qquad \Omega_{c} = {\f{a}{(\l^{2} + {a^{2}})}\, \Big(1+\f{\l^{2}}{R_{(c)}^{2}} \big(1- \f{R_{(c)}^{2}}{\l^{2}}\big) \Big)}\, ,\ee
where $K$ is a non-zero constant that, as remarked above, can vary from one Kerr-de Sitter solution to another, i.e., can depend on $m$ and $a$. The Killing field $\ub{t}^{a}$ has two interesting properties:\\
1. It is time-like in a (large) neighborhood of $\scriml$ within $\Ml$; up to constant rescalings, it is the \emph{only} Killing field in the Kerr-de Sitter space-time with this property.  \\
2. Irrespective of the choice of the constant $K$, its surface gravity on $\scrimr$ is \emph{non-zero}. Therefore, the equivalence class $[\ub{t}^{a}]$ endows $\scrimr$ with the structure of a non-extremal WIH structure. It then follows that $[\ub{t}^{a}]$ must vanish on one and only one cross section of $\scrimr$. That cross-section turns out to be precisely $\izl$. This implies that, irrespective of the  choice of the non-zero constant $K$,  the affine parameter $\ub{v}$ of  the restriction of $\ub{t}^{a}$ to $\scrimr$ runs from $-\infty$ to $\infty$ on $\scriml$ as well as on $\scripl$.
  \smallskip
 
Since surface gravity is non-zero, we can simply fix the constant $K$ such that the surface gravity of $\ub{t}^{a}$ has the desired value $\kappa_{\ul{\l}}$. With this choice, $\ub{t}^{a}$ satisfies all three desired conditions. Note that $\ub{t}^{a}$ \emph{is the precise analog of the standard time-translation Killing vector $t^{a}$ in Kerr space-time} in the following sense. First,  $\ub{t}^{a}$ is null on $\scriml \cup \scripl$ just as the $t^{a}$ is on $\scrimz \cup \scripz$.  Second, $\scriml$ and $\scripl$ are both complete w.r.t the affine parameter of $\ub{t}^{a}$, just as $\scrimz$ and $\scripz$ are complete with respect to the affine parameter of $t^{a}$ in the $\Lambda=0$ case. Finally, the neighborhood of $\scriml \cup \scripl$ in which $\ub{t}^{a}$ is time-like is completely analogous to the neighborhood of $\scrimz \cup \scripz$ in which $t^{a}$ is time-like  --both extend up to the ergoregion that surround the black hole horizons.

Since $t^{a}$ is the Killing field that defines the mass in Kerr space-time with $\Lambda=0$ case, it is natural to use $\ul{\l}^{a}$ to define mass at $\scriml$ in the $\Lambda>0$ case. This is exactly what our general procedure of section \ref{s5} leads us to do. The resulting mass (determined by $\Psi_{2}$ on $\scriml$ as in section \ref{s5}) is then given by:
\be \label{kerrM}  M = \f{R_{(c)}}{2G}\,\, \big(1 - \f{R_{(c)}^{2}}{\l^{2}} \big) \equiv \,m \Bigg[\,\f{\big(1 +\f{a^{2}}{\l^{2}} - \f{a^{2}}{R_{(c)}^{2}}\big)^{\f{1}{2}} }{(1 + \f{a^{2}}{\l^{2}})^{2}}\,\Bigg] \, .\ee
In the Schwarzschild-de Sitter space-time,  we have $ 0\le M=m\le (\l/3\sqrt{3})$. In Kerr-de Sitter space-time, on the other hand, $M<m$ if $a\not=0$, and for a given value of the parameter $m$, the mass $M$ decreases as $a$ increases.  For the full Kerr-de Sitter family we again have  $0 \le M \le  \l/(3\sqrt{3})$; the minimum value, $M=0$, is reached for de Sitter space-time $m=a=0$, and numerical evaluations show that the maximum value $M = \l /(3 \sqrt{3})$ is again reached at the Nariai solution.\smallskip

Results of section \ref{s5} also enable us to calculate the angular momentum. Recall first that the rotational Killing fields are normalized by asking that their affine parameter should run in the interval $[0,\, 2\pi)$. Therefore, the presence of a positive cosmological constant does not introduce any complications in identifying the Killing field with which to associate angular momentum: It is just $\varphi^{a}$.  The angular momentum $J_{\varphi}$, given by setting $K^{a} =\varphi^{a}$ in Eq. (\ref{QK}), can now be expressed as: 
\be \label{kerrJ} J_{\varphi} =   {-} 
\f{Ma}{\Big(1 + \f{a^{2}}{\l^{2}} - \f{a^{2}}{R_{(c)}^{2} }\Big)^{\f{1}{2}} }\, = {-} \f{ma}{(1+\f{a^{2}}{\l^{2}})^{2}} .\ee
In the Schwarzschild-de Sitter space-time we have $a=0$ whence $J_{\varphi}$ vanishes, as it must.  For the  full Kerr-de Sitter family, in the limit $\Lambda\to 0$ we  have $\l \to \infty$ and $R_{(c)} \to \infty$, whence we obtain $J_{\varphi} \to { -Ma}$, as in the Kerr space-time with $\Lambda=0$. Thus, both the mass $M$ and angular momentum $J_{\varphi}$ reduce to the expected results in the two independent limits, $a\to 0$ and $\Lambda \to 0$.  For the full Kerr-de Sitter family,  Eqns (\ref{kerrM}) and (\ref{kerrJ})  are simply the evaluations of Hamiltonians, discussed in section \ref{s5},  that generate motions along $\ub{t}^{a}$ and $\varphi^{a}$  for all permissible values of $m,\,a,\,\l$. 

We will conclude with a discussion of how these notions of mass and angular momentum are related to those defined at $\scri^{\pm}$. In the $\Lambda=0$ case, the Killing vector field $t^{a}$  becomes unit and hypersurface orthogonal at infinity. Since it defines a time-translation in an asymptotically non-rotating frame, we associate mass with $t^{a}$. But in the $\Lambda >0$ case, the physical norm of the Killing field $t^{a}$ diverges at infinity \emph{and} it fails to be hypersurface orthogonal even asymptotically. The combination of these two facts create an unforeseen complication in defining mass of Kerr-de Sitter space-time at $\scri^{\pm}$.

More precisely, we have the following.  In the Schwarzschild-de Sitter space-time (which can again be obtained by setting the parameter $a=0$ in the metric), $t^{a}$ is hypersurface orthogonal. Therefore, in the Schwarzschild  de Sitter space-time, motions along $t^{a}$ can be regarded as time-translations in the asymptotic frame that is non-rotating and we can use it to define mass \cite{abk1}.%
\footnote{\label{11}This interpretation holds only in a generalized sense, since $t^{a}$ is space-like on $\scri$ rather than time-like as in the $\Lambda=0$ case. But this generalization is inescapable because   $\scri^{\pm}$ are themselves space-like, and \emph{every} space-time Killing field must be tangential to $\scri^{\pm}$,  whence space-like.}
In  the Kerr-de Sitter case, $t^{a}$ is not hypersurface orthogonal but one may hope that it would become hyperspace orthogonal asymptotically, as in the Kerr solution. To investigate if this happens, let us carry out a conformal completion of the Kerr-de Sitter space-time using $\Omega= 1/r$ as the conformal factor. Then the conformally rescaled metric $\h{g}_{ab} = \Omega^{2} g_{ab}$ is smooth at  the boundaries $\scri^{\pm}$. The intrinsic 3-metric $\h{q}_{ab}$  on  the space-like boundaries $\scri^{\pm}$ of the conformally completed space-time is given by \cite{abk1}:
\be 
\h{q}_{ab}\, {\rm d}x^{a}\, {\rm d}x^{b} = \frac{1}{\left(1 + \frac{a^2}{l^2} \right)^2} {\rm d}t^{2} -  \frac{2 a \sin^2 \theta}{\left(1 + \frac{a^2}{l^2} \right)^2}\, {\rm d}t {\rm d}\varphi + \frac{l^2}{1 + \frac{a^2}{l^2} \cos^2 \theta} {\rm d}\theta^{2} + \frac{l^2 \sin^2 \theta}{1+ \frac{a^2}{l^2}} {\rm d}\varphi^{2},
\ee
Thus, $t^{a}$ fails to be hypersurface orthogonal \emph{even on} $\scri^{\pm}$.  There \emph{is} a  Killing field,  unique up to constant rescalings, that is hypersurface orthogonal at $\scri^{\pm}$,  but it is given by a (constant) linear combination of $t^{a}$ and $\phi^{a}$:
\be \label{ttilde}  \t{t}^{a} =  \big( t^{a} +  \f{a}{a^{2}+\l^{2}}\, \varphi^{a}  \big)\, ;\ee 
 It is this $\t{t}^{a}$ that generates `time-translations' (in the generalized sense of footnote \ref{11}) in the frame that is non-rotating  at infinity.  In the $\Lambda=0$ case, one fixes the rescaling freedom in the analog of $\t{t}^{a}$ by requiring that the norm of the vector field (with respect to the physical metric) should tend to  $-1$ at infinity. In the $\Lambda >0$ case, the norm diverges as one approaches $\scri^{\pm}$. Therefore, without a new, extra input, we cannot eliminate the freedom to rescale $\t{t}^{a}$ by a constant,  and furthermore this constant can depend on $m$ and $a$, i.e.,  can vary from one phase space point to another.  As far as we know the issue of finding the `correct' normalization has not been discussed in the Kerr-de Sitter case. However,  in the case of Kerr \emph{anti-}de Sitter space-times, this freedom is generally fixed by requiring that the first law of black hole mechanics should hold (see, e.g. \cite{gpp}).  Although there are no cosmological horizons in Kerr anti-de Sitter space-time, the main ideas can be carried over also to the cosmological horizon $\scriml$  in the  Kerr-de Sitter family.  The required rescaling leads us to rescale $\t{t}^{a}$ as:
\be \h{t}^{a} =  \big(1+ \f{a^{2}}{\l^{2}}\big) \, \t{t}^{a}, \qquad  {\rm so\,\, that}  \qquad \l^{a} = \h{t}^{a} +\h{\Omega}_{c}\, \varphi^{a} \,\,\, {\rm with}\,\,\,  \h\Omega_{c}  = \f{a}{R_{(c)}^{2}} (1- \f{R_{(c)}^{2}}{\l^{2}}) \, \, ,\ee
is a null normal to $\scriml$.  Thus,  the `correct' expression of energy is given by  the `charge' associated with the time-translation Killing field,  $\h{t}^{a}$; it is  is now the $\Lambda >0$ analog of $t^{a}$ in the $\Lambda=0$ case.  One can use the structure at $\scri^{\pm}$ to define this charge $Q_{\h{t}^{a}}$, and angular momentum $Q_{\varphi}$ associated with $\varphi^{a}$  for the Kerr d Sitter family\cite{abk1}:
\be Q_{\h{t}^{a}}  = \f{m}{(1+\f{a^{2}}{\l^{2}})^{2}} \equiv  \f{M}{\big(1 +\f{a^{2}}{\l^{2}} - \f{a^{2}}{R_{(c)}^{2}}\big)^{\f{1}{2}} }  \qquad  {\rm and}  \qquad Q_{\varphi}  = { -} \f{ma}{(1+\f{a^{2}}{\l^{2}})^{2}}  \equiv J_{\varphi} \, .  \ee
(These are direct analogs of the mass and angular momentum used in the discussion of the first law in Kerr anti-de Sitter space-time \cite{gpp}.) Then we have the familiar-looking first law: $\delta Q_{\h{t}^{a}} = (1/8\pi G)\, \kappa_{\l} \,\delta A { -} \h\Omega\, \delta Q_{\varphi}$, where $\kappa_{\l}$ is the surface gravity of the null normal $\l^{a}$ and $A$, the area of any 2-sphere cross section of $\scriml$. 

This discussion brings out the fact that the parameters $m,\, a$ that enter the metric are not as directly related to the mass and angular momentum as they are in the $\Lambda = 0$ case, \emph{irrespective of} whether one defines the mass --or, the charge associated with the time-translation symmetry-- using structures available at $\scriml$, as in our main text, or at $\scrip$. 
\\

\emph{Remarks:}

1.  As we saw in section \ref{s5}, a first law of horizon mechanics holds on $\scriml$ for the entire class $\CLi$  of space-times considered in this paper: $\delta M = (1/8\pi G) \, \kappa_{\ul\l}\, \delta A$ where $A$ is again the area of any 2-sphere cross section of $\scriml$. In particular, the law holds for our Kerr-de Sitter family and, as we saw, $M$ is associated with the time-translation generated by the Killing field $\ub{t}^{a}$ that coincides with $\ul\l^{a}$ on $\scriml$. As we just discussed, in the Kerr-de Sitter family, one can use the Killing vectors, define charges associated with them using structures at $\scri^{\pm}$, and arrive at another first law, with a more familiar form, $\delta Q_{\h{t}^{a}} = (1/8\pi G)\, \kappa_{\l} \,\delta A - \Omega \,\delta Q_{\varphi}$.\vskip0.1cm

2. The emergence of two distinct first laws may seem surprising at first. But this is in fact a general feature of the WIH framework, where we have an infinite family of first laws, each associated with a (so-called ``permissible'') vector field  that generates horizon symmetries  \cite{afk,abl1,aabkrev}. Furthermore, there is an interesting interplay with the Hamiltonian theory: a first law emerges if and only if the 1-parameter family of diffeomorphisms generated by these vector fields induces a Hamiltonian flow on the covariant phase space (of all solutions to field equations that admit a WIH as a boundary). 

The salient differences in the two distinct first laws we discussed are the following: (i)  the null normal $\ul\l^{a}$ used in the first version is distinct from the null normal $\l^a$ used in the second. They are proportional to each other on $\scriml$ and they both vanish on $\izl$. However, the proportionality factor varies from one Kerr-de Sitter space-time to another, whence $\kappa_{\ul\l} \not= \kappa_{\l}$.  (ii) In the first version, mass $M$ is the charge associated with $\ub{t}^{a}$ (which is null on $\scriml$) and evaluated using fields at $\scriml$. In the second version, the charge $Q_{\h{t}^{a}} $ is associated with the vector field $\h{t}^{a}$ (which is space-like on $\scriml$) and evaluated using fields on $\scri^{\pm}$ \cite{abk1}. (iii) Finally, the angular velocities  --$\Omega_{c}$ in the first version and $\h\Omega_{c}$ in the second version-- are also different. \vskip0.1cm

3.  General space-times in the class $\CLi$ considered in this paper  do not admit \emph{any} Killing field. Yet, as we saw in section \ref{s5},  structure naturally available on $\scriml$ enables us to introduce a notion of mass (and an 
angular momentum vector) because $\scriml$ is analogous to $\scrimz$ in the asymptotically flat case. The fact that there is also a first law is an added bonus arising from the fact that,  $\scriml$ is also a WIH.

\section{Miscellaneous Issues}
\label{a2}

In this Appendix we introduce the Newman-Penrose tetrads and specify the corresponding components of various geometric fields used in the main text; prove a key identity (\ref{repsi2}) used in section \ref{s5}; and discuss conserved charges associated with the generators of the symmetry group $\G$.\bigskip

\centerline{\it  The Newman-Penrose tetrads} 
\bigskip

Let $\l^{a}$ denote a null normal to $\scrimr$. Then, given any 2-sphere cross-section $C$ of $\scrimr$ we introduce three vector fields $n^{a}, m^{a}$ and $\b{m}^{a}$ on $C$ to obtain a Newman Penrose null tetrad: $n^{a}$ is the other null normal to $C$ satisfying $g_{ab} \l^{a}n^{b} =-1$; $m^{a}$ is a complex null vector field tangential to  $C$;\, and $\b{m}^{a}$, its complex conjugate, such that $g_{ab}m^{a}\bar{m}^{b} =1$. Thus, the only non-zero scalar products between these tetrad vectors are $\l^{a}n_{a}$ and $m^{a}\b{m}_{a}$. Generally we need these tetrad vectors only on $C$ but they can also be extended away from $C$ by demanding that they be parallel transported along $\l^{a}$. Occasionally we specialize $\l^{a}$ so that it belongs to the canonical equivalence class of normals $[\lo^{a}]$ on $\scrimr$ (where $\lo^{a} \sim \lo^{\prime\,a}$ iff $\lo^{\prime\, a} = c \lo^{a}$ where $c$ is a positiver constant),  or to the canonical null normal $\ul\l^{a}$ on $\scriml$, selected by the cross-section $\izl$ and the normalization condition (motivated by the Kerr-de Sitter solution).  

Geometrical fields we used that refer to this null tetrad are: the intrinsic 2-metric and the area 2-form on $C$:
\be \b{q}_{ab} = 2m_{(a}\b{m}_{b)} \qquad {\rm and} \qquad  \b{\epsilon}_{ab}  = 2m_{[a} \b{m}_{b]}\, ;  \ee
the expansion and the shear of $\l^{a}$ and $n^{a}$
\be \Theta_{(\l)} = \b{q}^{ab} \nabla_{a} \l_{b}  (=:  -2\rho); \qquad {\rm and}\qquad   \Theta_{(n)} = \b{q}^{ab} \nabla_{a} n_{b}  (=: 2\mu)\, ; \label{spincoeff1} \ee
\be \sigma^{(\l)}_{ab} = (\b{q}_{a}^{c}\, \b{q}_{b}^{d}\, -\f{1}{2} \b{q}_{ab}\,\b{q}^{cd})  \nabla_{c} \l_{d} \,\, (=: -\sigma \b{m}_{a} \b{m}_{b}); \qquad 
\sigma^{(n)}_{ab} = (\b{q}_{a}^{c}\, \b{q}_{b}^{d}\, -\f{1}{2} \b{q}_{ab}\,\b{q}^{cd})  \nabla_{c} n_{d}\,\,  (=: \lambda m_{a}m_{b})\, ;  \label{spincoeff2} \ee
and six components of the Weyl tensor, given by
\be \Psi_{0} = C_{abcd}\, \l^{a} m^{b}\l^{c} m^{d},  \qquad \Psi_{1} = C_{abcd}\, \l^{a} m^{b}\l^{c} n^{d},  \quad {\rm and} \quad \Psi_{2} = C_{abcd}\, \l^{a} m^{b}\b{m}^{c} n^{d} . \ee
In Eq. (\ref{spincoeff1})  and (\ref{spincoeff2}) the scalars  ($\rho, \, \mu,\, \sigma,\, \lambda$)  in parenthesis refer to the commonly used Newman-Penrose notation for spin coefficients. Finally,  because $\scrimr$ is a non-expanding horizon (NEH), $\Psi_{0}$ and $\Psi_{1}$ vanish identically and the real and imaginary parts of $\Psi_{2}$, 
\be \Re\, \Psi_{2} = \f{1}{2}\, C_{abcd} \l^{a}n^{b} \l^{c} n^{d} \qquad {\rm and} \qquad  \Im\,\Psi_{2} = \f{1}{2}\, {}^{\star}\!C_{abcd} \l^{a}n^{b} \l^{c} n^{d}  \ee
are insensitive to the choice of the null normal $\l^{a}$ and $n^{a}$ to the cross section $C$.   \bigskip\bigskip \goodbreak  

\centerline{\it Derivation of Eq. (\ref{repsi2})}
\bigskip

Fix a space-time $(M, g_{ab})$ and a 2-dimensional space-like sub-manifold $S$  in $M$. Denote by $\b{q}_{ab}$ the intrinsic metric on $S$. There is a general identity that relates the 4-dimensional curvature $R_{abcd}$ of $g_{ab}$ to the intrinsic curvature of $S$ which is completely determined by its scalar curvature ${}^{2}\R$. This is the 2+2 analog of the more familiar Gauss equation that relates curvature of $g_{ab}$ with that  of the induced metric on a 3-dimensional submanifold, which leads to the familiar Hamiltonian constraint of general relativity.

Let $V^{a}$ be a vector field in $M$ that is tangential to $S$. Then, the action of the intrinsic (torsion-free) derivative operator $\b{D}_{a}$ on $S$, compatible with $\b{q}_{ab}$ is related to the action of the (torsion-free) derivative operator $\nabla_{a}$ on $M$, compatible with $g_{ab}$ via: \, $D_{a}V_{b} = q_{a}^{m}\, q_{b}^{n}\, \nabla_{m} V_{n}$. Using this fact and the definition of the curvature tensor, we can relate the Riemann tensor ${}^{2}\R_{abcd}$ of $\b{q}_{ab}$ with the Riemann tensor $R_{abcd}$ of $g_{ab}$ and the extrinsic curvatures of $S$ in $M$. These extrinsic curvatures can be expressed conveniently using any two null normals $\l^{a}$ and $n^{a}$ to $S$ such that $g_{ab}\l^{a} n^{a} =-1$. Then the extrinsic curvature terms can be expressed in terms of the shear and expansion of the null vectors $\l^{a}$ and $n^{a}$ and one obtains:
\be \b{q}^{ac}\, \b{q}^{bd}\, R_{abcd} = {}^{2}\R \,+\,\Theta_{(n)} \Theta_{(\l)} \, - 2 \sigma^{(n)}_{ab} \sigma^{(\l)}_{cd} \,
\b{q}^{ac} \b{q}^{bd}\, . \ee
We can now decompose the 4-dimensional Riemann tensor in terms of its Weyl and Ricci parts to simplify the left side:
\be \b{q}^{ac}\, \b{q}^{bd}\, R_{abcd} = -2 C_{abcd} \l^{a}n^{b}\l^{c}n^{d} + 2 G_{ab} \l^{a}n^{b} - \f{1}{3} R \ee
where as usual $G_{ab}$ denotes the Einstein tensor.  The last two equations are just differential geometric identities that hold on \emph{any} space-like 2-manifold $S$ in \emph{any} 4-dimensional space-time $(M,g_{ab})$.  Let us now use Einstein's equation $G_{ab} + \Lambda g_{ab} = 8\pi G T_{ab}$ to arrive at an equation that relates the intrinsic curvature ${}^{2} \R$ of $\b{q}_{ab}$ to $\Re\,\Psi_{2}$, $\Lambda$, $T_{ab}$ and the extrinsic curvatures:
\be {}^{2}\R = -4 \Re\Psi_{2} +\f{2}{3} \Lambda \,+\, 8\pi G (2 T_{ab}\l^{a} n^{b} + \f{1}{3} T)\, +\,  2\sigma^{(n)}_{ab} \sigma^{(\l)}_{cd}\,\b{q}^{ac}\b{q}^{bd} -\, \Theta_{(n)} \Theta_{(\l)}  \label{general} \ee
Note that the right side is insensitive to the choice of null normals $\l^{a}$ and $n^{a}$ to $S$ so long as they satisfy 
$\l^{a} n^{b} g_{ab} = -1$. This local equality holds for any 2-dimensional space-like surface $S$ in a solution to Einstein's equation with a cosmological constant $\Lambda$. Let us now restrict $S$ to be a cross-section of $\scrimr$. Because $\scrimr$ is an NEH, $\Theta_{(\l)} =0$ and $\sigma^{(\l)}_{ab} =0$. Hence the last two terms in  (\ref{general}) vanish and we obtain  Eq (\ref{repsi2}) used in the main text.\bigskip \bigskip

\centerline{\it The symmetry group $\G$ on $\scrimr$ and conserved charges}
\bigskip

In sections \ref{s3} and \ref{s4} we considered $\Mr$ as well as  $\Ml$ as portions of space-time of interest. Their past boundaries are $\scrimr$ and $\scriml$, respectively. The symmetry group $\G$ of $\scrimr$ is infinite dimensional. However, that of the portion $\scriml$ --or its complement,\,\, $\scrimr\setminus \scriml$--  is just a seven dimensional subgroup $\G_{7}$ of $\G$. In order to make contact with $\scrimz$ in the asymptotically flat case, in section \ref{s5} we focused on $\scriml$ and introduced the charges $E_{t}$ corresponding to the time-translation subgroup $\T_{1}$ of $\G_{7}$, and $Q_{K}$ corresponding to the Lorentz subgroup $\Lor$. They arose as Hamiltonians generating the action of these groups on the covariant phase space $\Gcov$, tailored to the natural WIH structure induced on $\scriml$  by the null normals $[\l^{a}]$ (selected by $\izl$). We will now return to $\scrimr$ and seek charges $Q_{\xi}$ associated with the generators $\xi^{a}$ of $\G$. 

Recall from Eq. (\ref{xi}) that, if we choose a fiducial null normal $\lo^{a}$ in the equivalence class $[\lo^{a}]$ that $\scrimr$ is naturally equipped with, and an affine parameter $\vo$ of this $\lo^{a}$, then any $\xi^{a}$ in the Lie algebra $\LG$ of $\G$ can be expressed as
\be \xi^{a} = \big[ f(\theta,\varphi) + \kappa \vo \big] \lo^{a} + \bar{K}^{a}   = V^{a} + \b{K}^{a}\label{xi3}\, .\ee
Here, in the first step, $\kappa$ is a constant, and $\b{K}^{a}$ is tangential to the $v_{o} = {\rm const}$ cross-sections and a conformal Killing field of the round 2-sphere metrics $\mathring{\b{q}}_{ab}$ thereon, and in the second step  we have simply grouped together the first two vector fields which are vertical. Recall that on $\scriml$, we have a canonical foliation. If we choose the affine parameter $\vo$ so that $\vo = {\rm const}$ 2-spheres used in (\ref{xi3}) are the leaves of this preferred  foliation, then the vector fields $(\kappa \vo) \lo^{a} + \bar{K}^{a}$ span a sub-Lie-algebra $\LG_{7}$ of $\LG$ that we used  to obtain the charges $E_{t}$ and $Q_{K}$ (in sections \ref{s5.2.2} and \ref{s5.2.3}). We will now introduce a natural extension of that procedure to general $\xi^{a}$ of the form (\ref{xi3}).

Because we are now interested in $\scrimr$ as a whole, let us drop reference to $\izl$ (and therefore to $\scriml$) and let the foliation be general. Consider the rotation 1-form $\omegaz_{a}$ defined by $D_{a} \lo^{b} = \omegaz_{a} \lo^{b}$ (which is associated with the full equivalence class $[\lo^{a}]$ since it is insensitive to constant rescalings of $\lo^{a}$). Since $\lo^{a}$ provides an extremal WIH structure on $\scrimr$, we have: $\omegaz_{a} \lo^{a} = 0$ and $\Lie_{\lo}\, \omegaz_{a}=0$.  Therefore $\omegaz_{a}$ is the pull-back to $\scrimr$ of a 1-form $\mathring{\t{\omega}}_{a}$ on the space $\t\scri^{-}_{\rm Rel}$  of integral curves of $\lo^{a}$. By the very definition of $\lo^{a}$,  the 1-form $\mathring{\t{\omega}}_{a}$ is divergence-free on $\t\scri^{-}_{\rm Rel}$ (with respect to the metric $\t{q}_{ab}$ thereon). Hence, the pull-back $\mathring{\b\omega}_{a}$ of $\omegaz_{a}$ to the leaves of our foliation is also divergence-free. Therefore we repeat the procedure used in section \ref{s5.2.3} to define $Q_{\b{K}}$. We first note that, being a vector field tangential to the 2-spheres $\vo ={\rm const}$, we can expand $\b{K}^{a}$ as
\be \b{K}^{a} = \b\epsilon^{ab} \b{D}_{b} \b{f}\, +\, \b{q}^{ab} \b{D}_{b} \b{g}  \label{Kbar}\ee
for some functions $\b{f} (\theta,\phi) $ and $\b{g}(\theta,\phi)$, where $\b{q}_{ab}$ and $\b\epsilon_{ab}$ are the pull-backs to the leaves of the foliation of the physical metric $q_{ab}$ and the area 2-form $\epsilon_{ab}$ on $\scrimr$.%
\footnote{Note that each $\xi^{a}$ admits a natural projection $\t{K}^{a}$ to the 2-sphere $\t{S}$ of generators of $\scrimr$ since $\Lie_{\lo} \xi^{a} \propto \lo^{a}$. The natural diffeomorphism between $\t{S}$ and any  $\vo = {\rm const}$ 2-sphere sends $\b{K}^{a}$ to $\t{K}^{a}$ and vice versa. Therefore we can express $\t{K}^{a}$ as $\t{K}^{a} = \t\epsilon^{ab} \t{D}_{b} \t{f}\, +\, \t{q}^{ab} \t{D}_{b} \t{g}$, and use pull-backs of $\t{f}$ and $\t{g}$ as $\b{f}$ and $\b{g}$ in (\ref{Kbar}). Then we have $\Lie_{\lo} \b{f} =0$ and $\Lie_{\lo} \b{g} =0$, whence $\b{f}, \b{g}$ are functions only of $(\theta,\phi)$.}
 Following the procedure used in section \ref{s5.2.3} we are led to express $Q_{\b{K}}$ as an integral over a leaf $C$ of the foliation:
\ba Q_{\b{K}}  &=& - \f{1}{8\pi G} \oint_{C} \mathring{\b\omega}_{a} \b{K}^{a}\, \rmd^{2}V \nonumber \\
&=&  \f{1}{8\pi G} \oint_{C} \b{f}\, \b\epsilon^{ab} \b{D}_{a} \mathring{\b\omega}_{b} \, \rmd^{2}V \nonumber \\
&=& - \f{1}{4\pi G} \oint_{C} \b{f}\, \Im\, \Psi_{2} \, \rmd^{2}V \, \, , \label{QbarK}\ea
where in the second step we have  used (\ref{Kbar}) and carried out an integration by parts, and in the third step used (\ref{impsi2}). Although we have expressed $Q_{K}$ as an integral over a 2-sphere $v=v_{o}$, the final result is independent of this choice.  Indeed, since $\xi^{a}$ projects down unambiguously to the base space $\t{S}$, $Q_{\b{K}}$ can be expressed entirely using an integral on the base space $\t{S}$ without reference to any foliation at all. The angular momentum charges on $\scrimr$ are the same as those we obtained in section \ref{s5.2.3} on $\scriml$.

Let us next consider the vertical part $V^{a} =\big(f(\theta,\phi) + \kappa \vo\big)\, \lo^{a}$ of $\xi^{a}$. To begin with let us suppose that $\kappa$ is non-zero.  Since $V^{a}$ is a null normal to $\scrimr$ with surface gravity $\kappa$, from now on we will replace $\kappa$ with $\kappa_{V}$. Since $\kappa_{V}$ is a non-zero constant, it follows that $V^{a}$ endows $\scrimr$ with the structure of a non-extremal WIH. (As expected, $V^{a}$ vanishes precisely at one cross section of $\scrimr$, given by $\vo= - (1/\kappa_{V})\, f(\theta,\phi)$; it is a complete vector field on either side of this  cross-section; and is future directed on one side and past directed on the other.) Therefore, we can use the covariant phase space $\Gcov$ that is available for space-times admitting a non-extremal WIH as a boundary \cite{afk}. The issue then is whether the diffeomorphism generated by a space-time vector field preserves the symplectic structure; if it does, the Hamiltonian generating the corresponding canonical transformation would provide the charge $Q_{V}$. However, as has been explained in detail in the literature, there is a subtlety: we need to specify what we mean by the `same' vector field in different solutions of Einstein's equations that constitute $\Gcov$. This step can be carried out by specifying the surface gravity of the vector field $V^{a}$ as a function of the horizon area. Indeed, we used this strategy on $\scriml$ by encoding in surface gravity the `correct normalization' (which in turn was determined by taking the $\Lambda \to 0$ of the Kerr family).  One can argue that the same strategy should be used in the general case. Then,  the same arguments that were used in section \ref{s5.2.2} lead us to the Hamiltonian
\be E_{V} =  - \, \f{1}{8\pi G}\, \oint_{C}  R_{(c)} \,C_{abpq} \,\,\ul{l}^{a}\ub{n}^{b}\, V^{p}\, \ub{n}^{q} \, \rmd^{2}V \,\, 
\equiv    - \, \f{\kappa_{V}}{4\pi G \kappa_{\ul{\l}}}\, \oint_{C}  R_{(c)}  \Re\, \Psi_{2}\,  \rmd^{2}V  \label{EV} \ee
where, as before $\kappa_{\ul{\l}} = (1/2R_{(c)})\, \big(1 - 3 (R_{(c)}^{2}/\l^{2})\big)$.  (The only difference between (\ref{Et}) of section \ref{s5.2.2} and (\ref{EV}) is that $\l^{a}$ is now replaced by $V^{a}$.)

So far we have restricted ourselves to the vertical vector fields $V^{a}$ for which $\kappa_{V} \not=0$. However, because the charge $E_{V}$ is a linear map from the space of vertical vector fields $V$ to $\mathbb{R}$, $E_{V}$ of (\ref{EV}) admits a unique extension to all $V$. Suppose $V^{a} = \big(f(\theta,\phi) + \kappa \vo\big)\lo^{a}$ and $V^{\prime\,a} = \big(f^{\prime}(\theta,\phi) + \kappa \vo\big)\lo^{a}$, we have $E_{V} = E_{V^{\prime}}$, with $\kappa\not=0$. Then the above prescription can be applied to both $V^{a}$ and $V^{\prime\, a}$, whence by linearity we arrive at a rather surprising result that $E_{V}=0$ if $V^{a} = f(\theta,\phi) \lo^{a}$, i.e., if $V^{a}$ is any supertranslation. Therefore, from the above Hamiltonian perspective, supertranslations have to be regarded as `gauge transformations' and the space of genuine symmetries is then the quotient $\G/\S = \G_{7}$. 

This perspective is natural from the WIH framework where the Hamiltonian framework is based on \emph{non-extremal}  WIH structures.  On the other hand, $\scrimr$ is naturally endowed with an \emph{extremal} WIH structure through its equivalence class of null normals $[\lo^{a}]$ and there may well be other perspectives that emphasize the \emph{extremal} WIH structures. Indeed, as pointed out in Remark 3 at the end of section \ref{s5.2.1}, the notion of mass $M$ can be introduced using these extremal null normals.  The first equality in (\ref{EV}) suggests a natural strategy to define supermomenta. Suppose we could select a preferred $\ul{\lo}^{a} \in [\lo^{a}]$. Then, given a supertranslation $S^{a}= f(\theta,\phi) \,\ul{\lo}^{a}$, using Eq (\ref{EV}) as motivation we could set 
\ba  Q_{{S}}&=& \f{1}{8\pi G} \oint_{C}  R_{(c)}\, C_{abpq} \, \ul{\lo}^{a}\ul{\no}^{b}\, S^{p}\, \ul{\no}^{q} \, \rmd^{2}V \nonumber\\
&=&  \f{1}{8\pi G} \oint_{C} f(\theta, \phi) R_{(c)}\, C_{abpq} \, \ul{\lo}^{a}\ul{\no}^{b}\, \ul{\lo}^{p} \ul{\no}^{q} \,\rmd^{2}V\,\, \equiv\,\,   \f{1}{4\pi G} \oint_{C} f(\theta, \phi)\,  \Re\Psi_{2}\, \rmd^{2}V\, ,\label{smo} \ea
so that $f(\theta,\phi)$ serves as a `weighting function in the last step. Indeed, this is precisely how supermomentum is defined on $\scrimz$ in the asymptotically flat context (in absence of incoming radiation): Given any Bondi conformal frame, we  obtain a preferred null normal $\ul\lo^{a}$ (rather than an equivalence class $[\ul\lo^{a}]$) and supermomentum is defined precisely as the limit of (\ref{smo}) as  $C$ approaches a cross-section of $\scrimz$.  (This $\ul\lo^{a}$ is the limit to $\scrimz$ of a \emph{unit} time-translation.) Thus, if  there were a physically motivated and/or mathematically natural procedure to select a preferred $\ul\lo^{a}$, on $\scrimr$ we would at least have a candidate expression. We could then investigate if it arises as a Hamiltonian generating the canonical transformation induced by the supertranslation $S^{a}$. But for this strategy to work, we do need a preferred $\ul{\lo}^{a} \in [\lo^{a}]$. For, under a constant rescaling $\lo^{a} \to k \lo^{a}$, we have $\no^{a}\to (1/k) \no^{a}$ and $f(\theta,\phi) \to  (1/k) f(\theta,\phi)$,\, whence the right hand side of (\ref{smo}) would be multiplied by $1/k$, giving us a different value of $Q_{S}$ on the same $\scrimr$! Now, in the non-extremal case, we \emph{could} select a canonical $\ul\l^{a}\in [\l^{a}]$ by fixing its surface gravity. In the extremal case, this avenue is not available because $\kappa_{\lo} =0$ for all $\lo^{a} \in [\lo^{a}]$.  And  the shear and expansion of each $\lo^{a}$ also vanish because $\scrimr$ is an NEH. Thus, it seems difficult to select a canonical $\lo^{a} \in [\lo^{a}]$, i.e. to write down an unambiguous candidate expression for supermomentum on $\scrimr$.\smallskip

There is also a deeper conceptual obstruction to selecting a preferred $\ul\lo^{a} \in [\ul\lo^{a}]$. What principle would one use to `correctly normalize' $\ul\lo^{a}$ on $\scrimr$? On $\scriml$ we chose the `correctly' normalized  $\ul\l^{a}$ by making appeal to the $\Lambda\to 0$ limit, in which a neighborhood of $\scriml$ of the Kerr-de Sitter family becomes a neighborhood of $\scrimz$ of the Kerr solution, and we know what the correct normalization is for the time-translation Killing field in the Kerr space-time. As we saw in section \ref{s3.2},  already for the Schwarzschild-de Sitter family, full $\scrimr$ does not have a well-defined limit as $\Lambda\to 0$.%
\footnote{ Following considerations suggest that this will happen more generally. In the $\Lambda=0$ case, the asymptotic region of the physical space-time is the intersection of the causal past of $i^{+}$ with the causal future of $i^{-}$. In the $\Lambda>0$ case, this intersection is just $\Ml$, whose past outer boundary is $\scriml$ and future outer boundary is $\scripl$. In the limit $\Lambda \to 0$, they will tend to $\scrim_{o}$ and $\scrip_{o}$ respectively. Thus, as in de Sitter space-time discussed in section \ref{s3.1} and Schwarzschild-de Sitter space-time discussed in section \ref{s3.2}$, \scrimr \setminus 
\scriml$ will simply disappear in the limit.}
Consequently, there is no guidance as to what the correct normalization of $\lo^{a}$ should be. Indeed, if supertranslations $S^{a}$ were to be  regarded as genuine symmetries of $\scrimr$, at least in the Kerr-de Sitter family one would expect them to tend to a symmetry of $\scrimz$ in the limit $\Lambda\to 0$. But this does not seems possible because: (i) for each $\Lambda >0$ the supertranslations fail to leave $\scriml$ invariant; (ii) the expression of $S^{a}$ makes no reference to $\Lambda$; and (iii)  $\scrimz$ is the limit of $\scriml$. 
\bigskip

\end{appendix}
\pagebreak


\begin{thebibliography}{99}

\bibitem{abk1} A.~Ashtekar, B.~Bonga and A.~Kesavan, Asymptotics with a positive 	
	cosmological constant: I. Basic framework, Class. Quant. Grav. \textbf{32}, 025004 (2015). 

\bibitem{bondi} H.~Bondi, M.~van der Burg, and A.~Metzner, Gravitational waves in genera1 relativity VII. Waves from axi-symmetric isolated systems, Proc. R. Soc. (London) A\textbf{269}, 21 (1962);\\
H.~Bondi, Radiation from an isolated system, In: \textit{Proceedings on theory of gravitation}, edited by L.~Infeld (Gauthier-Villars, Paris, PWN Editions Scientific de Pologne, Warszawa and Pergamon Press, Oxford, 1964).

\bibitem{sachs} R.~K.~Sachs, Gravitational waves in general relativity VIII. Waves in asymptotically flat space-times Proc. R. Soc. (London) A \textbf{270}, 103 (1962).

\bibitem{rp1} R.~Penrose, Zero rest mass fields including gravitation: asymptotic behavior, Proc. R. Soc. (London) A 284, 159-203 (1965);\\
R.~Penrose and E.~T.~Newman, New conservation laws for zero rest mass fields in asymptotically flat space-times, Proc. R. Soc. (London) A\textbf{305}, 175-204 (1968). 

\bibitem{kp} P.~Krtou\v{s} and J.~Podolsk\'y, Asymptotic directional structure of radiative fields in spacetimes with a cosmological constant, Class. Quantum Grav. \textbf{21} R233-R273 (2004).

\bibitem{rp2}  R.~Penrose, On cosmological mass with positive $\Lambda$, Gen. Rel. Grav. \textbf{43}, 3355-3366  (2011).

\bibitem{abk2} A.~Ashtekar, B.~Bonga and A.~Kesavan, Asymptotics with a positive cosmological constant: II.  Linear fields on de Sitter space-time, Phys. Rev. D\textbf{92}, 044011 (2015).

\bibitem{abk3} A.~Ashtekar, B.~Bonga and A.~Kesavan, Asymptotics with a positive cosmological constant: III. The quadrupole formula, Phys. Rev. D\textbf{92}, 1

\bibitem{ae}A.~Einstein, N\"aherungsweise Integration der Feldgleichungen der Gravitation,  Sitzungsberichte der K\"{o}niglich Preussischen Akademie der Wissenschaften, Berlin, (1916);\\
 \"{U}ber Gravitationswellen,  Sitzungsberichte der K\"{o}niglich Preussischen Akademie der Wissenschaften, Berlin, (1918). 
 
 \bibitem{abklett} A.~Ashtekar, B.~Bonga and A.~Kesavan, Gravitational waves from isolated systems: Surprising consequences of a positive cosmological constant, Phys. Rev. Lett. \textbf{116}, 051101 (2016). 

\bibitem{bbthesis} B. ~Bonga,   Gravitational radiation with a positive cosmological constant, ph. D. Dissertation, Penn State (2017) p 102-103;  https://etda.libraries.psu.edu/catalog/14161bpb165

\bibitem{aa-yau}A.~Ashtekar, Geometry and physics of null infinity, In: \emph{One hundred years of general relativity}, edited by  L.~Bieri and S.~T.~Yau (International press, Boston, 2015), pp 99-122; \texttt{arXiv:1409.1800}. 

\bibitem{hf} H.~Friedrich, On the global existence and the asymptotic behavior of solutions to the Einstein-Maxwell-Yang-Mills equations, J. Diff. Geo. \textbf{34}, 275-345 (1991).

\bibitem{dcsk}D.~Christodoulou and S.~Klainnerman, \textit{The global non-linear stability of Minkowski space} (Princeton University Press, Princeton, 1993).

\bibitem{pced} P.~T.~Chrusciel and E.~Delay, Existence of non-trivial, vacuum, asymptotically simple space-times, Class. Quant. Grav. \textbf{19}  L71-L80 (2002). 
 
 \bibitem{lb}  L.~Bieri,  \textit{Extensions of the stability theorem of the Minkowski space in general relativity. Solutions of the Einstein vacuum equations.} (AMS, International Press, 2009); \\
An Extension of the stability theorem of the Minkowski space in general relativity,  Journal of Differential Geometry.  \textbf{86}.  no.1. (2010). 17-70.  

\bibitem{nz} N.~Zipser, \textit{Extensions of the stability theorem of the Minkowski space in general relativity. Solutions of the Einstein-Maxwell equations.} (AMS, International Press, 2009).

\bibitem{hr} H.~Ringstr\"om, \emph{On the Topology and Future Stability of the Universe} (Oxford University Press, Oxford, 2013).

\bibitem{afk} A.~ Ashtekar, S. ~Fairhurst, and B. ~Krishnan, Laws Governing isolated horizons:
inclusion of distortion, Phys. Rev. D\textbf{62}, 104025 (2000).

\bibitem{abl1} A. ~Ashtekar, C. ~Beetle and J. ~Lewandowski, Geometry of generic isolated horizons, Class.  Quant. Grav. \textbf{19}, 1195-1225 (2002).

\bibitem{aabkrev} A.~Ashtekar and B.~Krishnan, Isolated and dynamical horizons and their applications, Liv. Rev. (Rel) \textbf{7}:10  (2004). 

\bibitem{np} E.~T.~Newman and R.~Penrose, Note on the Bondi Metzner Sachs group, J. Math. Phys. \textbf{7} 863-870 (1965).

\bibitem{aa-radmodes} A.~Ashtekar, Radiative degrees of freedom of the gravitational field in exact general relativity, J. Math. Phys. \textbf{22}, 2885-2895 (1981).

\bibitem{aa-ropp} A. ~Ashtekar, Implications of a positive cosmological constant for general relativity, Rep. Prog. Phys \textbf{80}, 102901 (2017). 

\bibitem{hv} P. ~Hintz and A. ~Vasy, The global non-linear stability of the Kerr-de sitter family of black holes, arXiv:1606.04014 [math.DG],  (2016).

\bibitem{ib} I.~Booth, Black hole boundaries,  Can.J.Phys. \textbf{83}, 1073-1099 (2005). 

\bibitem{gj} E. ~Gourgoulhon and J.~L. ~Jaramillo, A 3+1 perspective on null hypersurfaces and isolated horizons, Phys. Rept. \textbf{423}, 159-294 (2006). 

\bibitem{aabk1} A. ~Ashtekar and B. ~Krishnan, Dynamical horizons: Energy, angular momentum,
fluxes and balance laws, Phys. Rev. Lett. \textbf{89}, 261101 (2002)

\bibitem{aabk2} A. ~Ashtekar and B. ~Krishnan, Dynamical horizons and their properties, Phys.
Rev. D\textbf{68}, 104030 (2003).

\bibitem{aepv} A. ~Ashtekar, J. ~Engle, T. ~Pawlowski and C. ~Van Den Broeck, Multipole moments of isolated horizons, Class. Quant. Grav. \textbf{21}, 2549-2570 (2004).

\bibitem{gt} A. ~Gomberoff and C. ~Teitelboim, de Sitter black holes with either of the two horizons as a boundary, Phys. Rev. D\textbf{67}, 104024 (2003).

\bibitem{cg} A. ~Corichi and A. ~Gomberoff, Black holes in de Sitter space: Mass, energies and entropy upper bounds, Phys. Rev. \textbf{69}, 064016 (2004).

\bibitem{hn} H. ~Nariai, On a new cosmological solution of Einstein's field equations of gravity, The science reports of Tohoku university, Series I, number 1,  (1951). Reprinted in Gen. Relativ. Grav. \textbf{31}, 963 (1999).

\bibitem{aank} A. ~Ashtekar and N. ~Khera, Non expanding horizons, the BMS group and black hole `hair' (in preparation)

\bibitem{abl2} A. ~Ashtekar, C. ~Beetle and J. ~Lewandowski, Mechanics of rotating isolated horizons,  Phys. Rev. D\textbf{64} 044016 (2001). 

\bibitem{jltp}  J.~Lewandowski and  T.~Pawlowski, Symmetric non-expanding horizons, Class. Quant. Grav.  \textbf{23} 6031 (2006).

\bibitem{cfp} V.~Chandrasekharan, E.~Flanagan and K.~Prabhu, Symmetries and charges of general relativity at null boundaries, JHEP \textbf{11}, 125 (2018).

\bibitem{adm} R.~Arnowitt, S.~Deser and C.~W.~Misner, The dynamics of general relativity, in \emph{Gravitation: An Introduction to Current Research}, edited by L.~Witten (1962, N.Y., Wiley),  and references therein.

\bibitem{bk} B.~ Krishnan, The spacetime in the neighborhood of a general isolated black hole, Class. Quant. Grav.  \textbf{29},  205006 (2012).

\bibitem{jlcl} J.~Lewandowski and C.~Li, Space-time near Kerr isolated horizon,  \texttt{arXiv:1809.0471}

\bibitem{abbott} L.~F.~Abbott and S. Deser, Stability of gravity with a positive cosmological constant, Nucl. Phys. B\textbf{195}, 76-96 (1982).

\bibitem{km} W.~R.~Kelly and D.~Marolf, Phase space of asymptotically de Sitter cosmologies, Class. Quant. Grav.  \textbf{29},  205013 (2012).

\bibitem{aarh} A.~Ashtekar and R.~O.~Hansen, Unified treatment of null and spatial infinity in general relativity. I. Universal structure, asymptotic symmetries and conserved quantities at spatial infinity, J. Math. Phys.  \textbf{19}, 1542-1566 (1978).

\bibitem{Kennefick:book} D.~Kennefick, Traveling at the speed of thought: Einstein and the quest for gravitational waves, (Princeton university press, Princeton (2007)).

\bibitem{ycb} Y. Four\`es Bruhat, Th\'eor\`em d'existence pour certains syst\`emes d'\'equations aux d\'eriv\'ees partielles non lin\'eaires,  Acta Mathematica \textbf{88},  141-225 (1952).

\bibitem{lc} T.~Levi-Civita, $ds^{2}$ einsteiniani in campi netwoniani, Rend. Acc. Nazl. Lincei \textbf{27}, 343 (1918).

\bibitem{aatd} A.~Ashtekar and T.~Dray, On the Existence of Solutions to Einstein's Equation with Non-Zero Bondi News, Commun. Math. Phys. \textbf{79}, 581-589 (1981).

\bibitem{num1} M.~Shibata, K.~Nakao, T.~Nakamura, and K.~Maeda,
Dynamical evolution of gravitational waves in asymptotically de Sitter spacetime, Phys. Rev. D\textbf{50}, 708-119 (1994).

\bibitem{num3} M.~Zilhao, V.~Cardoso, L.~Gualtieri, C.~Herdeiro, U.~Sperhake, and H.~Witek, Dynamics of black holes in de Sitter spacetimes, Phys. Rev. D\textbf{85}, 104039 (2012).  

\bibitem{schlue} V.~Schlue, Global results for linear waves on expanding Kerr and Schwarzschild de Sitter cosmologies, Commun. Math. Phys. \textbf{334} 977-1023 (2015);\\
Decay of the Weyl curvature in expanding black hole cosmologies, e-Print: arXiv:1610.0417


\bibitem{ew} E.~Witten, A new proof of the positive energy theorem,
    Commun. Math. Phys. \textbf{80}, 381 (1981).

\bibitem{bc} B.~Carter, in \emph{Les Astre Occulus}, edited by B. DeWitt and C. M. DeWitt (Gordon and Breach, N.Y., 1973);  Reproduced in Gen. Relativ. Gravit. \textbf{41}, 2873Ð2938 (2009);  pages 2915-17.

\bibitem{am} S.~Akcay and R.~A.~Matzner, The Kerr de Sitter universe,  Class.  Quant.Grav. \textbf{28}, 085012 (2011). 

\bibitem{gpp} G.~W. ~Gibbons, M.~J.~Perry and C.~N.~Pope,  The first law of thermodynamics in Kerr-Anti-de-Sitter black holes, Class. Quant. Grav. \textbf{22} 1503-1526 (2005).  


\end{thebibliography}
\end{document}